\documentclass[12pt]{JHEP3}

\usepackage{amsmath}
\usepackage{amsfonts}
\usepackage{mathrsfs}
\usepackage{cite}
\usepackage{epsfig}
\usepackage{picture,epic}

\newcommand{\beqs}{\begin{equation*}}
\newcommand{\beq}{\begin{equation}}

\newcommand{\eeqs}{\end{equation*}}
\newcommand{\eeq}{\end{equation}}

\newcommand{\beqas}{\begin{eqnarray*}}
\newcommand{\beqa}{\begin{eqnarray}}

\newcommand{\eeqas}{\end{eqnarray*}}
\newcommand{\eeqa}{\end{eqnarray}}

\newcommand{\eq}[2]{\begin{equation} #1 \label{#2} \end{equation}}

\newcommand{\eps}{\varepsilon}
\newcommand{\al}{\alpha}
\newcommand{\be}{\beta}
\newcommand{\ga}{\gamma}
\newcommand{\de}{\delta}

\newcommand{\ka}{\kappa}
\newcommand{\la}{\lambda}
\newcommand{\si}{\sigma}

\newcommand{\Ga}{\Gamma}
\newcommand{\De}{\Delta}

\newcommand{\La}{\Lambda}

\newcommand{\blist}{\begin{itemize}}

\newcommand{\elist}{\end{itemize}}

\providecommand{\href}[2]{#2}

\DeclareFontFamily{OT1}{rsfs}{}
\DeclareFontShape{OT1}{rsfs}{m}{n}{ <-7> rsfs5 <7-10> rsfs7 <10->rsfs10}{} 
\DeclareMathAlphabet{\mycal}{OT1}{rsfs}{m}{n}

\DeclareMathOperator{\extdm}{d}
\newcommand{\extd}{\extdm \!}

\newcommand{\eom}{EOM}

\newcommand{\volint}{\!\extd^3y}

\title{AdS$_3$/LCFT$_2$ -- Correlators in Cosmological Topologically Massive Gravity}

\author{Daniel Grumiller\\
           Institute for Theoretical Physics, 
           Vienna University of Technology,\\
           Wiedner Hauptstr. 8--10/136,
           A-1040 Vienna, Austria\\
           and\\ 
           Center for Theoretical Physics, Massachusetts Institute of Technology,\\
	   77 Massachusetts Ave, Cambridge, MA 02139, USA \\
           Email: \email{grumil@hep.itp.tuwien.ac.at}}

 \author{Ivo Sachs\\
          Arnold Sommerfeld Center for Theoretical Physics,\\
	  Theresienstrasse 37,
	  D-80333 Munich, Germany\\
          Email: \email{Ivo.Sachs@physik.uni-muenchen.de}}

\abstract{
For cosmological topologically massive gravity at the chiral point we calculate 
momentum space 2- and 3-point correlators of operators in the postulated dual CFT on the cylinder. These operators are sourced by the bulk and boundary gravitons. Our correlators are fully consistent with the proposal that cosmological topologically massive gravity at the chiral point is dual to a logarithmic CFT. In the process we give a complete classification of normalizable and non-normalizeable left, right and logarithmic solutions to the linearized equations of motion in global AdS$_3$.
}


\preprint{MIT--CTP 4079, LMU--ASC 45/09\\ TUW--09--13, ESI 2188}

\begin{document}

\section{Introduction}

Cosmological topologically massive gravity \cite{Deser:1982sv} (CTMG) is a 3-dimensional theory of gravity that exhibits gravitons \cite{Deser:1982vy,Deser:1982wh} 
and black holes \cite{Banados:1992wn}. Its action is given by
\eq{
S_{\textrm{\tiny CTMG}} = -\frac{1}{\ka^2}\,S_{\textrm{\tiny EH}} -\frac{1}{\ka^2}\, S_{\textrm{\tiny CS}}
}{eq:f1}
The Einstein--Hilbert action with negative cosmological constant $\La=-1/\ell^2$
\eq{
S_{\textrm{\tiny EH}} = \int\volint\sqrt{-g}\,\Big[R+\frac{2}{\ell^2}\Big]
}{eq:f2}
is supplemented by the Chern--Simons action for the Christoffel-connection
\eq{
S_{\textrm{\tiny CS}} = \frac{1}{2\mu}\,\int\volint\,\epsilon^{\la\mu\nu}\Ga^\rho{}_{\si\la}\Big[\partial_\mu\Ga^\si{}_{\rho\nu}+\frac23 \,\Ga^\si{}_{\kappa\mu}\Ga^\kappa{}_{\si\nu}\Big]
}{eq:f3}
If the coupling constant $\mu$ and the AdS radius $\ell$ satisfy the condition
\eq{
\mu\ell = 1
}{eq:f3.5}
the theory is called ``CTMG at the chiral point'' (CCTMG). The condition \eqref{eq:f3.5} is special because one of the central charges of the asymptotic isometry group vanishes, $c_L=0$, $c_R\neq 0$ \cite{Kraus:2005zm}. 

This observation together with the fact that CTMG supports asymptotically AdS solutions was the motivation for  Li, Song and Strominger to consider CTMG at the chiral point, dubbed ``chiral gravity'' \cite{Li:2008dq}. They conjectured that CTMG at the chiral point is dual to a chiral CFT. However, there are examples of CFTs that have vanishing left-moving central charge without being chiral, namely logarithmic CFTs (LCFTs), see \cite{Gurarie:1993xq,Flohr:2001zs,Gaberdiel:2001tr} and references therein. The defining property of a LCFT is that the Virasoro generator $L_0$ is not diagonalizable. For instance,
\eq{ 
L_0 \left(\begin{array}{c} \psi^{\rm log} \\ \psi^L
\end{array}\right) = \left(\begin{array}{c@{\quad}c}
2 & \frac12 \\
0 & 2
\end{array}\right) \left(\begin{array}{c} \psi^{\rm log} \\ \psi^L \end{array}\right)
}{eq:cg79} 
In the parlance of LCFT literature the mode $\psi^{\rm log}$ is the logarithmic partner of the mode $\psi^L$. Interestingly, precisely the form \eqref{eq:cg79} was found for CTMG at the chiral point \cite{Grumiller:2008qz}. The mode $\psi^L$ is the left-moving boundary graviton and its logarithmic partner $\psi^{\rm log}$ is essentially the bulk graviton, a propagating spin-2 excitation that is present for all (finite) values of $\mu$ and $\ell$ \cite{Carlip:2008jk,Grumiller:2008pr,Carlip:2008qh}. Moreover, it was shown that $\psi^{\rm log}$ is compatible with asymptotic AdS behavior \cite{Grumiller:2008qz}. This was confirmed independently in \cite{Grumiller:2008es,Henneaux:2009pw,Maloney:2009ck,Skenderis:2009nt}. For additional recent literature on CTMG cf.~e.g.~\cite{Hotta:2008yq,Li:2008yz,Park:2008yy,Sachs:2008gt,Lowe:2008ye,Myung:2008ey,Carlip:2008eq,Lee:2008gta,Sachs:2008yi,Gibbons:2008vi,Anninos:2008fx,Giribet:2008bw,Strominger:2008dp,Compere:2008cv,Myung:2008dm,deHaro:2008gp,Stevens:2008hv,Deser:2008rm,Hotta:2008xt,Quevedo:2008ry,Oh:2008tc,Garbarz:2008qn,Kim:2008bf,Mann:2008rx,Blagojevic:2008bn,Nam:2009dd,Hellerman:2009bu,Sezgin:2009dj,Anninos:2009zi,Compere:2009zj,Hotta:2009zn,Anninos:2009jt,Carlip:2009ey,Chow:2009km,Becker:2009mk,Blagojevic:2009ek,Vasquez:2009mk,Duncan:2009sq,Andrade:2009ae,Miskovic:2009kr,Skenderis:2009kd,Ertl:2009ch,Afshar:2009rg}.

Given that \eqref{eq:cg79} is realized in CCTMG, it appears that the dual CFT is not chiral but logarithmic \cite{Grumiller:2008qz}, although a chiral CFT might be obtained as a consistent truncation \cite{Maloney:2009ck}. So far no good gravity duals for LCFTs are known, see \cite{Ghezelbash:1998rj,Myung:1999nd,Kogan:1999bn,Lewis:1999qv, MoghimiAraghi:2004ds}. 
If an AdS$_3$/LCFT$_2$ dictionary could be established, we can use CCTMG as a gravity dual for certain (strongly coupled) LCFTs, with potential applications in condensed matter physics where LCFTs are applied. It is thus of importance to provide further evidence for the proposal\footnote{This possibility was pointed out first by John McGreevy during a talk by Andy Strominger at MIT in May 2008, four days before the posting of Ref.~\cite{Grumiller:2008qz}.} \cite{Grumiller:2008qz} that the CFT dual to CCTMG, if it exists, is a LCFT.

The LCFT conjecture can be tested as follows: calculate correlators on the gravity side, relate them by the purported AdS$_3$/LCFT$_2$ correspondence to correlators on the CFT side and check if these correlators really have the properties as required by a LCFT. Conformal symmetry poses particularly stringent constraints on 2- and 3-point correlators \cite{Ginsparg:1988ui}, so if the conjecture is true then these correlators must have an essentially unique form in CCTMG.  On the other hand, if the conjecture is wrong then it is suggestive that some of the stringent LCFT constraints may be violated already at the level of 2- or 3-point correlators. The calculation of 2- and 3-point correlators on the gravity side is thus a major step towards establishing the LCFT conjecture. Such a check was carried out recently by Skenderis, Taylor and van Rees \cite{Skenderis:2009nt} for 2-point correlators. They found perfect agreement with the LCFT 2-point correlators, which supports the conjecture that CTMG at the chiral point is dual to a LCFT. 

In this paper we provide the basis for the calculation of arbitrary correlators on the gravity side. In particular, we construct all regular non-normalizable left, right and logarithmic modes in global coordinates in terms of elementary functions and show how these modes are organized in $SL(2,\mathbb{R})$ representations. We then plug these modes into the second and third variation of the action to calculate 2- and 3-point correlators. 
We find perfect agreement with the behavior expected from LCFT correlators. Thus, we corroborate the conjecture that CCTMG is dual to a LCFT.

This paper is organized as follows: In section \ref{se:2} we calculate the first three variations of the action \eqref{eq:f1}. In section \ref{se:lin} we discuss generic solutions of the linearized equations of motion, find all regular non-normalizable left, right and logarithmic modes in global coordinates, and discuss their properties. In section \ref{se:corr} we calculate on the gravity side 2- and 3-point correlators in CCTMG. We conclude with a discussion in section \ref{se:5}, where we address the status of CCTMG as a possible gravity dual to LCFTs and as a tentative toy model for quantum gravity. We mention also spin-offs and generalizations of our calculations.

Before starting we mention some of our conventions, which coincide with the conventions used in \cite{Grumiller:2008qz}. Our signature is $(-,+,+)$. The overall sign in front of the CTMG action \eqref{eq:f1} is irrelevant in the present work, but for sake of completeness we mention that we have chosen it such that black hole solutions have positive energy and graviton excitations negative energy \cite{Li:2008dq}. In three dimensions the Riemann tensor is determined uniquely from the Ricci tensor $R_{\mu\nu}=R^\si{}_{\mu\si\nu}$ as follows: $R_{\si\rho\mu\nu} = \big(R_{\si\mu}g_{\rho\nu} + R_{\rho\nu}g_{\si\mu} - R_{\si\nu}g_{\rho\mu} - R_{\rho\mu}g_{\si\nu}\big) - \frac12\,R\,\big(g_{\si\mu}g_{\rho\nu} - g_{\si\nu}g_{\rho\mu}\big)$. The sign of the Ricci tensor is defined by $R_{\mu\nu}=\partial_\si\Ga^\si{}_{\mu\nu}+\dots$ The Levi-Civita symbol is denoted by $\epsilon^{\al\be\ga}$ (its sign is fixed by $\epsilon^{\tau\phi\rho}=+1$), and the Levi-Civita tensor by $\varepsilon^{\al\be\ga}=\epsilon^{\al\be\ga}/\sqrt{-g}$. Our gravitational coupling constant $\ka$ is related to Newton's constant $G_N$ by $\ka^2=16\pi\,G_N$.

\section{Perturbative expansion of the action}\label{se:2}

In this section we calculate the first three variations of the CTMG action \eqref{eq:f1}. To fix our remaining notations we briefly review the first variation of the CTMG action \eqref{eq:f1}, modulo boundary terms, which will be taken into account in due course: 
\eq{
\de S_{\textrm{\tiny CCTMG}} = -\frac{1}{\ka^2}\,\int\volint\sqrt{-g}\,\de g^{\mu\nu} \Big[G_{\mu\nu} + \frac1\mu\,C_{\mu\nu}\Big] 
}{eq:f68}
Setting $\de S_{\textrm{\tiny CCTMG}}=0$ leads to the equations of motion ({\eom})
\eq{
G_{\mu\nu}+\frac1\mu\,C_{\mu\nu} = 0
}{eq:f37}
Here $G_{\mu\nu}$ is the modified Einstein tensor
\eq{
G_{\mu\nu}=R_{\mu\nu}-\frac12\, g_{\mu\nu} R - \frac{1}{\ell^2}\, g_{\mu\nu}
}{eq:f4}
and $C_{\mu\nu}$ is the Cotton tensor
\eq{
C_{\mu\nu}= \varepsilon_\mu{}^{\kappa\si}\nabla_\kappa \big(R_{\si\nu}-\frac14\, g_{\si\nu} R\big)
}{eq:f5}

We are interested in the second and third variation of the action with respect to metric variations of the form
\eq{
\de g_{\mu\nu} = h_{\mu\nu}\qquad \de g^{\mu\nu} = - h^{\mu\nu}
}{eq:f9}
Indices of $h$ are raised and lowered with the background metric $g$. In addition to the second variation of the metric
\eq{
\de^{(2)} g_{\mu\nu}=0\qquad \de^{(2)} g^{\mu\nu} = -\de h^{\mu\nu} = 2 h^\mu_\al h^{\al\nu}
}{eq:f12}
we are going to need the variation of various geometric quantities, which we collect in the following. The variation of the volume element is given by
$2\,\de\sqrt{-g} = \sqrt{-g}\, g^{\mu\nu} h_{\mu\nu} = \sqrt{-g}\, g_{\mu\nu} h^{\mu\nu}$.
The variations of the Riemann tensor 
\eq{
\de R^\al{}_{\mu\be\nu} = \nabla_\be \de\Ga^\al{}_{\mu\nu} - \nabla_\nu\de\Ga^\al{}_{\be\mu}
}{eq:f111}
and the Ricci tensor $\de R_{\mu\nu} = \de R^\al{}_{\mu\al\nu}$
are determined from the variation of the Christoffel connection. It is useful to have a formula valid for arbitrary variations of the Christoffel connection:
\eq{
\de^{(n)}\Ga^\rho{}_{\si\la} = \frac n2 \big(\nabla_\si h_{\la\kappa}+\nabla_\la h_{\si\kappa}-\nabla_\kappa h_{\si\la}\big)\,\de^{(n-1)} g^{\rho\kappa} 
}{eq:f13}
The variation of the Cotton tensor density is given by
\eq{
\de \big(\sqrt{-g}\,C_{\mu\nu}\big) = \sqrt{-g}\,\big(\de g_{\mu\la}\,C^\la{}_\nu\big) - g_{\mu\la}\,\epsilon^{\la\si\kappa} \big(\de(\nabla_\kappa R_{\si\nu})-\frac14\,(\de g_{\si\nu} \partial_\kappa R + g_{\si\nu}\partial_\kappa\,\de R)\big)
}{eq:f11a}
For our purposes we are going to need only the first and second variation of various quantities, including the second variation of the Ricci tensor
\eq{
\de^{(2)} R_{\mu\nu} = \nabla_\al \de^{(2)}\Ga^\al{}_{\mu\nu} - \nabla_\mu\de^{(2)}\Ga^\al{}_{\nu\al} + 2\de\Ga^\ka{}_{\mu\nu}\,\de\Ga^\la{}_{\ka\la} - 2\de\Ga^\ka{}_{\la\mu}\,\de\Ga^\la{}_{\ka\nu} 
}{eq:f14}
and of the Cotton tensor density
\eq{
\de^{(2)}\big(\sqrt{-g}\,C_{\mu\nu}\big) = 2 \,\de g_{\mu\la}\,\de \big(\sqrt{-g}\,C^\la{}_\nu\big) - g_{\mu\la}\,\epsilon^{\la\si\kappa}\,\de^{(2)}\big(\nabla_\kappa R_{\si\nu}\big)  +  \frac14\,\Xi_{\mu\nu}
}{eq:f14a}
where
$\Xi_{\mu\nu} = 2g_{\mu\la}\,\epsilon^{\la\si\kappa}\,\de g_{\si\nu}\,\partial_\kappa\,\de R + g_{\mu\la}g_{\nu\si}\,\epsilon^{\la\si\kappa}\,\partial_\kappa\,\de^{(2)}R$.  
In the formula for the second variation of the Cotton tensor density \eqref{eq:f14a} we have separated two contributions in the tensor $\Xi_{\mu\nu}$, because they vanish either after contraction with a symmetric tensor like $\de g^{\mu\nu}$ or due to background and gauge fixing identities. We describe now these identities in detail.

\subsection{Post-variational identities}

The identities above are valid generically in three dimensions. Now we consider identities that can be used only after performing all variations.

The first set of such identities comes from assuming that the background metric is pure AdS
\eq{
\extd s^2 = g_{\mu\nu}\extd x^\mu \extd x^\nu = \ell^2\,\big(\extd\rho^2-\frac14\,\cosh^2{\!\!\rho}\, (\extd u+\extd v)^2 +\frac14\,\sinh^2{\!\!\rho}\,(\extd u-\extd v)^2\big)
}{eq:cg20}
and therefore maximally symmetric.
\eq{
R=-\frac{6}{\ell^2}\qquad\qquad R_{\mu\nu} = -\frac{2}{\ell^2}\, g_{\mu\nu} 
\qquad\qquad C_{\mu\nu} = 0
}{eq:f15}
The second set of identities comes from assuming that the linear fluctuations obey the transverse-traceless gauge conditions \cite{Liu:1998bu,Arutyunov:1999nw,Li:2008dq}:
\eq{
h_{\mu\nu} g^{\mu\nu} = 0 \qquad \nabla^\mu h_{\mu\nu} = 0
}{eq:f16}
Both sets together imply further useful identities, which we collect here:
\eq{
h^{\mu\nu} R_{\mu\nu} = 0 \qquad g^{\mu\nu} \de R_{\mu\nu} = 0 \qquad \delta R=0 \qquad \de\Ga^\al{}_{\nu\al} = 0 \qquad \de\sqrt{-g}=0
}{eq:f17}
The second derivative of the metric variation simplifies to
\eq{
\nabla_\kappa\nabla_\si h^\kappa_\la = \left[\nabla_\kappa,\nabla_\si\right] h^\kappa_\la = - \frac{3}{\ell^2}\,h_{\si\la}
}{eq:f18}
The variations of the Ricci tensor
$\de R_{\mu\nu} = \nabla_\al\de\Ga^\al{}_{\mu\nu} = -\frac12 \big(\nabla^2 h_{\mu\nu}+\frac{6}{\ell^2}\,h_{\mu\nu}\big)$
and of the modified Einstein tensor
\eq{
\de G_{\mu\nu} = -\frac12 \big(\nabla^2 h_{\mu\nu}+\frac{2}{\ell^2}\,h_{\mu\nu}\big)
}{eq:f20}
also simplify considerably. The second variation of the modified Einstein tensor contracted with the fluctuation $h^{\mu\nu}$ simplifies to 
\eq{
h^{\mu\nu}\,\de^{(2)} G_{\mu\nu} = h^{\mu\nu}\,\de^{(2)} R_{\mu\nu} = h^{\mu\nu}\nabla_\al\,\de^{(2)}\Ga^{\al}{}_{\mu\nu}-h^{\mu\nu}\nabla_\mu\,\de^{(2)}\Ga^\al{}_{\nu\al}-2h^{\mu\nu}\,\de\Ga^\ka{}_{\la\mu}\,\de\Ga^\la{}_{\ka\nu}
}{eq:f49}
Finally, the variation of the Riemann tensor can be expressed in terms of variations of the Einstein tensor and the metric.
\begin{multline}
\de R_{\si\rho\mu\nu} = \de G_{\si\mu}\,g_{\rho\nu} + \de G_{\rho\nu}\,g_{\si\mu} - \de G_{\si\nu}\,g_{\rho\mu} - \de G_{\rho\mu}\,g_{\si\nu} \\
- \frac{1}{\ell^2}\,\big(\de g_{\si\mu}\,g_{\rho\nu} + \de g_{\rho\nu}\,g_{\si\mu} - \de g_{\si\nu}\,g_{\rho\mu}- \de g_{\rho\mu}\,g_{\si\nu}\big)
\label{eq:f110}
\end{multline}

\subsection{Second variation of the action}\label{se:2.2}

It is convenient for later applications to denote the two variations of the metric by $\de g_{\mu\nu}$ and by $h_{\mu\nu}$, respectively. For our purposes we can neglect terms that vanish on the background. 
The second variation of the action
\eq{
\de^{(2)} S_{\textrm{\tiny CCTMG}} = -\frac{1}{\ka^2}\,\int\volint\sqrt{-g}\,\de g^{\mu\nu} \Big[\de G_{\mu\nu}(h) + \frac1\mu\,\de C_{\mu\nu}(h)\Big] =  \frac{1}{\ka^2}\,\int\volint\,\de {\cal L}^{(2)} 
}{eq:f69}
leads to the linearized {\eom} for the fluctuations $h_{\mu\nu}$ in the gauge \eqref{eq:f16}
\eq{
\de G_{\mu\nu}(h) +\frac{1}{\mu}\, \de C_{\mu\nu}(h) = -\frac12\,\big(\nabla^2 +\frac{2}{\ell^2}\big)\big(h_{\mu\nu}+\frac{1}{\mu}\,\varepsilon_\mu{}^{\al\be}\nabla_\al h_{\be\nu}\big) = 0
}{eq:f21}
Defining the mutually commuting first order operators
\eq{
\big({\cal D}^M\big)_\mu^\be = \de_\mu^\be + \frac{1}{\mu}\,\varepsilon_\mu{}^{\al\be}\nabla_\al \qquad \big({\cal D}^{L/R}\big)_\mu^\be = \de_\mu^\be \pm \ell \,\varepsilon_\mu{}^{\al\be}\nabla_\al
}{eq:f22}
the  linearized {\eom} can be reformulated as \cite{Li:2008dq}
\eq{
 ({\cal D}^M\,\de G(h))_{\mu\nu} = ({\cal D}^M{\cal D}^L{\cal D^R} h)_{\mu\nu} = 0
}{eq:22.5} 
Here we have taken advantage of the identities $\de G_{\mu\nu}(h) +\frac{1}{\mu}\, \de C_{\mu\nu}(h) = ({\cal D}^M\,\de G(h))_{\mu\nu}$ and
\eq{
2 \ell^2 \,\de G_{\mu\nu}(h) = ({\cal D}^L {\cal D}^R h)_{\mu\nu} 
}{eq:f22a}
A mode annihilated by ${\cal D}^M$ (${\cal D}^L$) [${\cal D^R}$] is called massive (left-moving) [right-moving] and is denoted by $\psi^M$ ($\psi^L$) [$\psi^R$].

\subsection{Third variation of the action}\label{se:2.3}

In analogy with section \ref{se:2.2}  we parameterize the third variation in terms of three independent fields $\de g$, $h$ and $k$. The third variation of the action is then given by
\eq{
\de^{(3)} S_{\textrm{\tiny CCTMG}} = -\frac{1}{\ka^2}\,\int\volint\sqrt{-g}\,\de g^{\mu\nu} \Big[\de^{(2)} R_{\mu\nu}(h,k) + \frac1\mu\,\de^{(2)} C_{\mu\nu}(h,k)\Big]
}{eq:f70}
where we have used \eqref{eq:f49}. In \eqref{eq:f70} and all formulas below we always assume that the metric is given by the AdS background \eqref{eq:cg20}, all fluctuations around it are transverse-traceless \eqref{eq:f16}, and solve the linearized {\eom} \eqref{eq:f21}. Therefore, we exploit all the identities derived above to simplify expressions like the second variation of the Cotton tensor density \eqref{eq:f14a}.
The second variation of the Cotton tensor
\eq{
\de^{(2)} C_{\mu\nu}(h,k) = h_{\mu\la} \,g^{\la\si}\, \de C_{\si\nu}(k) + k_{\mu\la} \,g^{\la\si}\, \de C_{\si\nu}(h) - \varepsilon_\mu{}^{\si\kappa} \de^{(2)} \big(\nabla_\kappa R_{\si\nu}\big)(h,k)
}{eq:f71}
can be re-expressed using the linearized {\eom} \eqref{eq:f21} and the second variation of the Ricci-tensor \eqref{eq:f14}:
\eq{
\de^{(2)} C_{\mu\nu}(h,k) =  \mu\,\De_{\mu\nu}(h,k) - \varepsilon_\mu{}^{\si\kappa} \nabla_\kappa \,\de^{(2)} R_{\si\nu}(h,k) 
}{eq:f53}
with
\eq{
\De_{\mu\nu}(h,k)=-\frac{1}{2\ell^2}\,k_\mu^\si\, \big({\cal D}^L{\cal D}^R h\big)_{\si\nu} + \frac{1}{2\mu\ell^2}\,\varepsilon_\mu{}^{\si\kappa}\,\de\Ga^\al{}_{\kappa\nu}(k)\,\big({\cal D}^L{\cal D}^R h\big)_{\si\al} + h \leftrightarrow k
}{eq:f76}
We have exploited the identity \eqref{eq:f22a} to bring $\De_{\mu\nu}(h,k)$ into the form above. The second variation of the Ricci tensor yields
\eq{
\de^{(2)}R_{\mu\nu}(h,k) = h^\al_\be \,\de R^\be{}_{\mu\nu\al}(k) +\frac14\big(2\nabla_\la h_\mu^\ka \nabla^\la k_{\ka\nu}-2\nabla_\la h_\mu^\ka \nabla_\ka k_\nu^\la+\nabla_\mu h_\la^\ka\nabla_\nu k_\ka^\la\big) 
 +h \leftrightarrow k
}{eq:f75}

The definition \eqref{eq:f22} of the linear operator ${\cal D}^M$ allows us to provide a convenient re-formulation of the third variation of the action:
\begin{align}
\de^{(3)} S_{\textrm{\tiny CCTMG}}(\de g,h,k) &= -\frac{1}{\ka^2}\,\int\volint\sqrt{-g}\,\de g^{\mu\nu} \Big[\big({\cal D}^M\,\de^{(2)} R(h,k)\big)_{\mu\nu} + \De_{\mu\nu}(h,k)\Big]\nonumber \\
& = \frac{1}{\ka^2}\,\int\volint\,\de{\cal L}^{(3)} 
\label{eq:f73}
\end{align}
With the formula for the variation of the Riemann tensor \eqref{eq:f110} and the definitions \eqref{eq:f22}, \eqref{eq:f76} and \eqref{eq:f75} we obtain explicitly:
\begin{multline}
\de {\cal L}^{(3)}=-\de g^{\mu\nu} \Big[\big({\cal D}^M\big)_\mu{}^\be \big(\frac14\,(2\nabla_\la h_\be^\ka \nabla^\la k_{\ka\nu}-2\nabla_\la h_\be^\ka \nabla_\ka k_\nu^\la+\nabla_\be h_\la^\ka\nabla_\nu k_\ka^\la)-\frac{1}{\ell^2}\,h_\be^\al k_{\al\nu}\big)\\
+\frac{1}{2\mu\ell^2}\,\varepsilon_\mu{}^{\sigma\kappa}\big(
\nabla_\sigma (k_\kappa^\alpha ({\cal D}^L {\cal D}^R h)_{\alpha\nu})-\delta\Gamma_{\nu\kappa\alpha}(k) ({\cal D}^L {\cal D}^R h)_\sigma^\alpha \big)+ h \leftrightarrow k \Big]\,\sqrt{-g}
\label{eq:f74}
\end{multline}
Our final result for the third variation of the action, \eqref{eq:f73} with \eqref{eq:f74}, requires the definitions  \eqref{eq:f13}, \eqref{eq:f22} but otherwise is explicit in the three variations $\delta g, h$ and $k$. If $h$ and $k$ are both linear combinations of only left- and right-moving modes the $\varepsilon$-term in the second line of \eqref{eq:f74} vanishes. 

\section{Linearized solutions}\label{se:lin}

According to the standard AdS/CFT dictionary CFT correlators are obtained upon insertion of non-normalizable solutions of the linearized {\eom} into variations of the action \cite{Aharony:1999ti}. We have considered the first three variations of the action in the previous section. In this section we find all solutions to the linearized {\eom} \eqref{eq:22.5}. We start by classifying them in section \ref{se:class}. We review normalizable modes, primaries and their descendants in section \ref{se:norm}. We then discuss the massive branch in section \ref{se:mass}, since the corresponding solutions encompass all other solutions as special (sometimes singular) limits. We construct explicitly all regular non-normalizable left and right modes in sections \ref{se:L}, and show how to obtain regular non-normalizable logarithmic modes in section \ref{se:log}. We also unravel their algebraic properties. For simplicity we set $\ell=1$ from now on.

\subsection{Classification of linearized solutions}\label{se:class}

The linearized {\eom} \eqref{eq:22.5} contain three mutually commuting first order operators \eqref{eq:f22}. If they are non-degenerate we can build the general solution from three branches: massive, left and right branches, whose modes are annihilated, respectively, by ${\cal D}^M$, ${\cal D}^L$, and ${\cal D}^R$. At the chiral point ${\cal D}^M$ and ${\cal D}^L$ degenerate with each other and we obtain instead the following three branches: logarithmic, left and right, where the logarithmic modes are annihilated by $({\cal D}^L)^2$ but not by ${\cal D}^L$.

For each branch the linearized solutions can be regular or singular at the origin $\rho=0$. A mode $\psi$ is called singular if at least one of its components diverges at $\rho=0$; clearly, perturbation theory breaks down for such a solution near $\rho=0$ since the AdS background metric remains bounded at the origin, and thus the linearized solution no longer is a small perturbation there. In the absence of point particles or black holes the singular modes should be discarded for consistency. Of main interest to us are therefore regular modes. 

The asymptotic (large $\rho$) behavior allows us to classify modes $\psi$ into normalizable and non-normalizable ones. This classification is very simple in Gaussian normal coordinates\footnote{Below we are not going to use Gaussian normal coordinates, but rather modes $\psi$ in transverse-traceless gauge. The associated coordinate transformations are very simple asymptotically. We are not going to provide them explicitly. See appendix \ref{app:bc} for a summary of boundary conditions.} for the perturbed metric $\bar g=g+\psi$, where $g$ is the background metric \eqref{eq:cg20}
\eq{
\big(g_{\mu\nu}+\psi_{\mu\nu}^{\textrm{\tiny GNC}}\big) \extd x^\mu\extd x^\nu = \extd\rho^2+\big(g_{ij}(x^k,\rho)+\psi_{ij}(x^k,\rho)\big)\extd x^i\extd x^j
}{eq:GNC}
For modes that allow a Fefferman--Graham expansion (all left and right modes)
\eq{
\psi_{ij}(x^k,\rho) = \psi_{ij}^{(0)}(x^k)\,e^{2\rho} + \psi_{ij}^{(1)}(x^k) \rho + \psi_{ij}^{(2)}(x^k) + \dots
}{eq:FG}
normalizability means $\psi_{ij}^{(0)}(x^k)=0$. The non-normalizable left and right modes have a non-zero leading term in the Fefferman--Graham expansion \eqref{eq:FG}, $\psi_{ij}^{(0)}\neq 0$. According to the AdS/CFT dictionary these non-normalizable modes act as sources for the operators associated with the left- and right-moving boundary gravitons. Massive modes in general are not compatible with the expansion \eqref{eq:FG}. The logarithmic modes discovered in \cite{Grumiller:2008qz} are compatible with the Fefferman--Graham expansion \eqref{eq:FG} with $\psi^{(0)}_{ij}=0$ and thus they are normalizable. One of the goals of this section is to find their non-normalizable counterparts, because they are needed for correlators involving insertions of logarithmic modes \cite{Skenderis:2009nt}.

\subsection{Normalizable modes, primaries and descendants}\label{se:norm}

Li, Song and Strominger exploited the $SL(2,\mathbb{R})_L\times SL(2,\mathbb{R})_R$ isometry algebra of the AdS$_3$ background \eqref{eq:cg20} in their construction of normalizable regular primaries for the massive, left and right branches. The normalizable modes are descendants of primaries 
with respect to the isometry algebra. The properties of the isometry algebra will also be useful for the non-normalizable modes constructed in section \ref{se:L}. We summarize now briefly the relevant formulas.

The $SL(2,\mathbb{R})_L$ generators read [we recall that $u=t+\phi$, $v=t-\phi$, see \eqref{eq:cg20}]
\begin{align}
L_0 &= i\partial_u \label{eq:cg21}\\
L_+ &= ie^{-iu}\,\Big(\frac{\cosh{2\rho}}{\sinh{2\rho}}\partial_u-\frac{1}{\sinh{2\rho}}\partial_v + \frac i2 \partial_\rho\Big)\label{eq:cg22}\\
L_- &= ie^{iu}\,\Big(\frac{\cosh{2\rho}}{\sinh{2\rho}}\partial_u-\frac{1}{\sinh{2\rho}}\partial_v - \frac i2 \partial_\rho\Big) \label{eq:cg23}
\end{align}
with algebra
\eq{
\big[L_0,L_{\pm}\big]=\pm L_{\pm}\,,\qquad \big[L_-,L_+\big]=2L_0
}{eq:cg24}
and quadratic Casimir
\eq{
L^2=\frac12 \,\big(L_- L_+ + L_+ L_-\big)-L_0^2\,.
}{eq:cg25}
The $SL(2,\mathbb{R})_R$ generators $\bar L_0$, $\bar L_+$, $\bar L_-$ satisfy the same algebra and are given by \eqref{eq:cg21}-\eqref{eq:cg23} with $u\leftrightarrow v$ and $L\leftrightarrow \bar L$. The equivalence
\eq{
\big({\cal D}^L{\cal D}^R\psi\big)_{\mu\nu}=0 \qquad\leftrightarrow\qquad(L^2+\bar L^2+2)\psi_{\mu\nu}=0
}{eq:EHop}
is useful to construct solutions of the linearized Einstein equations. Namely, for primaries $L_-\psi_0=0$, $\bar L_-\psi_0=0$ we can generate new solutions of the linearized {\eom} by acting on $\psi_0$ with the ladder operators $L_+$, $\bar L_+$. It should be noted that general solutions to the linearized {\eom} are not descendants of primaries. The latter correspond to modes that are regular and normalizable. Similar remarks apply to the logarithmic modes. Starting from the equivalence\footnote{The linearized equation in \eqref{eq:MGop} contains not only linearized solutions of CCTMG, but also linearized solutions of New Massive Gravity at a chiral point \cite{Bergshoeff:2009hq}. However, for the same reason that left and right modes do not mix, the logarithmic modes do not mix: CCTMG solutions always generate other CCTMG solutions when acting on them with ladder operators $L_\pm$, $\bar L_{\pm}$.}
\eq{
\big(({\cal D}^L)^2({\cal D}^R)^2\psi\big)_{\mu\nu}=0 \qquad\leftrightarrow\qquad(L^2+\bar L^2+2)^2\psi_{\mu\nu}=0
}{eq:MGop}
all regular normalizable logarithmic modes can be written as descendants of the logarithmic primary algebraically. We recall now in a bit more detail how this works.

Starting point is the separation Ansatz
\eq{
\psi_{\mu\nu}(h,\bar h)=e^{-ihu-i\bar h v}\left(\begin{array}{ccc}
F_{uu}(\rho) & F_{uv}(\rho) & F_{u\rho}(\rho) \\
             & F_{vv}(\rho) & F_{v\rho}(\rho) \\
             &              & F_{\rho\rho}(\rho)
\end{array}\right)
}{eq:mass2}
so that the modes $\psi$ are Eigenvectors of $L_0$ and $\bar L_0$:
\eq{
L_0 \psi = h\psi\qquad \bar L_0 \psi = \bar h \psi
}{eq:eigenL0}
The Eigenvalues $h,\bar h$ are the weights of the state $\psi$ if it is a primary, and otherwise they are sums of weights and levels. For sake of brevity we shall always refer to them as ``weights'', even when $\psi$ is not a primary. Periodicity in the angular coordinate requires that the difference of the weights, the angular momentum, is an integer. We shall always assume that this is the case. If additionally the sum of the weights (and therefore the weights) are integer then additionally periodicity in time is guaranteed. We do not necessarily assume that this is the case. 

We focus now on the regular normalizable left branch, ${\cal D}^L\psi^L=0$. For primaries, $L_-\psi=0=\bar L_-\psi$, it is then required that $h=2$, $\bar h=0$. The angular momentum equals to 2 and the excitation is a (boundary) graviton. The corresponding primary $\psi^L$ is then given by 
\eq{
\psi^L_{\mu\nu}(2,0) = \frac{e^{-2iu}}{\cosh^4{\!\!\rho}}\left(\begin{array}{c@{\quad}c@{\quad}c}
\frac14\,\sinh^2{\!(2\rho)} & 0 & \frac i2 \sinh{(2\rho)} \\
0 & 0 & 0 \\
\frac i2 \sinh{(2\rho)} & 0 & -1
\end{array}\right)_{\!\!\!\mu\nu} 
}{eq:cg46}
Descendants of the primary are obtained by acting on it repeatedly with $L_+$ and/or $\bar L_+$. For instance we have
\eq{
\psi^L_{\mu\nu}(2+n,0) = \big((L_+)^n\psi^L(2,0)\big)_{\mu\nu} \propto e^{-inu}\tanh^n\!\rho\,\psi^L_{\mu\nu}(2,0)
}{eq:nolabel}
The action of $\bar L_+$ on the left primary is more complicated. For later purposes we note that the $v\mu$-components vanish for the primary \eqref{eq:cg46} and consequently the first $\bar L_+$ descendant has a vanishing $vv$-component, but all further descendants have $\psi_{vv}^L\neq 0$:
\eq{
\psi_{vv}^L(h,0)=\psi_{vv}^L(h,1)=0\qquad \psi_{vv}^L(h,\bar h)\neq 0\;\textrm{if}\;\bar h\geq 2
}{eq:FvvL}

The right modes are obtained from the left modes by exchanging $u\leftrightarrow v$ and $h\leftrightarrow \bar h$. We do not address them separately. Both the left and the right modes are pure gauge in the bulk and therefore do not constitute a local physical degree of freedom. The only physical bulk degree of freedom comes therefore from the massive or the logarithmic branch. The massive modes will be discussed extensively in section \ref{se:mass} below.

The regular normalizable logarithmic modes, $({\cal D}^L)^2\psi^{\rm log}=0$, are related to the regular normalizable left modes by \cite{Grumiller:2008qz}
\eq{
\psi^{\rm log}_{\mu\nu}= -\frac12\,(i(u+v)+\ln{\cosh^2\!{\rho}})\,\psi^L_{\mu\nu}
}{eq:cg26}
A particular example for a logarithmic mode is the normalizable regular logarithmic primary:
$
\psi^{\rm log}_{\mu\nu}(2,0)=-\frac12\,(i(u+v)+\ln{\cosh^2\!{\rho}})\, 
\psi^L_{\mu\nu}(2,0)
$.
By construction, the logarithmic primary 
$\psi^{\rm log}_{\mu\nu}(2,0)$
is annihilated by $L_-$ and $\bar L_-$. Acting on it with $L_+$ and $\bar L_+$ produces a tower of descendants. The logarithmic modes obtained in this way are not Eigenstates of $L_0$ and $\bar L_0$, but only of their difference [cf.~\eqref{eq:cg79}]:
\eq{
L_0\,\psi^{\rm log}_{\mu\nu} = h \psi^{\rm log}_{\mu\nu}+\frac12\,\psi^L_{\mu\nu} \qquad \bar L_0\,\psi^{\rm log}_{\mu\nu} = \bar h \psi^{\rm log}_{\mu\nu} + \frac12\,\psi^L_{\mu\nu}
}{eq:cg31}

All the modes above are regular at the origin $\rho=0$. In appendix \ref{app:bc} we discuss point particle modes as particular examples of singular modes. We do not further dwell on this case and consider from now on exclusively regular modes. Having classified the normalizable modes, we now describe the most general set of regular modes. To this end it is sufficient to consider the massive branch and extract the other branches as certain limits thereof.

\subsection{Generic massive solutions}\label{se:mass}

In this subsection we discuss generic solutions to the {\eom} that are not necessarily descendants of the primaries. This is tantamount to giving up the normalizability condition. Non-normalizable modes play the role of sources in the AdS/CFT dictionary. 

We make again the separation Ansatz \eqref{eq:mass2} and solve the equation ${\cal D}^M\psi=0$, viz.,
\eq{
\psi_{\mu\nu} + \frac{1}{\mu}\,\eps_\mu{}^{\al\be}\nabla_\al \psi_{\be\nu} = 0
}{eq:mass3}
If $\mu=1$ ($\mu=-1$) we obtain left (right) solutions.
Note that solutions of \eqref{eq:mass3} are necessarily traceless and transversal. The six independent {\eom} \eqref{eq:mass3} are sufficient to determine all components $F_{\mu\nu}$ for any given set of weights $h,\bar h$. Four of these equations are algebraic:
\begin{subequations}
\label{eq:alg}
\begin{align}
& \bar h F_{uu} - h F_{uv} = \frac{\mu-1}{4i}\,\sinh{(2\rho)}\,F_{u\rho} \\
& \bar h F_{uv} - h F_{vv} = \frac{\mu+1}{4i}\,\sinh{(2\rho)}\,F_{v\rho} \\
& \bar h F_{u\rho} - h F_{v\rho} = \frac{i}{\sinh{(2\rho)}}\,\big(F_{vv}(\mu+1)+F_{uu}(\mu-1)-2\mu\,\cosh(2\rho)\,F_{uv}\big) \\
& F_{\rho\rho} = \frac{4}{\sinh^2(2\rho)}\,\big(2\cosh(2\rho)F_{uv}-F_{uu}-F_{vv}\big)
\end{align}
\end{subequations}
The remaining two provide a coupled set of linear first order differential equations for $F_{uv}$ and $F_{vv}$:
\begin{subequations}
\label{eq:ODEs}
\begin{align}
\frac{\extd F_{uv}}{\extd\rho} &= \frac{\mu+1}{\sinh(2\rho)}\Big(F_{uv}\big(\frac{4h\bar h}{(\mu+1)^2}-\cosh(2\rho)\big)+F_{vv}\big(1-\frac{4h^2}{(\mu+1)^2}\big)\Big) \label{eq:ODEa} \\
\frac{\extd F_{vv}}{\extd\rho} &= -\frac{\mu+1}{\sinh(2\rho)}\Big(F_{vv}\big(\frac{4h\bar h}{(\mu+1)^2}-\cosh(2\rho)\big)+F_{uv}\big(1-\frac{4\bar h^2}{(\mu+1)^2}\big)\Big)  \label{eq:ODEb}
\end{align}
\end{subequations}
 For later purposes we parameterize the coupling constant $\mu$ by
\eq{
\frac{\mu+1}{2}=1-\eps
}{eq:mass1}
At the moment $\eps$ need not be small or positive. We also define 
\eq{
x:=\cosh(2\rho)
}{eq:mass2a}
Decoupling the two differential equations leads to a second order equation for $F_{vv}$ (prime denotes derivative with respect to $x$):
\eq{
F_{vv}''+\frac{2x}{x^2-1}\,F_{vv}'-\frac{\al x^2-2h\bar h\, x + h^2+\bar h^2-\al}{(x^2-1)^2}\,F_{vv} = 0
}{eq:mass3a}
with $\al=(1-\eps)+(1-\eps)^2$. The differential equation \eqref{eq:mass3a} can be transformed easily into a hypergeometric differential equation. Its most general solution is given by
\begin{multline}
F_{vv} = a_1 (x-1)^{(\bar h-h)/2}(x+1)^{(\bar h + h)/2}\,_2F_1\big(2+\bar h-\eps,\,-1+\bar h+\eps,\,1+\bar h -h;\,\frac{1-x}{2}\big) \\
+ a_2 (x-1)^{(h-\bar h)/2}(x+1)^{(h + \bar h)/2}\,_2F_1\big(2+h-\eps,\,-1+h+\eps,\,1+h-\bar h;\,\frac{1-x}{2}\big)
\label{eq:mass4}
\end{multline}
provided the difference between the weights is not integer, $h-\bar h\not\in\mathbb{Z}$. 
Useful identities for the Gauss hypergeometric function $_2F_1$ are collected in appendix \ref{app:hyper}.

We consider now regularity at the origin $x=1$. Near the origin we can expand $_2F_1=1+O(x-1)$. The singular modes are those where $a_2=0,a_1\neq 0$ and the regular modes are those where $a_1=0,a_2\neq 0$. We are exclusively interested in regular modes and therefore set $a_1=0$. The regular solution for $F_{vv}$ is given by
\begin{equation}
F_{vv} = a_2 (x-1)^{(h-\bar h)/2}(x+1)^{(h+\bar h)/2}\, _2F_1(2+h-\eps, -1+h+\eps,1+h-\bar h; \frac{1-x}{2})
\label{eq:mass5}
\end{equation}  
Similar considerations yield the regular solution for $F_{uv}$.
\begin{equation}
F_{uv} = \tilde a (x-1)^{(h-\bar h)/2}(x+1)^{-(h+\bar h)/2}\, _2F_1(-\bar h+\eps, 1-\bar h-\eps,1+h-\bar h; \frac{1-x}{2})
\label{eq:mass6}
\end{equation}
The constant $\tilde a$ is determined uniquely by the choice of the overall normalization $a_2$, see appendix \ref{app:C} for explicit results. The solutions for $F_{uu}$ and $F_{uv}$, \eqref{eq:mass5} and \eqref{eq:mass6}, also solve the first order system \eqref{eq:ODEs}, apart from certain degenerate cases that we shall address separately. All other components of $F_{\mu\nu}$ are obtained from the algebraic relations \eqref{eq:alg}.

The case $\bar h> h$ is recovered upon exchanging $a_1\leftrightarrow a_2$ and $h\leftrightarrow \bar h$ in the discussion and the formulas above. The case $h=\bar h$ leads to one regular mode given in \eqref{eq:mass5} and another mode that turns out to be singular at $x=1$. The same issue arises if the difference between the weights is integer, $h-\bar h\in\mathbb{Z}$. Thus, the most general regular case is covered by \eqref{eq:mass5}. 

\subsection{Non-normalizable left and right solutions}\label{se:L}

The solutions of ${\mathcal D}^L\psi=0$ are recovered from the massive solutions above in the limit $\eps\to 0$. If the weights are integers then we obtain elementary functions instead of hypergeometric ones. We assume that this is the case. For concreteness we demand 
\eq{
h>\bar h
}{eq:inequality}
and address the other cases in the end. Solutions of ${\cal D}^R\psi=0$ are obtained from the left modes by replacing everywhere $u\leftrightarrow v$, $h\leftrightarrow\bar h$ and $L$ with $R$.

We classified solutions into normalizable (all components $F_{\mu\nu}$ are bounded for large $x$) and non-normalizable ones (not all components $F_{\mu\nu}$ are bounded for large $x$). Since we require regularity there is no freely adjustable parameter anymore in our solution. For any given set of weights the component $F_{vv}$ must take the form \eqref{eq:mass5}. Thus, for any given set of weights only three possibilities exist: there is only a normalizable mode, there is only a non-normalizable mode, or there is both a normalizable and a non-normalizable mode and the former has $F_{vv}=0$. We have mentioned in equation \eqref{eq:FvvL} that normalizable modes with non-vanishing component $F_{vv}$ exist for any $\bar h\geq 2$. Therefore, a necessary condition for non-normalizable regular modes is the inequality $\bar h \leq 1$. In the following paragraph we establish conditions that are necessary and sufficient.

For $\bar h=1$ or $\bar h=0$ we find that there are no regular non-normalizable solutions, see the end of appendix \ref{app:hyper}. Modes with weights $h\leq-2$ and negative $\bar h$ are normalizable. Thus, regular non-normalizable modes exist if and only if the weights obey the inequalities 
\eq{
h\geq -1\qquad\qquad\bar h\leq-1 
}{eq:necsuf}
The only-if-part of the statement is clear from the previous discussion. The if-part will be shown explicitly by constructing all the modes. Note that the inequalities \eqref{eq:necsuf}, if saturated, include the mode $h=\bar h=-1$, which turns out to be non-normalizable and regular \eqref{eq:hm1}. We keep this mode in the discussion below, even though it does not obey the strict inequality \eqref{eq:inequality}.

The results for $F_{vv}$ and $F_{uv}$ can be extracted from \eqref{eq:hyper7} and \eqref{eq:hyper11} plugged into \eqref{eq:mass5} and \eqref{eq:mass6} (with $\eps=0$), respectively. All other components follow algebraically from \eqref{eq:alg}. The results are presented in detail in appendix \ref{app:L}. As an example we present here the result for the non-normalizable left moving boundary graviton:
\eq{
\psi^L_{\mu\nu}(1,-1)=e^{-iu+iv}\,\left(\begin{array}{ccc}
0 & 0 & 0 \\
0 & x-1 & -2i\sqrt{\frac{x-1}{x+1}} \\
0 & -2i\sqrt{\frac{x-1}{x+1}} & -\frac{4}{x+1}
\end{array}\right)_{\!\!\mu\nu}
}{eq:Lgrav}
For generic values $h\geq-1\geq\bar h$ the results in appendix \ref{app:C} lead to the following asymptotic expansion:
\begin{subequations}
\label{eq:Fasymptotic}
\begin{align}
F_{vv}&= x + h \bar h + {\cal O}\big(\frac1x\big)\\
F_{uv}&= 1-h^2+{\cal O}\big(\frac{\ln{x}}{x}\big)\\
F_{uu}&= -\frac{h}{\bar h}\,(h^2-1)+{\cal O}\big(\frac{\ln{x}}{x}\big)\\
F_{v\rho}&= -2ih+{\cal O}\big(\frac{\ln{x}}{x}\big)\\
F_{u\rho}&= \frac{2i(1-h^2)}{\bar h\,x}\big(h+\bar h-2h\bar h\,(\ln{\frac x2}-\psi(h)-\psi(1-\bar h)-2\gamma)\big)+{\cal O}\big(\frac{\ln{x}}{x^2}\big)\\
F_{\rho\rho}&= \frac{4(1-2h^2)}{x}+{\cal O}\big(\frac{\ln{x}}{x^2}\big)
\end{align}
\end{subequations}
The quantity $\ga$ is the Euler--Mascheroni constant and $\psi=\Ga^\prime/\Ga$ is the digamma-function. We note in passing that the asymptotic expansion \eqref{eq:Fasymptotic} is valid for non-integer values of $h$, $\bar h$ as well. The $uu$-component is non-polynomial in the weight $\bar h$. It will become clear in section \ref{se:corr} how this non-locality is related to the dynamics of CFT.

Let us now discuss algebraic properties of the non-normalizable left modes. These properties are very useful to relate solutions of different weights, analogous to the discussion in section \ref{se:norm} for normalizable modes. We could again exploit the property \eqref{eq:EHop}, but we use a slightly stronger statement here. Namely, when acting on left modes the operator ${\cal D}^L$ commutes with the generators $L_\pm$ \eqref{eq:cg22}-\eqref{eq:cg23} and $\bar L_\pm$:\footnote{The quickest way to show this is as follows: Take the tensor identity ${\cal D}(\bar g)^L\psi^L=0$ and make a coordinate change ${\cal D}(\bar g+{\cal L}_\xi\bar g)^L(\psi^L+{\cal L}_\xi \psi^L)=0$, where ${\cal L}_\xi$ is the Lie-derivative along a vector field $\xi$. If $\xi$ is one of the Killing vectors $L_\pm$, $\bar L_\pm$ we obtain ${\cal D}(\bar g)^L({\cal L}_\xi \psi^L)=0$, which is equivalent to the statement in \eqref{eq:alg0}. We thank Niklas Johansson for providing this argument.}
\eq{
\big[{\cal D}^L,\,L_\pm\big] \psi^L = 0\qquad \big[{\cal D}^L,\,\bar L_\pm\big] \psi^L = 0
}{eq:alg0}
Starting from a regular non-normalizable left mode we can therefore create algebraically further left modes (not necessarily regular or non-normalizable). We find the algebraic relations
\begin{subequations}
\label{eq:algebraic}
\begin{align}
 L_0\, \psi^L(h,\bar h) &= h\,\psi^L(h,\bar h) \\
 L_-\, \psi^L(h,\bar h) &= (h+1)\,\psi^L(h-1,\bar h) + \de_{h,-1}\,N(-2,\bar h)\\
 L_+\, \psi^L(h,\bar h) &= (h-1)\,\psi^L(h+1,\bar h) \\
 \bar L_0\,\psi^L(h,\bar h) &= \bar h\,\psi^L(h,\bar h) \\
 \bar L_-\, \psi^L(h,\bar h) &= (\bar h-1)\,\psi^L(h,\bar h-1) \\
 \bar L_+\, \psi^L(h,\bar h) &= (\bar h+1)\,\psi^L(h,\bar h+1) + \de_{\bar h,-1}\,N(h,0)
\end{align}
\end{subequations}
where $N(h,\bar h)$ with $|h|>1$ are left modes that are regular and normalizable (if $|h|\leq 1$ the quantity $N(h,\bar h)$ vanishes) and $\psi^L(h,\bar h)$ are regular non-normalizable modes. We have fixed the normalization of all left modes in such a way that it is compatible with the asymptotic expansion \eqref{eq:Fasymptotic}. Note that the normalizable modes never mix with the non-normalizable modes: if the right hand side of one of the relations \eqref{eq:algebraic} contains a normalizable contribution, the non-normalizable one automatically vanishes. 

To prove the relations \eqref{eq:algebraic} it is essentially sufficient to consider the action of $L_\pm$ on the $vv$-component and the action of $\bar L_\pm$ on the $uu$-component. This produces immediately the relations above, but without the terms proportional to normalizable modes $N$. As long as it is non-vanishing, it is sufficient to consider the $vv$-component ($uu$-component) and compare it with the corresponding component of regular non-normalizable modes: since we know that the modes $\psi^L$ must be solutions of ${\cal D}^L\psi^L=0$ knowledge of one non-vanishing component determines the other components algebraically. The only caveat is that a normalizable mode with the same weights $h,\bar h$ could mix with the regular non-normalizable mode, provided the former has a zero $vv$-component ($uu$-component). However, for generic weights no such modes exist; the $vv$-component ($uu$-component) of normalizable modes is non-vanishing, with only a few exceptions. These exceptions lead precisely to the terms proportional to the modes $N$ in the algebraic relations \eqref{eq:algebraic}.

\begin{figure}
\begin{center}
\setlength{\unitlength}{1mm}
\begin{picture}(100,100)(-10,-10)
\thicklines
\matrixput(0,30)(10,0){4}(0,10){6}{\circle{2}} 
\matrixput(0,0)(10,0){5}(0,10){3}{\circle*{2}} 
\matrixput(40,60)(10,0){5}(0,10){3}{\circle*{2}} 
\matrixput(50,0)(10,0){4}(0,10){6}{\circle{2}} 
\put(38.6,28.6){X} 
\put(38.6,38.6){X} 
\put(38.6,48.6){X} 
\dottedline{2}(-5,55)(85,55) 
\dottedline{2}(-5,25)(85,25) 
\dottedline{2}(35,25)(35,85) 
\dottedline{2}(45,55)(45,-5) 
%
%
\put(40,-10){\vector(0,1){100}} 
\put(-10,40){\vector(1,0){100}} 
\put(40,95){\makebox(0,0){$h$}}
\put(95,40){\makebox(0,0){$\bar h$}}
\thinlines
%
\matrixput(1,60)(10,0){8}(0,10){3}{\vector(1,0){8}}
\matrixput(1,30)(10,0){3}(0,10){3}{\vector(1,0){8}}
\matrixput(51,30)(10,0){3}(0,10){3}{\vector(1,0){8}}
\matrixput(1,0)(10,0){4}(0,10){3}{\vector(1,0){8}}
\matrixput(51,0)(10,0){3}(0,10){3}{\vector(1,0){8}}
%
\matrixput(79,60)(-10,0){4}(0,10){3}{\vector(-1,0){8}}
\matrixput(29,60)(-10,0){3}(0,10){3}{\vector(-1,0){8}}
\matrixput(79,30)(-10,0){3}(0,10){3}{\vector(-1,0){8}}
\matrixput(29,30)(-10,0){3}(0,10){3}{\vector(-1,0){8}}
\matrixput(79,0)(-10,0){8}(0,10){3}{\vector(-1,0){8}}
%
\matrixput(0,1)(10,0){9}(0,10){2}{\vector(0,1){8}}
\matrixput(0,31)(10,0){5}(0,10){2}{\vector(0,1){8}}
\matrixput(0,61)(10,0){5}(0,10){2}{\vector(0,1){8}}
\matrixput(50,21)(10,0){4}(0,10){6}{\vector(0,1){8}}
%
\matrixput(0,79)(10,0){4}(0,-10){8}{\vector(0,-1){8}}
\matrixput(40,79)(10,0){5}(0,-10){2}{\vector(0,-1){8}}
\matrixput(40,49)(10,0){5}(0,-10){2}{\vector(0,-1){8}}
\matrixput(40,19)(10,0){5}(0,-10){2}{\vector(0,-1){8}}
\end{picture}
\caption{Display of the ladder operators $L_+$ (up), $L_-$ (down), $\bar L_+$ (right) and $\bar L_-$ (left). Notation: $\bullet=$~normalizable regular, $\circ=$~non-normalizable regular, X~$=$~singular}
\label{fig:alg}
\end{center}
\end{figure}
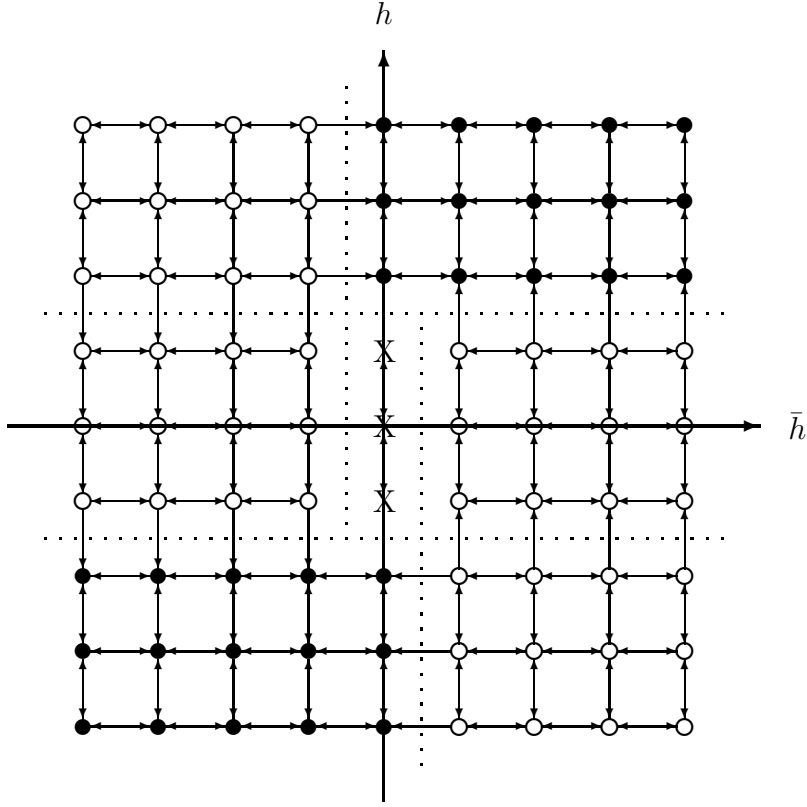

A simple example is provided by the boundary graviton \eqref{eq:Lgrav}. Inserting it into the algebraic relations \eqref{eq:Lgrav} we obtain
\begin{subequations}
\label{eq:algex}
\begin{align}
 L_0\, \psi^L(1,-1) &= \psi^L(1,-1) \\
 L_-\, \psi^L(1,-1) &= 2\,\psi^L(0,-1) \\
 L_+\, \psi^L(1,-1) &= 0 \\
 \bar L_0\,\psi^L(1,-1) &= -\psi^L(1,-1) \\
 \bar L_-\, \psi^L(1,-1) &= -2\,\psi^L(1,-2) \\
 \bar L_+\, \psi^L(1,-1) &= 0
\end{align}
\end{subequations}
The validity of the relations \eqref{eq:algex} can be checked explicitly with the formulas in appendix \ref{app:L}. Note that the boundary graviton is annihilated by $L_+$ and $\bar L_+$. In that sense it is the non-normalizable analog of the primary \eqref{eq:cg46}. However, as opposed to the situation in the normalizable case we cannot generate from it all non-normalizable modes: the modes with $h>1$ are not accessible by repeatedly acting with the generators $L_{\pm}$, $\bar L_{\pm}$ on the boundary graviton. We obtain in this way only modes with weights $(1,\bar h)$, $(0,\bar h)$ and $(-1,\bar h)$. The modes with $h\geq 2$ can be obtained algebraically starting from the mode with weights $(2,-1)$ given explicitly in \eqref{eq:Lgen}.

We have started the discussion in this subsection by assuming $h>\bar h$ \eqref{eq:inequality}. If we assume instead $h=\bar h$ then nearly all modes are regular and normalizable. There are three exceptions. If $h=\bar h=0$ we obtain singular modes. Additionally, there are two non-normalizable regular modes, namely $h=\bar h=-1$ \eqref{eq:hm1} and $h=\bar h=1$. We have included the former already in the discussion above, while the latter is included in the following case. If we assume $h<\bar h$ we just have to replace $h\to-h$ and $\bar h\to -\bar h$ in the discussion in this subsection.
We recall that the right modes are obtained from the left modes by replacing everywhere $u\leftrightarrow v$, $h\leftrightarrow\bar h$ and $L$ with $R$. Additionally, we assumed that both weights are integer. If only their difference is an integer, but not their sum, then we cannot express the hypergeometric functions appearing in \eqref{eq:mass5} and \eqref{eq:mass6} in terms of more elementary functions. Moreover, the algebra analog to \eqref{eq:algebraic} will no longer be ``osmotic'' between non-normalizable and normalizable modes, i.e., the figure \ref{fig:alg} does not apply to this more general situation.

In conclusion, the only regular non-normalizable left modes must have weights $(h\geq-1,\bar h\leq-1)$ or $(h\leq1,\bar h\geq 1)$. The general result for regular non-normalizable left modes is given in appendix \ref{app:L}. The asymptotic expansion for general regular non-normalizable left modes is provided in \eqref{eq:Fasymptotic}. The algebraic properties \eqref{eq:algebraic} depicted in figure \ref{fig:alg} relate the modes and allow to generate all regular left modes starting e.g.~from \eqref{eq:Lgen}.

\subsection{Non-normalizable logarithmic solutions}\label{se:log}

The logarithmic modes emerge from the linear combination \cite{Grumiller:2008qz}
\eq{
\psi^{\rm log}_{\mu\nu} = \lim_{\eps\to 0}\frac{\psi^M_{\mu\nu}(\eps)-\psi^L_{\mu\nu}}{\eps} = \frac{\extd \psi^M_{\mu\nu}}{\extd\eps}\Big|_{\eps=0}
}{eq:log1}
For finite values of $\eps$ the middle expression in \eqref{eq:log1} is annihilated by ${\cal D}^M{\cal D}^L$, but not by either of ${\cal D}^{M/L}$. After taking the limit $\eps\to 0$ all three expressions in \eqref{eq:log1} are annihilated by $({\cal D}^L)^2$, but not by ${\cal D}^L$. This is the defining property of logarithmic modes.

We focus again first on the case $h\geq \bar h$. The discussion is analogous to the one of left modes. Generic non-normalizable logarithmic modes exist for the same weights as non-normalizable left modes, $h\geq-1\geq\bar h$. One can use relations between contiguous functions similar to \eqref{eq:contiguous} to establish recursion relations resembling \eqref{eq:recursion}. This procedure allows to express all hypergeometric functions appearing in the logarithmic modes in terms of rational functions and logarithms. However, there is a simpler way to obtain generic logarithmic modes: make the shifts $h\to h+\eps$, $\bar h\to \bar h +\eps$ and define $F_{\mu\nu}=F_{\mu\nu}^L+\eps F_{\mu\nu}^{\rm log}+ {\cal O}(\eps^2)$, where $F_{\mu\nu}^L$ is a non-normalizable left mode. Expanding the algebraic equations \eqref{eq:alg} to ${\cal O}(\eps)$ yields
\begin{subequations}
\label{eq:alglog}
\begin{align}
& \bar h F_{uu}^{\rm log} - h F_{uv}^{\rm log} = F_{uv}^L \big(1-h+\frac{x-h}{\bar h}\big)+F_{vv}^L\,\frac{h^2-1}{\bar h} \\
& \bar h F_{uv}^{\rm log} - h F_{vv}^{\rm log} = \frac{\sqrt{x^2-1}}{2i}\,F_{v\rho}^{\rm log}+F_{vv}^L(h+1)-F_{uv}^L(\bar h+1) \\
& \bar h F_{u\rho}^{\rm log} - h F_{v\rho}^{\rm log} = \frac{2i}{\sqrt{x^2-1}}\,\big(F_{vv}^{\rm log}-x F_{uv}^{\rm log}-F_{uu}^L-F_{vv}^L+2x F_{uv}^L\big)+F_{u\rho}^L-F_{v\rho}^L \\
& F_{\rho\rho}^{\rm log} = \frac{4}{x^2-1}\,\big(2xF_{uv}^{\rm log}-F_{uu}^{\rm log}-F_{vv}^{\rm log}\big)
\end{align}
\end{subequations}
The differential equations \eqref{eq:ODEs} yield to ${\cal O}(\eps)$
\begin{subequations}
\label{eq:ODElog}
\begin{align}
\frac{\extd F_{uv}^{\rm log}}{\extd x} &= \frac{1}{x^2-1}\Big(F_{uv}^{\rm log}(h\bar h-x)+F_{vv}^{\rm log}(1-h^2)+F_{uv}^L(x+h+\bar h +h \bar h)-F_{vv}^L(1+h)^2\Big) \label{eq:ODEloga} \\
\frac{\extd F_{vv}^{\rm log}}{\extd x} &= \frac{1}{1-x^2}\Big(F_{vv}^{\rm log}(h\bar h-x)+F_{uv}^{\rm log}(1-\bar h^2)+F_{vv}^L(x+h+\bar h+h\bar h)-F_{uv}^L(1+\bar h)^2\Big)  \label{eq:ODElogb}
\end{align}
\end{subequations}
We know the homogeneous solutions to the system \eqref{eq:ODElog}: they are precisely the non-normalizable left modes provided in appendix \ref{app:L}. 
Solving the differential equations \eqref{eq:ODElog} is straightforward (if lengthy) and yields the components $F_{uv}^{\rm log}$ and $F_{vv}^{\rm log}$, up to two integration constants. One of the integration constants parameterizes the expected ambiguity corresponding to an addition of left modes to the logarithmic modes. We may fix this ambiguity for instance by demanding that $F_{vv}^{\rm log}$ asymptotically has no contribution linear in $x$. The other integration constant is fixed by demanding regularity at $x=1$. The remaining components $F_{\mu\nu}^{\rm log}$ follow algebraically from \eqref{eq:alglog}. The logarithmic mode for weights $h\geq-1\geq\bar h$ is then given by
\eq{
\psi_{\mu\nu}^{\rm log} = i(u+v)\,\psi^L_{\mu\nu}-F_{\mu\nu}^{\rm log}\,e^{-ihu-i\bar h v}
}{eq:loggeneric}
We have fixed the overall normalization constant in a convenient way.
We provide explicit results for logarithmic modes in appendix \ref{app:log}. As an example we present here the logarithmic partner of the non-normalizable left moving boundary graviton \eqref{eq:Lgrav}:
\begin{subequations}
\label{eq:NNgrav}
\begin{align}
\psi^{\rm log}_{vv}(1,-1) &= e^{-iu+iv}\,(i(u+v)+\ln\frac{x+1}{2})\,(x-1) \\
\psi^{\rm log}_{uv}(1,-1) &= 4e^{-iu+iv}\,\big(1-2\frac{\ln\frac{x+1}{2}}{x-1}\big) \\
\psi^{\rm log}_{uu}(1,-1) &= -4e^{-iu+iv}\,\big(1-2\frac{\ln\frac{x+1}{2}}{x-1} \big) \\
\psi^{\rm log}_{v\rho}(1,-1) &= -2ie^{-iu+iv}\, \Big((i(u+v)+\ln\frac{x+1}{2})\sqrt{\frac{x-1}{x+1}}-2\,\frac{x^2-4x+3+4\ln\frac{x+1}{2}}{(x-1)\sqrt{x^2-1}}\Big) \\
\psi^{\rm log}_{u\rho}(1,-1) &= 16i e^{-iu+iv}\,\frac{x-1-(x+1)\ln{\frac{x+1}{2}}}{(x-1)\sqrt{x^2-1}}\\
\psi^{\rm log}_{\rho\rho}(1,-1) &=  -4\, \frac{e^{-iu+iv}}{x+1}\,\Big(i(u+v)+\ln\frac{x+1}{2}- 4\,\frac{(2x+1)(x-1-2\ln\frac{x+1}{2})}{(x-1)^2} \Big)
\end{align}
\end{subequations}
It is a straightforward exercise to check that $\psi^{\rm log}_{\mu\nu}(1,-1)$ is indeed annihilated by $({\cal D}^L)^2$. Acting on it with ${\cal D}^L$ only once we obtain
\eq{
\big({\cal D}^L\psi^{\rm log}(1,-1)\big)_{\mu\nu} = (L_0+\bar L_0)\psi^{\rm log}_{\mu\nu}(1,-1) = -2\,\psi^L_{\mu\nu}(1,-1)
}{eq:DLlog}
where the regular non-normalizable left moving boundary graviton $\psi^L_{\mu\nu}(1,-1)$ is given explicitly in \eqref{eq:Lgrav}.
The logarithmic mode \eqref{eq:NNgrav} has angular momentum 2: 
\eq{
(L_0-\bar L_0)\psi^{\rm log}_{\mu\nu}(1,-1) = 2\,\psi^{\rm log}_{\mu\nu}(1,-1) 
}{eq:NNspin} 
Therefore, we call the mode \eqref{eq:NNgrav} ``non-norma\-lizable bulk graviton''. 

It is straightforward to obtain an asymptotic expansion for $F_{\mu\nu}^{\rm log}$ by solving asymptotically the differential equations \eqref{eq:ODElog}. This obtains
\begin{subequations}
\label{eq:Flogasy}
\begin{align}
F_{vv}^{\rm log} &= -(x+h\bar h)\,\ln{x}+h+\bar h+h\bar h+{\cal O}\big(\frac{\ln{x}}{x}\big)\\
F_{uv}^{\rm log} &=(h^2-1)\ln{x}+(1-3h)(h+1)+{\cal O}\big(\frac{\ln{x}}{x}\big)
\end{align}
\end{subequations}
Inserting this result (and the asymptotic expansions \eqref{eq:hyper12}, \eqref{eq:hyper13} for the left modes) into the algebraic relations \eqref{eq:alglog} yields asymptotic results for the non-normalizable logarithmic modes with arbitrary weights $h\geq -1\geq \bar h$:\footnote{For $h=0,\pm 1$ 
the following limits are needed: $\lim_{h\to 0} h\, \psi(h-1)=\lim_{h\to\pm 1} (h\mp 1)\, \psi(h-1)=-1$.} 
\begin{subequations}
\label{eq:genlogasy}
\begin{align}
\psi_{vv}^{\rm log} &\sim e^{-i(hu+\bar h v)}\,\big((x+h\bar h)(\ln x+i(u+v))-h-\bar h-h\bar h\big)+{\cal O}\big(\frac{\ln{x}}{x}\big) \\
\psi_{uv}^{\rm log} &\sim -e^{-i(hu+\bar h v)}\,\big((h^2-1)(\ln x+i(u+v))+(1-3h)(h+1)\big)+{\cal O}\big(\frac{\ln{x}}{x}\big)\\
\psi_{uu}^{\rm log} &\sim -e^{-i(hu+\bar h v)}\,\Big(\frac{h}{\bar
h}(1-h^2)(\ln x - i(u+v)) \nonumber \\
&\qquad +2\,\frac{h}{\bar h}\,(h^2-1)\,\big(\psi(h-1)+\psi(-\bar
h)-\frac32+2\ga+\ln{2}\big)\Big)+{\cal O}\big(\frac{\ln x}{x}\big)\\
\psi_{v\rho}^{\rm log} &\sim -2i\,e^{-i(hu+\bar h v)}\,\big(h\,(\ln x+i(u+v))-(h+1)\big)+{\cal O}\big(\frac{\ln{x}}{x}\big)\\
\psi_{u\rho}^{\rm log} &\sim {\cal O}\big(\frac{\ln x}{x}\big)\\
\psi_{\rho\rho}^{\rm log} &\sim -\frac{4}{x}\,e^{-i(hu+\bar h v)}\,\big((2h^2-1)\,(\ln x+i(u+v))+2(1-3h)(h+1)\big)+{\cal O}\big(\frac{\ln{x}}{x^2}\big)
\end{align}
\end{subequations}
The notation $\sim$ indicates that all equalities above are true up to the addition of non-normalizable left modes with the same weights. The $vv$-component grows asymptotically like $x\ln{x}\sim e^{2\rho}\rho$, which is logarithmically stronger growth as compared to the Fefferman-Graham expansion \eqref{eq:FG}. As anticipated, the non-normalizable logarithmic modes are not asymptotically AdS (see appendix \ref{app:bc}). Particularly the ${\cal O}(1)$ term of the $uu$-component will be important for the 2-point correlators in section \ref{se:corr}, since it is non-polynomial in the weights. 

We consider now algebraic properties of logarithmic modes, starting with the identities
\eq{
\big[({\cal D}^L)^2,\,L_\pm\big]\psi^{\rm log} = 0\qquad \big[({\cal D}^L)^2,\,\bar L_\pm\big]\psi^{\rm log} = 0
}{eq:alg1}
We find the algebraic relations
\begin{subequations}
\label{eq:logalgebraic}
\begin{align}
 L_0\,\psi^{\rm log} (h,\bar h) &= h\, \psi^{\rm log} (h,\bar h) - \psi^L (h,\bar h) \\
 L_-\, \psi^{\rm log}(h,\bar h) &\sim (h+1)\,\psi^{\rm log}(h-1,\bar h) + \de_{h,-1}\,N(-2,\bar h)\\
 L_+\, \psi^{\rm log}(h,\bar h) &\sim (h-1)\,\psi^{\rm log}(h+1,\bar h) \\
 \bar L_0\,\psi^{\rm log} (h,\bar h) &= \bar h\, \psi^{\rm log} (h,\bar h) - \psi^L (h,\bar h) \\
 \bar L_-\, \psi^{\rm log}(h,\bar h) &\sim (\bar h-1)\,\psi^{\rm log}(h,\bar h-1) \\
 \bar L_+\, \psi^{\rm log}(h,\bar h) &\sim (\bar h+1)\,\psi^{\rm log}(h,\bar h+1) + \de_{\bar h,-1}\,N(h,0)
\end{align}
\end{subequations}
where $N(h,\bar h)$ with $|h|>1$ are normalizable logarithmic modes (if $|h|\leq 1$ the quantity $N(h,\bar h)$ vanishes). The sign $\sim$ denotes equivalence up to the addition of left modes, the standard ambiguity for logarithmic modes. 
The relations above can be proven in a similar way as for the left modes.  The algebraic properties \eqref{eq:logalgebraic} relate the modes and allow to generate all regular logarithmic modes starting e.g.~from \eqref{eq:loggen}.

\section{Correlators}\label{se:corr}

We have now collected nearly all ingredients  to calculate all momentum space 2- and 3-point correlators of operators in the postulated dual CFT on the cylinder: in section \ref{se:2} we presented the first three variations of the bulk action and reviewed the linearized {\eom}. In section \ref{se:lin} we constructed the regular non-normalizable solutions of the linearized {\eom} that act as sources for the operators corresponding to bulk and boundary gravitons. The only missing ingredient for the calculation of correlators are boundary terms in the action and their variation. However, we shall prove that for 3-point correlators these boundary terms are not needed, while for the 2-point correlators we shall employ a convenient short-cut. In section \ref{se:41} we derive all 2-point correlators and in section \ref{se:42} we derive all 3-point correlators (in two cases qualitatively, in the other eight cases exactly). We work in Lorentzian signature and postpone a comparison to the more familiar Euclidean LCFT correlators in the short distance limit to section \ref{se:43}.

\subsection{Two-point correlators}\label{se:41}

We recall first 2-point correlators in a CFT with Lorentzian signature on the cylinder $S^1\times \mathbb{R}$ (see e.g.~\cite{Boyanovsky:1989jb} for a review). Next we explain how to obtain 2-point correlators in the momentum representation on the gravity side for Einstein gravity. Finally we derive all 2-point correlators in cosmological topologically massive gravity at the chiral point.

\subsubsection{Einstein gravity}\label{se:referee} 

As a warm-up as well as to fix the notation let us consider the stress tensor correlation function in a CFT dual to Einstein gravity in the momentum representation. 
The details of this CFT are irrelevant for this correlator since it is uniquely determined by the conformal Ward-identities. 
We shall refer to this CFT as ``Einstein-CFT'' below. Consider the time ordered $2$-point function 
\begin{eqnarray}
G_F(u,v)=\langle T\{T_{uu}(u)T_{uu}(0)\} \rangle 
\end{eqnarray}
We can obtain its momentum representation by noting that, quite generally, we have for $k\in \mathbb{Z}$, $\Delta \in \mathbb{N}$  and $\omega>0$
\begin{eqnarray}\label{ft1}
&& \int\limits_{-2\pi}^0 \!\extd\phi\int\limits_{-\infty}^\infty\!\extd t\,\frac{e^{i\omega t+ik\phi}}{\sin^{2\Delta}(\frac{t+\phi}{2} -i\epsilon\, \hbox{sgn} (t))}\\
&=&2^{2\Delta}\,2\pi i\,\frac{(i\omega)^{2\Delta-1}}{(2\Delta-1)!}\,\sum\limits_{n\geq 0}e^{i\omega \,2\pi n}\,\int\limits_{-2\pi}^0\!\extd\phi\; e^{i(k-\omega)\phi}\nonumber\\
&=&\frac{(-4)^\Delta 2\pi i}{\Gamma(2\Delta)}\; \frac{\omega^{2\Delta-1}}{2\bar h}
\end{eqnarray}
where we have regularized the $\phi$-integral through $\omega \to \omega+i\,0^+$. We recall that the coordinates $\phi$ and $t$ are related to the light-cone coordinates by $u=t+\phi$, $v=t-\phi$. The momentum and frequency are related to $h$ and $\bar h$ through $h= \frac{\omega+k}{2}$ and $\bar h= \frac{\omega-k}{2}$. Now, since the 2-point function of the stress tensor in Einstein-CFT dual to Einstein gravity on global AdS is given by \cite{Boyanovsky:1989jb}
\begin{eqnarray}\label{T_ll}
\langle T\{T_{uu}(u)T_{uu}(0)\}\rangle=\Big(\frac{{\cal{C}}}{4\sin^{4}(\frac{t+\phi}{2} -i\epsilon \,\hbox{sgn} (t))}+ \frac{{\cal{C}}}{6\sin^{2}(\frac{t+\phi}{2} -i\epsilon \,\hbox{sgn} (t))}\Big)
\end{eqnarray}
we can apply the general formula above to get 
\begin{eqnarray}\label{G_Ff}
\tilde  G_F(h,\bar h)&=& 2\pi i\,\frac{{\cal{C}}}{3}\; \frac{\omega^{3}-\omega}{\bar h}\nonumber\\
&=&2\pi i\,\frac{{\cal{C}}}{3}\;  \frac{h^{3}-h}{\bar h} \qquad+\quad \hbox{contact terms} \label{eq:ECFT}
\end{eqnarray}
The constant ${\cal{C}}$ is related to the central charge $c_{BH}$  through 
${\cal{C}}=\frac{c_{BH}}{8(2\pi)^2}$, with the Brown--Henneaux central charge given by \cite{Brown:1986nw}
\eq{
c_{BH} = \frac{3}{2\,G_N}
}{eq:BH}

Let us now compare the result \eqref{eq:ECFT} with the momentum space 2-point correlator obtained from the Einstein action with cosmological constant \eqref{eq:f2}. The source terms for $T_{uu}$ are the left-moving non-normalizable boundary gravitons in section (\ref{se:L}). These modes are not only solutions of linearized
CCTMG, but also of linearized Einstein gravity around AdS$_3$.  We thus substitute these into the second variation of the
Einstein--Hilbert action, together with some as yet unspecified normalization constant $\alpha$. The second
variation of the Einstein--Hilbert action $\de^{2}S_{\textrm{\tiny EH}} $ is given by the first term in
\eqref{eq:f69}
\eq{
\de^{(2)} S_{\textrm{\tiny EH}}(\psi^L, \psi^L) = -\frac{\alpha^2}{16\pi\,G_N}\,\int\volint\sqrt{-g}\,\psi^{\mu\nu\,\ast}_L \de G_{\mu\nu}(\psi^L) + {\rm boundary\;terms}
}{eq:2p3}
with the boundary terms determined by demanding that the second order action leads to a well-defined variational principle. 
Including these terms the on-shell action is then given by 
\eq{
\de^{2}S_{\textrm{\tiny EH}}(\psi^L, \psi^L) =
\frac{\alpha^2}{32 \pi\, G_N}\,\lim_{\rho\to\infty}\int\limits_{t_0}^{t_1}\!\extd
t\int\limits_0^{2\pi}\!\extd \phi\sqrt{-g}\,\psi^L_{ij}{}^\ast(h,\bar
h)\,g^{ik}g^{jl}\nabla_\rho \psi^L_{kl}(h',\bar h')
}{eq:EH1}
Here $g_{ij}$ is the induced metric at the boundary and $g$ its determinant. We explain now how to fix the normalization constant $\alpha$. We demand standard coupling of the metric to the stress tensor:
\eq{
S(\psi^{u\,L}_v,T^v_u)=  \frac12\,\int\!\extd t\extd\phi \sqrt{-g^{(0)}}\,\psi^{uu}_{L}T_{uu} = \int\!\extd t\extd\phi\,e^{-ihu-i\bar h v}\,T_{uu}
}{eq:canonicalcoupling}
Here $S$ is either some CFT action with background metric $g^{(0)}$ or a dual gravitational action with boundary metric $g^{(0)}$. The non-normalizable mode $\psi^L$ is the source for the energy-momentum flux component $T_{uu}$. The requirement \eqref{eq:canonicalcoupling} leads to the normalization $\alpha=\frac14$.

If $h\neq h'$ the integrand in the second variation of the on-shell action \eqref{eq:EH1} has an oscillating factor in $\phi$ that
integrates to zero. Similarly, if $\bar h\neq \bar h'$ the integrand has an
oscillating factor in $t$, which vanishes if integrated over a periodic
time interval (in the non-compact case one can argue that it vanishes in a
distributional sense). Therefore, the weights must match. 
\eq{
h=h'\qquad \bar h = \bar h'
}{eq:EH2}
Then the oscillating terms cancel precisely. Now we use the asymptotic
expansion \eqref{eq:Fasymptotic} together with the definition
\eqref{eq:mass2a} and keep only the leading terms:
\eq{
\de^{2}S_{\textrm{\tiny EH}}(\psi^L, \psi^L) =
\frac{\alpha^2}{16\, G_N}\,\int\limits_{t_0}^{t_1}\!\extd t \,\frac 14
e^{2\rho}\,\psi_v^{L\,u}{}^*(h,\bar h)\,\partial_\rho\psi_u^{L\,v}(h,\bar
h)\, (1+{\cal O}(e^{-2\rho}))
}{eq:EH3}
Collecting all factors, inserting $\alpha=\frac14$, taking the limit and replacing $G_N$ by means of the Brown--Henneaux result \eqref{eq:BH} yields
\eq{
\de^{2}S_{\textrm{\tiny EH}}(\psi^L, \psi^L) = \frac{c_{BH}}{24}\,\frac{h}{\bar
h}(h^2-1)\int\limits_{t_0}^{t_1}\extd t
}{eq:EH4}
In comparing (\ref{eq:EH4}) with the Einstein-CFT result \eqref{eq:ECFT} we should note that an extra factor of $2\pi \int\!\extd t$ arises due to the fact that the gravitational computation corresponds to a double Fourier transform with respect to both coordinates of the 2-point function. The extra factor $i$ in \eqref{eq:ECFT} comes about because we work in Lorentzian signature, where the action is multiplied by $i$. The result on the gravity side \eqref{eq:EH4} agrees therefore exactly with the result on the Einstein-CFT side \eqref{eq:ECFT}, provided the central charge takes the Brown--Henneaux value \eqref{eq:BH}. This is of course well-known.

For the right modes the same calculation goes through, upon
exchanging  $u\leftrightarrow v$ and $h\leftrightarrow \bar h$.
For mixed correlators, i.e., correlators that contain one left and one
right mode, the result for the correlator turn out to vanish up to a polynomial in the weights which corresponds to contact terms. Thus, all left-right 2-point
correlators in Einstein gravity vanish up to contact terms in agreement with the predictions of the postulated dual Einstein-CFT.

\subsubsection{Cosmological topologically massive gravity at the chiral
point}

At the chiral point the left-moving central charge $c_L$ vanishes  \cite{Kraus:2005zm}, whereas the right-moving central charge $c_R=2c_{BH}$ is twice that of the Einstein-CFT. From the CFT point of view the 2-point function (\ref{T_ll}) should thus vanish while its right moving companion should take twice the value compared to Einstein gravity. This structure is precisely encoded in CCTMG as we shall now see. 

Generically the 2-point correlators on the gravity side between two modes
$\psi^1(h,\bar h)$ and $\psi^2(h',\bar h')$ in momentum space are
determined by
\eq{
\langle \psi^1(h,\bar h)\, \psi^2(h',\bar h')\rangle = \frac12
\,\big(\de^{(2)} S_{\textrm{\tiny CCTMG}}(\psi^1,\psi^2)+ \de^{(2)}
S_{\textrm{\tiny CCTMG}}(\psi^2,\psi^1)\big)
}{eq:2pointdef}
where  $\langle\psi^1\,\psi^2\rangle$  stands for the correlation function of the CFT operators dual to the graviton modes $\psi^1$ and $\psi^2$.
On the right hand side one has to plug the non-normalizable modes $\psi^1$
and $\psi^2$ into the second variation of the on-shell action
\eqref{eq:f69}  and symmetrize with respect to the
two modes. However, one should be careful and collect all boundary terms,
because the whole contribution to the correlator \eqref{eq:2pointdef}
turns out to be a boundary term evaluated at the asymptotic boundary. This
was done in great detail in \cite{Skenderis:2009nt}. We use instead a
short-cut to calculate the 2-point correlators in momentum space that does
not require the construction of these boundary terms. To see how this works we take the second variation of the on-shell action \eqref{eq:f69} which can be written as 
\eq{
\de^{(2)} S_{\textrm{\tiny CCTMG}} =
-\frac{1}{16 \pi\, G_N}\,\int\volint\sqrt{-g}\,\big({\cal D}^L\de
g\big)^{\mu\nu} \de G_{\mu\nu}(h) + {\rm boundary\;terms}
}{eq:2p1}
We then see that the bulk term on the right hand side has the same form as in Einstein theory with $\delta g$ replaced by ${\cal D}^L\de g$. Now, consider the variation of this action. Possible obstructions to a well-defined boundary value problem can come only from the variation $\de G_{\mu\nu}(h)$. Thus any boundary terms appearing  in (\ref{eq:2p1}) containing normal derivatives must be identical with those in Einstein gravity. In addition there can be boundary terms which do not contain normal derivatives of the metric. However, from the asymptotic expansion \eqref{eq:Fasymptotic} we infer that such terms can at most lead to contact terms in the holographic computation of 2-point functions.\footnote{As mentioned in the last paragraph of section \ref{se:referee} in the momentum representation contact terms correspond to contributions which are polynomial in $h$ and $\bar h$.} Since we ignore such contributions these additional boundary terms are irrelevant (although they will be needed below in order to obtain a finite result). The upshot of this discussion is that we can reduce the calculation of all possible 2-point functions in CCTMG to the equivalent calculation in Einstein gravity with suitable source terms. To continue we go on-shell.
\eq{
{\cal D}^L\psi^L=0\qquad{\cal D}^L\psi^R=2\psi^R 
}{eq:2p2}
By virtue of the linearized {\eom} \eqref{eq:2p2} we obtain immediately
\begin{subequations}
\label{eq:correlatorsI}
\begin{align}
&\langle \psi^R(h,\bar h)\, \psi^R(h',\bar h') \rangle_{\textrm{\tiny
CCTMG}} \sim 2 \langle \psi^R(h,\bar h)\, \psi^R(h',\bar h')
\rangle_{\textrm{\tiny EH}} \\
&\langle \psi^L(h,\bar h)\, \psi^L(h',\bar h') \rangle_{\textrm{\tiny
CCTMG}} \sim 0 \\
&\langle \psi^L(h,\bar h)\, \psi^R(h',\bar h') \rangle_{\textrm{\tiny
CCTMG}} \sim \langle \psi^L(h,\bar h)\, \psi^R(h',\bar h')
\rangle_{\textrm{\tiny EH}} \sim 0 
\end{align}
\end{subequations}
The results on the gravity side \eqref{eq:correlatorsI} exactly match the LCFT prediction. 

Let us now consider the logarithmic modes. According to the AdS/CFT correspondence they should source an operator (see e.g.~\cite{Gurarie:1999yx,Kogan:2001ku}) $t=\frac{b}{c_L}\,T_{uu}+\frac{b}{2}\,\tilde T$ where $\tilde T$ is a second operator with conformal weight $\Delta=2$ such that $\langle T_{uu}(u)\,\tilde T(0)\rangle=0$. The parameter $b$ is fixed by the precise nature of the logarithmic CFT. Now, while $\langle T_{uu}(u) \,T_{uu}(0)\rangle$ vanishes for $c_L\to 0$, the correlator 
\begin{eqnarray}
\langle T_{uu}(u)\,  t(0)\rangle=b \,\frac{\langle T_{uu}(u)\, T_{uu}(0)\rangle}{c_L}
\end{eqnarray}
is finite and determines the parameter $b$. This structure is again beautifully realized in CCTMG. Indeed using again (\ref{eq:2p2}) together with the on-shell relation 
\eq{
{\cal D}^L\psi^{\rm
log}=-2\,\psi^L
}{eq:2p2i}
we obtain upon substitution in (\ref{eq:2p1}) immediately\footnote{We have to take care of the normalizations of modes. We found above that we should use $\alpha\psi^L$ for left (or right) modes by demanding that these modes be correctly normalized sources for the energy momentum tensor, with $\alpha=\frac14$. We make the Ansatz $\beta \psi^{\rm log}$ for the normalization of the logarithmic modes. At the end of our calculations we require a standard form of the LCFT 2-point correlators and find $\beta=\alpha=\frac14$.}
\begin{subequations}
\label{eq:correlatorsII}
\begin{align}
&\langle \psi^L(h,\bar h)\, \psi^{\rm log}(h',\bar h')
\rangle_{\textrm{\tiny CCTMG}} \sim -2 \, \langle \psi^L(h,\bar h)
\,\psi^L(h',\bar h') \rangle_{\textrm{\tiny EH}} 
\end{align}
and after a bit of calculation also
\begin{eqnarray}
&\langle \psi^R(h,\bar h) \,\psi^{\rm log}(h',\bar h')
\rangle_{\textrm{\tiny CCTMG}} \sim 0 
\end{eqnarray}
\end{subequations}
Again the results on the gravity side \eqref{eq:correlatorsII} exactly match the LCFT prediction.

Finally we address  the 2-point correlator with two logarithmic insertions. In principle one can again obtain the $\langle t\, t\rangle$-correlator using conformal Ward-identities in logarithmic CFT  by expanding the $\langle\tilde T\, \tilde T\rangle$-correlator about $c_L=0$ \cite{Gurarie:1999yx,Kogan:2002mg}. As a consequence of the conformal Ward-identities $\langle\tilde T\, \tilde T\rangle$ is again given by a generalization of (\ref{ft1}) for conformal weights  $(\Delta,\bar \Delta)=(2+\delta,\delta)$. The $\langle t\, t\rangle$-correlator is then obtained upon differentiating with respect to $\delta$ at $\delta=0$. In practice we face the problem that (\ref{ft1}) is valid for $\Delta$ and $\bar\Delta  \in \mathbb{N}$. Its analytic continuation to non-integer values of $\Delta$ is ambiguous and therefore not conclusive. However, below we will argue on general grounds that the result predicted by CCTMG is the correct one. In order to obtain it  we use again the on-shell relations \eqref{eq:2p2}, \eqref{eq:2p2i}. This does not lead to a correlator known from Einstein gravity, because there remains still one logarithmic
mode, and these modes do not exist in Einstein gravity. However, we can
still use the relation \eqref{eq:2p2i} to convert the second variation of
the CCTMG action \eqref{eq:2p1} into a variation that takes the form of
the second variation of the Einstein--Hilbert action \eqref{eq:2p3}. The
only missing ingredient are boundary counterterms that make the correlator
finite:
\eq{
\de^{(2)} S_{\textrm{\tiny CCTMG}}(\psi^{\rm log},\psi^{\rm log}) =
\frac{\be^2}{8 \pi\, G_N}\,\int\!\volint\sqrt{-g}\,\psi^{L\,\mu\nu}\de
G_{\mu\nu}(\psi^{\rm log}) + {\rm boundary\;terms}
}{eq:loglog1}
Again, it is clear from the asymptotic expansion \eqref{eq:genlogasy} that boundary counterterms can at most be contact terms.
Analogous to the Einstein--Hilbert case \eqref{eq:EH1} we obtain
\begin{multline}
\de^{(2)} S_{\textrm{\tiny CCTMG}}(\psi^{\rm log},\psi^{\rm log}) =
-\frac{4\be^2}{16 \pi\, G_N}\,\lim_{\rho\to\infty}\int\limits_{t_0}^{t_1}\!\extd
t\int\limits_0^{2\pi}\!\extd \phi\,\sqrt{-g}\,\psi^L_{ij}{}^*(h,\bar
h)\,g^{ik}g^{jl}\nabla_\rho \psi^{\rm log}_{kl}(h',\bar h')
\\
+ {\rm contact\;terms}
\label{eq:loglog2}
\end{multline}
To avoid oscillating integrals the weights must match \eqref{eq:EH2}.
Keeping only terms that do not vanish in the limit $\rho\to\infty$ we
get, modulo contact terms,
\begin{multline}
\de^{(2)} S_{\textrm{\tiny CCTMG}}(\psi^{\rm log},\psi^{\rm log})
\sim -\lim_{\rho\to\infty}\frac{4\be^2}{G_N}\,\int\limits_{t_0}^{t_1}\extd
t\,\big(\psi^L_{vv}{}^*\,\partial_\rho(\psi^{\rm
log}_{uu}e^{-2\rho})+\psi^L_{uu}{}^*\,\partial_\rho(\psi^{\rm
log}_{vv}e^{-2\rho})\big)
\label{eq:loglog3}
\end{multline}
Inserting $\beta=\frac14$ as well as the asymptotic expansions \eqref{eq:Fasymptotic} and
\eqref{eq:genlogasy} yields
\begin{multline}
\de^{(2)} S_{\textrm{\tiny CCTMG}}(\psi^{\rm log},\psi^{\rm log}) \sim
-\lim_{\rho\to\infty}\frac{1}{4\,G_N}\,\int\limits_{t_0}^{t_1}\extd
t\,\Big(\frac{h}{\bar h}\,(h^2-1)\, \big( \psi(h-1) + \psi(-\bar h) \big)
\\ + \big(\tilde\alpha(\rho+it)+\tilde\beta\big)\, \frac{h}{\bar h}\,(h^2-1) \Big) 
\label{eq:loglog4}
\end{multline}
valid for $h\geq -1, \bar h\leq -1$. Again $\psi$ is the digamma function and $\tilde\al, \tilde\be$ are weight-independent constants. The expression in the second line of \eqref{eq:loglog4} diverges as the cut-off for $\rho$ tends to infinity. To cancel it we could introduce an
appropriate boundary counterterm, as it was done in
\cite{Skenderis:2009nt}. However, there is an alternative possibility. The
factor $h(h^2-1)/\bar h$ is precisely the factor that arises in the
correlator between logarithmic and left modes, see \eqref{eq:correlatorsII}
with \eqref{eq:EH4}. For each finite value of the cutoff $\rho$ we
can exploit the shift ambiguity $\psi^{\rm log}\to\psi^{\rm
log}+\gamma\,\psi^L$ to cancel such terms. Therefore, they play no
physical role and can be dropped, even without introducing new boundary
counterterms. 
%
%
The term in the first line of \eqref{eq:loglog4} is finite, analytic in $h$, $\bar h$ and non-trivial. 
This is our final result for the 2-point correlator on the gravity side
between two logarithmic modes on the cylinder.  

In order to show that our result \eqref{eq:loglog4} provides the correct answer for a LCFT we first note that in the
short-distance limit we should recover the continuum result. To this end we evaluate the first line of
\eqref{eq:loglog4} in the limit of large weights $h\to\infty$, $\bar
h\to-\infty$ and obtain by virtue of \eqref{eq:hyper14} the asymptotic
result
\eq{
\lim_{h,-\bar h\to\infty}\de^{(2)} S_{\textrm{\tiny CCTMG}}(\psi^{\rm
log},\psi^{\rm log}) \sim -\frac{1}{2\,G_N}\,\frac{h^3}{\bar
h}\,\ln{\sqrt{-h\bar h}}\,\int\limits_{t_0}^{t_1}\extd t
}{eq:loglog5}
Here the sign $\sim$ means equality up to contact terms and up to
additional terms that can be absorbed by shifting $\psi^{\rm
log}\to\psi^{\rm log}+\gamma\,\psi^L$ with some constant $\ga$.

Comparison with 2-point correlators of logarithmic modes in a LCFT reveals that the momentum space correlator (\ref{eq:loglog4}) has the correct short-distance behavior (see also section \ref{se:43} below). For generic values of $h,\bar h$ with $\bar h h\neq 0$, CFT-momentum space correlators on a finite cylinder should have no cuts or poles. Our result  (\ref{eq:loglog4}) is the only such function with the correct asymptotic behavior. Of course, there is a freedom of adding a polynomial of degree 2 or less in $h$ to (\ref{eq:loglog4}) without spoiling the asymptotic behavior \eqref{eq:loglog5}. Equation (\ref{ft1}) then shows that such terms correspond to adding non-logarithmic operators of conformal weight $\Delta<2$ to $t$. Similarly a contribution of the form $\frac{f(\bar h)}{h}$ with analytic $f(\bar h)$ is compatible with conformal invariance. Such terms arise in the right moving sector involving $T_{vv}$. 

We summarize now the results of this section so far. All 2-point correlators in CCTMG match precisely with corresponding 2-point correlators in a LCFT, in agreement with the analysis in \cite{Skenderis:2009nt}. Many of the correlators could be reduced to correlators known from Einstein gravity by exploiting specific features of CCTMG and the second variation of its action. The most interesting correlator, the 2-point correlator between two logarithmic modes \eqref{eq:loglog4}, is analytic in $h$ and $\bar h$ and only develops a branch cut in the continuum limit \eqref{eq:loglog5}.

\subsection{Three-point correlators}\label{se:42}

\begin{figure}
\begin{center}
\epsfig{file=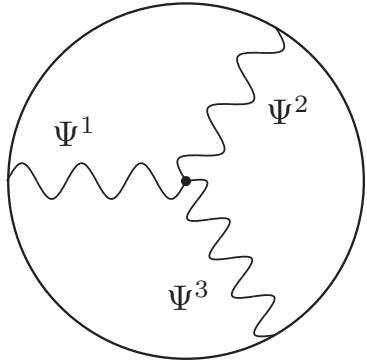,width=0.4\linewidth}
\caption{Witten diagram for three graviton correlator (4.25)}
\label{fig:1} 
\end{center}
\end{figure}

The 3-point correlators on the gravity side between three modes $\psi^1(h,\bar h)$, $\psi^2(h',\bar h')$ and $\psi^3(h'',\bar h'')$ in momentum space are determined by
\eq{
\langle \psi^1(h,\bar h)\, \psi^2(h',\bar h')\, \psi^3(h'',\bar h'')\rangle = \frac16 \,\big(\de^{(3)} S_{\textrm{\tiny CCTMG}}(\psi^1,\psi^2,\psi^3)+5\;{\rm permutations}\big)
}{eq:3pointdef}
On the right hand side one has to plug the non-normalizable modes $\psi^1$, $\psi^2$ and $\psi^3$ into the third variation of the on-shell action \eqref{eq:f73} and symmetrize with respect to all three modes. The Witten diagram corresponding to the correlator \eqref{eq:3pointdef} is depicted in Fig.~\ref{fig:1}.

\subsubsection{Boundary terms}

Before proceeding with calculations we prove that all boundary terms can be neglected. This considerably simplifies our calculations. Since all our modes are regular at the origin there are only asymptotic boundary terms. We can therefore exploit the asymptotic results \eqref{eq:Fasymptotic} for the left modes (and right modes upon exchanging $u\leftrightarrow v$, $h\leftrightarrow\bar h$) and the logarithmic modes \eqref{eq:genlogasy}. Actually, it is sufficient to keep track of the exponential behavior in $\rho$, so there is no essential difference between logarithmic and left modes for our proof. Thus, in this paragraph we do not discriminate between logarithmic and left modes. Whenever a statement is valid for left modes it is also valid for logarithmic modes (up to irrelevant polynomial terms in $\rho$). We only have to keep contributions that are not contact terms. As explained below \eqref{eq:2p1} this implies that each expression must contain at least one component $\psi_{uu}^L$ or one component $\psi_{vv}^R$. With one index raised these terms behave asymptotically like ${\cal O}(e^{-2\rho})$. On the other hand, each tri-linear expression of the form $\de g_\mu^\nu h_\nu^\la k_\la^\mu$ must contain at least two terms of order of unity and at most one term that decays asymptotically like $e^{-2\rho}$. This is so, because such tri-linear terms are multiplied by $\sqrt{-g}\sim e^{2\rho}$. If a tri-linear term decays faster than $e^{-2\rho}$ the corresponding boundary term vanishes in the limit $\rho\to\infty$. Consider for example the leading order contribution of two left and one right mode. The only expression that does not decay faster than $e^{-2\rho}$ is given by $\psi^{L\,u}_v\psi^{R\,v}_u\psi^{L\,u}_u$. However, this expression asymptotically is a contact term, since it does neither contain $\psi^{L\,v}_u$ nor $\psi^{R\,u}_v$: 
\eq{
\lim_{\rho\to\infty}\psi_\mu^{L\,\nu}\psi_\nu^{L\,\la}\psi_\la^{R\,\mu} = {\cal O}(e^{-4\rho}) + {\rm contact\;terms}
}{eq:3p1}
The same result applies for $L\leftrightarrow R$. Considering only left moving modes even leads to faster decay:
\eq{
\lim_{\rho\to\infty}\psi_\mu^{L\,\nu}\psi_\nu^{L\,\la}\psi_\la^{L\,\mu} = {\cal O}(e^{-4\rho})
}{eq:3p2}
The same result applies to right modes. Insertion of covariant derivatives, $\eps$-tensors or background metrics into tri-linear expressions does not change anything essential about the conclusions above. For instance, the only non-contact term that decays like $e^{-2\rho}$ constructed solely out of left modes schematically must be of the form $T^{\rho v}{}_u (\psi^{L\,u}_v\psi^{L\,u}_v\psi^{L\,v}_u)$, with some ${\cal O}(1)$ tensor $T$ constructed out of covariant derivatives, $\eps$-tensors or the background metric. Any such $T^{\rho v}{}_u$ is not ${\cal O}(1)$, but must decay at least like $e^{-2\rho}$. This is not completely obvious, but it can be shown straightforwardly by considering the asymptotic behavior of the Christoffel symbols and of the background metric. The same considerations apply to terms with any other combination of left and right modes. Generic boundary terms related to the bulk expression \eqref{eq:f73} are all of the type just discussed: they contain tri-linear expressions in the modes, possibly with some insertions of covariant derivatives, $\eps$-tensors or background metrics. 

In conclusion, all asymptotic boundary terms that emerge from partial integrations in 3-point correlators either vanish or yield contact terms. We are therefore free to partially integrate at will and to drop all boundary terms in the calculation of 3-point correlators.

\subsubsection{Correlators without log insertions}\label{se:422}

Partially integrating the first term in the third variation of the action \eqref{eq:f73} leads to a useful result:
\eq{
\de^{(3)} S_{\textrm{\tiny CCTMG}} \sim -\frac{1}{\ka^2}\,\int\volint\sqrt{-g}\,\Big[\big({\cal D}^L\de g\big)^{\mu\nu}\,\de^{(2)} R_{\mu\nu}(h,k) + \de g^{\mu\nu}\,\De_{\mu\nu}(h,k)\Big]
}{eq:3p3}
This expression allows us to use the same trick as for 2-point correlators: 3-point correlators that involve no logarithmic modes can be reduced to 3-point correlators calculated in Einstein gravity \cite{Arutyunov:1999nw}:
\eq{
\de^{(3)} S_{\textrm{\tiny EH}} = -\frac{1}{\ka^2}\,\int\volint\sqrt{-g}\,\de g^{\mu\nu}\,\de^{(2)} R_{\mu\nu}(h,k)
}{eq:3p4}
To show this we recall the properties \eqref{eq:2p2}, \eqref{eq:2p2i} and the fact that $\De_{\mu\nu}(h,k)=0$ if both $h$ and $k$ are left or right modes. Therefore, we have the following results:
\begin{subequations}
\label{eq:3cor0log}
\begin{align}
&\langle \psi^R(h,\bar h)\, \psi^R(h',\bar h')\,\psi^R(h'',\bar h'') \rangle_{\textrm{\tiny CCTMG}} \sim 2\, \langle \psi^R(h,\bar h)\, \psi^R(h',\bar h')\,\psi^R(h'',\bar h'') \rangle_{\textrm{\tiny EH}} \\
&\langle \psi^L(h,\bar h)\, \psi^R(h',\bar h')\,\psi^R(h'',\bar h'') \rangle_{\textrm{\tiny CCTMG}} \sim  -\frac{4}{3\ka^2}\,\int\!\volint\sqrt{-g}\,\psi^{R\,\mu\nu}\,\de^{(2)} R_{\mu\nu}(\psi^R,\psi^L) \sim 0 \\
&\langle \psi^L(h,\bar h)\, \psi^L(h',\bar h')\,\psi^R(h'',\bar h'') \rangle_{\textrm{\tiny CCTMG}} \sim -\frac{2}{3\ka^2}\,\int\!\volint\sqrt{-g}\,\psi^{R\,\mu\nu}\,\de^{(2)} R_{\mu\nu}(\psi^L,\psi^L) \sim 0 \\
&\langle \psi^L(h,\bar h)\, \psi^L(h',\bar h')\,\psi^L(h'',\bar h'') \rangle_{\textrm{\tiny CCTMG}} \sim 0
\end{align}
\end{subequations}
The first and the last result are compatible with the CCTMG values of the central charges. The other expressions reduce to terms that arise already in Einstein gravity, where they are found to be zero up to contact terms. 
The results \eqref{eq:3cor0log} coincide with the results in a LCFT.

\subsubsection{Single log insertions}\label{se:423}

Of course, the really interesting 3-point correlators contain one or more insertions of logarithmic modes. We discuss now correlators that contain exactly one such insertion.

We start with the correlator between two left and one logarithmic mode. If $\de g$ is the logarithmic mode we obtain from \eqref{eq:3p3}  the contribution
\eq{
\de^{(3)} S_{\textrm{\tiny CCTMG}}(\psi^{\rm log},\psi^L_1,\psi^L_2) \sim \frac{2}{\ka^2}\,\int\volint\sqrt{-g}\,\psi^{L\,\mu\nu}\,\de^{(2)} R_{\mu\nu}(\psi^L_1,\psi^L_2) 
}{eq:3p5}
where we have used the identity \eqref{eq:2p2i} and the vanishing of $\De_{\mu\nu}(\psi_1^L,\psi_2^L)=0$ \eqref{eq:f76}.
Up to the overall factor $-2$ this coincides precisely with the corresponding expression in Einstein gravity appearing in the 3-point correlator of three left modes. If $\de g$ is one of the left modes we obtain instead
\eq{
\de^{(3)} S_{\textrm{\tiny CCTMG}}(\psi^L_1,\psi^L_2,\psi^{\rm log}) \sim -\frac{1}{\ka^2}\,\int\volint\sqrt{-g}\,\psi_1^{L\,\mu\nu}\,\De_{\mu\nu}(\psi^L_2,\psi^{\rm log}) 
}{eq:3p6}
To calculate $\Delta_{\mu\nu}$ a helpful formula is
\eq{
\big({\cal D}^L (\psi^L\,\psi^L_2)\big)_{\mu\nu} = \eps_\mu{}^{\si\tau}\,\psi^{\al\,L}_\tau\,\nabla_\si\psi^L_{2\,\al\nu}
}{eq:3p7}
We obtain the intermediate result
\begin{multline}
\de^{(3)} S_{\textrm{\tiny CCTMG}}(\psi^L_1,\psi^L_2,\psi^{\rm log}) \sim \frac{2}{\ka^2}\,\int\volint\sqrt{-g}\,\psi_1^{L\,\mu\nu}\,\big(\frac12\,\eps_\mu{}^{\si\ka}\,\psi^{L\,\al}_\si\,(\nabla_{\nu}\psi^L_{2\,\al\ka}
-\nabla_{\al}\psi^L_{2\,\nu\ka}) \\
-\psi^{L\,\si}_{2\,\mu}\psi^L_{\si\nu}\big)
\label{eq:3p8}
\end{multline}
The definition \eqref{eq:f76} of $\Delta_{\mu\nu}$ makes it transparent why only left modes remain in the result \eqref{eq:3p8}.
Using on-shell manipulations like $\psi^L_{\al\be}=-\eps_\al{}^{\mu\nu}\nabla_\mu\psi^L_{\nu\be}$ or
\eq{
\psi_{2\,\mu}^{L\,\si}\,\psi^L_{\si\nu} = (\nabla_\al\psi_{2\,\mu\be}^L)(\nabla^\be\psi^{L\,\al}_\nu) - (\nabla_\al\psi_{2\,\mu\be}^L)(\nabla^\al\psi^{L\,\be}_\nu)
}{eq:3p9}
it is straightforward to show the identity [see \eqref{eq:f75}]
\begin{multline}
\de^{(3)} S_{\textrm{\tiny CCTMG}}(\psi^L_1,\psi^L_2,\psi^{\rm log}) + \de^{(3)} S_{\textrm{\tiny CCTMG}}(\psi^L_2,\psi^L_1,\psi^{\rm log}) \sim \\ 
 \frac{2}{\ka^2}\,\int\volint\sqrt{-g}\,\psi_1^{L\,\mu\nu}\,\de^{(2)} R_{\mu\nu}(\psi^L_2,\psi^L) +\frac{2}{\ka^2}\,\int\volint\sqrt{-g}\,\psi_2^{L\,\mu\nu}\,\de^{(2)} R_{\mu\nu}(\psi^L_1,\psi^L) 
\label{eq:3p10}
\end{multline}
The results \eqref{eq:3p5} and \eqref{eq:3p10} together imply that the correlator between two left and one logarithmic mode in CCTMG is reduced to the correlator between three left modes in Einstein gravity, multiplied by a factor $-2$. Similar considerations apply to correlators with one or two right modes instead of the left modes, where all expressions turn out to be contact terms. 
Though this result is simple and transparent from the CFT point of view, it is quite involved to derive it on the gravity side. In section \ref{se:424} we shall calculate this correlator explicitly in the limit of large weights, where considerable simplifications arise.
Let us summarize our results for 3-point correlators with one logarithmic insertion:
\begin{subequations}
\label{eq:3cor1log}
\begin{align}&\!\!\langle \psi^R(h,\bar h)\, \psi^R(h',\bar h')\,\psi^{\rm log}(h'',\bar h'') \rangle_{\textrm{\tiny CCTMG}} \sim 0 \\
&\!\!\langle \psi^L(h,\bar h)\, \psi^R(h',\bar h')\,\psi^{\rm log}(h'',\bar h'') \rangle_{\textrm{\tiny CCTMG}} \sim 0 \\
&\!\!\langle \psi^L(h,\bar h)\, \psi^L(h',\bar h')\,\psi^{\rm log}(h'',\bar h'') \rangle_{\textrm{\tiny CCTMG}} \sim -2\, \langle \psi^L(h,\bar h)\, \psi^L(h',\bar h')\,\psi^L(h'',\bar h'') \rangle_{\textrm{\tiny EH}} 
\end{align}
\end{subequations}
The results \eqref{eq:3cor1log} coincide with the results in a LCFT.

\subsubsection{Multiple log insertions and limit of large weights}\label{se:424}

The remaining correlators contain at least two logarithmic insertions and therefore are more complicated. In particular, the vanishing of the correlator between a right mode and two logarithmic ones is important for the consistency of the interpretation of CCTMG as gravity dual to some LCFT. We have not found a closed expression for these correlators, although we can evaluate them for any given set of weights. Nevertheless we can check the consistency with a LCFT on the infinite plane by considering the limit of small distance on the cylinder or, equivalently, large weights, $h,-\bar h\to\infty$ while keeping $h+\bar h$ finite. By the UV/IR connection \cite{Susskind:1998dq} the coincidence limit corresponds to the IR regime on the gravity side. It is therefore sufficient to substitute the asymptotic large-$x$ expansion of the non-normalizable modes (source terms) in the 3-point correlator \eqref{eq:3pointdef}. See also the discussion at the end of appendix \ref{app:hyper} affirming the UV/IR connection we are exploiting. 

We then proceed as follows. We start with the asymptotic expansions \eqref{eq:Fasymptotic} and \eqref{eq:genlogasy}  keeping only the leading and subleading terms in the large $x$ expansion. 
Next we evaluate by brute-force the missing three correlators $\langle\psi^{\rm log}\,\psi^{\rm log}\,\psi^{L/R/{\rm log}}\rangle$ with computer algebra \cite{grtensor}. We shall only keep terms that are of leading order in the weights, since by scale-invariance these should reproduce the momentum space correlator on the infinite plane. 
In addition, we are free to drop all polynomial terms in the weights as they correspond to contact terms. The remaining non-polynomial terms are then either rational functions in the weights or rational functions times logarithms in the weights, similar to our results for 2-point correlators. 

In order to avoid oscillating behavior of the integrand only two pairs of weights can be chosen freely, while the third one is determined by the others. For complex modes $\psi_1(h,\bar h)$, $\psi_2(h',\bar h')$, $\psi_3(h'',\bar h'')$ this condition is 
\eq{
h+h'+h'' = 0\qquad \bar h+\bar h'+\bar h'' = 0
}{eq:3p13}
For simplicity we consider at the moment the complex modes of appendix \ref{app:C}, rather than real ones.\footnote{An overall factor $\frac{1}{64}$ arises in all 3-point correlators due to the normalization factor $\frac14$ explained in section \ref{se:41}. Additional numerical factors arise if we use real instead of complex modes.} 
This means that the 3-point correlators calculated below will have a real and an imaginary part, corresponding to two specific linear combinations of 3-point correlators with real mode insertions. The main features of the 3-point correlators are captured by these calculations. For completeness we address correlators with real mode insertions in the end.

We need one more ingredient before we can start with the calculations. Namely, as we shall see below there are two types of rational functions that we can get. Either we obtain a sum of poles in all three weights (here and below quantities like $P(h,h^\prime,\bar h,\bar h^\prime)$ denote polynomials in the weights):
\eq{
\langle\psi_1(h,\bar h)\,\psi_2(h^\prime,\bar h^\prime)\,\psi_3(h'',\bar h'')\rangle \sim \frac{P(h,h^\prime,\bar h,\bar h^\prime)}{\bar h \bar h^\prime (\bar h+\bar h^\prime)}
\neq 0
}{eq:pole1}
Or we obtain a sum of poles in only two of the three weights:
\eq{
\langle\tilde\psi_1(h,\bar h)\,\tilde\psi_2(h^\prime,\bar h^\prime)\,\tilde\psi_3(h'',\bar h'')\rangle \sim \frac{\tilde P(h,h^\prime,\bar h,\bar h^\prime)}{\bar h \bar h^\prime} 
\sim 0
}{eq:pole2}
The latter case has the following simple interpretation in the dual CFT: All operators appearing in the operator product expansion (OPE) of the operators ${\cal O}_1$ and ${\cal O}_2$ sourced by $\psi_1$ and $\psi_2$ are such that their 2-point function with the third operator ${\cal O}_3$, sourced by $\psi_3$, is a contact term. 
Thus, we shall equate correlators of the form \eqref{eq:pole2} to zero modulo contact terms in the calculations below. 

To test this procedure we start with the calculation of correlators for which we have obtained exact expressions already. We use the relations \eqref{eq:3p13} between the weights to eliminate $h''$ and $\bar h''$ in terms of the other weights and find the following results 
\begin{subequations}
\label{eq:largeweightchecks}
\begin{align}
&\lim_{|\textrm{weights}|\to\infty}\langle \psi^L(h,\bar h)\, \psi^L(h',\bar h')\,\psi^L(h'',\bar h'') \rangle_{\textrm{\tiny CCTMG}} \sim 0 \\
&\lim_{|\textrm{weights}|\to\infty}\langle \psi^L(h,\bar h)\, \psi^L(h',\bar h')\,\psi^R(h'',\bar h'') \rangle_{\textrm{\tiny CCTMG}} \sim 0 \\
&\lim_{|\textrm{weights}|\to\infty}\langle \psi^L(h,\bar h)\, \psi^R(h',\bar h')\,\psi^R(h'',\bar h'') \rangle_{\textrm{\tiny CCTMG}} \sim 0 \\
&\lim_{|\textrm{weights}|\to\infty}\langle \psi^R(h,\bar h)\, \psi^R(h',\bar h')\,\psi^R(h'',\bar h'') \rangle_{\textrm{\tiny CCTMG}} \sim \frac{P^R(h,\bar h,h^\prime,\bar h^\prime)}{h h^\prime(h+h^\prime)} \\ 
&\lim_{|\textrm{weights}|\to\infty}\langle \psi^L(h,\bar h)\, \psi^L(h',\bar h')\,\psi^{\rm log}(h'',\bar h'') \rangle_{\textrm{\tiny CCTMG}} \sim \frac{P^L(h,\bar h,h^\prime,\bar h^\prime)}{\bar h \bar h^\prime(\bar h+\bar h^\prime)} \\
&\lim_{|\textrm{weights}|\to\infty}\langle \psi^L(h,\bar h)\, \psi^R(h',\bar h')\,\psi^{\rm log}(h'',\bar h'') \rangle_{\textrm{\tiny CCTMG}} \sim 0 \\
&\lim_{|\textrm{weights}|\to\infty}\langle \psi^R(h,\bar h)\, \psi^R(h',\bar h')\,\psi^{\rm log}(h'',\bar h'') \rangle_{\textrm{\tiny CCTMG}} \sim 0 
\end{align}
with the polynomials
\begin{align}
 P^R(h,\bar h,h^\prime,\bar h^\prime) &\propto
\big(h h^\prime(\bar h+\bar h^\prime)^2(\bar h^3+\bar h^{\prime\,3})-h(h+h^\prime)\bar h^{\prime\,2}(\bar h^3-(\bar h+\bar h^\prime)^3)\nonumber \\
& \quad -h^\prime(h+h^\prime)\bar h^2(\bar h^{\prime\,3}-(\bar h+\bar h^\prime)^3) 
\big)\\ 
 P^L(h,\bar h,h^\prime,\bar h^\prime) &= - P^R(h^\prime,\bar h^\prime,h,\bar h) \label{eq:lalapetz}
\end{align}
%
\end{subequations}
The polynomials are symmetric under the exchanges $h,\bar h\leftrightarrow h^\prime,\bar h^\prime$ and $h,\bar h\leftrightarrow h''=-(h+h^\prime),\bar h''=-(\bar h +\bar h^\prime)$. Of course, this must be the case by the trivial exchange symmetry $\langle \psi\, \psi^\prime \,\psi''\rangle=\langle\psi^\prime\,\psi\,\psi''\rangle$, so this merely provides a check on the correctness of the calculations. As evident from the last equation \eqref{eq:lalapetz} the polynomials are related to each other by an exchange of weights and by a proportionality constant $-1$. In coordinate space these properties correspond to an exchange $u\leftrightarrow v$ and a certain proportionality between the coefficients in the 3-point correlators. It is easy to check that these are precisely the properties required by the first correlator in \eqref{eq:3cor0log} and by the last correlator in \eqref{eq:3cor1log}, and that the proportionality constant $-1$ matches. 
The poles appearing in the non-vanishing correlators are located precisely where they should be, so the procedure above indeed works. On the other hand we should note that the polynomials $P^{R/L}$ are of degree seven,  whereas dimensional analysis shows that the correct degree is five. The origin of this mismatch is that we integrated over all of AdS$_3$ while the correct region of integration should be supported only at large $x$ in order to be consistent with the asymptotic $x$-dependence assumed for non-normalizable modes (source terms) inserted into the 3-point correlators \eqref{eq:3pointdef}. At a qualitative level this can be taken into account by imposing an infrared cut-off $x>\lambda^2$, where $\lambda$ is of the order of $h,\bar h, h',\dots$. The resulting polynomial is then $\sim \lambda^{-2}P^{L/R}$ which now has the correct dimension. Since we are interested only in 
the location of the poles we may ignore this issue at present.

Of course, we did not need to invoke this procedure for the correlators just discussed, since we have reduced their calculation to a calculation of similar correlators in Einstein gravity in sections \ref{se:422} and \ref{se:423}. So, the example just considered provides a consistency test for our approach to evaluate 3-point correlators involving logarithmic modes. 
The calculation of the last two correlators in \eqref{eq:largeweightchecks} is still somewhat involved, because one has to take into account the leading order term, whose integrand \eqref{eq:f74} decays only like $1/x$, and the next-to-leading order term, whose integrand \eqref{eq:f74} decays like $\ln{x}/x^2$, $t/x^2$ or $1/x^2$. All these terms independently turn out to be contact terms for large weights.

Comparing our results in the large weight limit \eqref{eq:largeweightchecks} with the exact results \eqref{eq:3cor0log} and \eqref{eq:3cor1log} shows that they agree in this limit in the following sense: we reproduce the vanishing of all correlators that should vanish and the correct location of the poles (but not their residues) of the correlators that do not vanish. 
This provides a consistency check that the large weight limit indeed produces the correct results for the correlators in the near-coincidence limit, in the sense just explained.

We consider now the correlator between one right and two logarithmic modes in the limit of large weights.\footnote{Details are available at \href{http://quark.itp.tuwien.ac.at/~grumil/Rloglog.html}{{\tt http://quark.itp.tuwien.ac.at/$\sim$grumil/Rloglog.html}}.} We find the following result
\begin{subequations}
\label{eq:angelinajolie}
\begin{align}
\lim_{|\textrm{weights}|\to\infty}& \langle \psi^{\rm log}(h,\bar h)\,\psi^{\rm log}(h',\bar h')\,\psi^R(h'',\bar h'')\,\rangle_{\textrm{\tiny CCTMG}} \sim \int\extd x\int\extd t \,\Big[\frac{f_1}{x}+\frac{f_2\,\ln^2{x}}{x^2}\nonumber\\
& +\frac{f_3\,t\,\ln{x}}{x^2}+\frac{f_4\,t^2}{x^2} +\frac{f_5\,\ln{x}}{x^2}+\frac{f_6\,t}{x^2}+\frac{f_7}{x^2} + {\cal O}\big(\frac{\ln^2{x}}{x^3}\big)\Big]
\end{align} 
where $f_1, f_2, f_3, f_4\sim 0$. The expressions $f_5, f_6, f_7$ (up to contact terms) are given by
\eq{
f_5\propto f_6 \propto \frac{h^3\bar h^{\prime\,2}}{\bar h}+\frac{\bar h^2 h^{\prime\,3}}{\bar h^\prime}\qquad f_7\propto \frac{h^3\bar h^{\prime\,2}}{\bar h}\,\ln{(-h\bar h)} + \frac{\bar h^2 h^{\prime\,3}}{\bar h^\prime}\,\ln{(-h^\prime\bar h^\prime)} + \alpha f_5
}{eq:angelinajolie2}
\end{subequations}
where $\alpha$ is a constant. The expressions \eqref{eq:angelinajolie} contain poles in the weights $\bar h$ and $\bar h^\prime$, but not in $h''=-h-h^\prime$. By the reasoning below equation \eqref{eq:pole2} these contributions are equal to zero up to contact terms. {\em Therefore the correlator between one right and two logarithmic modes vanishes up to contact terms.}

We consider next qualitatively the correlator between one left and two logarithmic modes in the limit of large weights. We find the following result
\begin{subequations}
\label{eq:Lloglog}
\eq{
\lim_{|\textrm{weights}|\to\infty} \langle \psi^{\rm log}(h,\bar h)\,\psi^{\rm log}(h',\bar h')\,\psi^L(h'',\bar h'')\,\rangle_{\textrm{\tiny CCTMG}} \sim \frac{P^{\rm log}(h,h^\prime,\bar h,\bar h^\prime)}{\bar h \bar h^\prime(\bar h+\bar h^\prime)}
}{eq:Lloglog1}
where the quantity $P^{\rm log}$ (up to contact terms) is given by
\eq{
P^{\rm log}(h,h^\prime,\bar h,\bar h^\prime)= P_1\,(h,h^\prime,\bar h,\bar h^\prime) \, \,\ln{(-h\bar h)} + P_1\,(h^\prime,h,\bar h^\prime,\bar h) \,\,\ln{(-h^\prime\bar h^\prime)} + P_2\,(h,h^\prime,\bar h,\bar h^\prime) \,
}{eq:withoutlabel}
\end{subequations}
The polynomials $P_1$ and $P_2$ are again of degree seven in the weights.\footnote{Details are available at \href{http://quark.itp.tuwien.ac.at/~grumil/Lloglog.html}{{\tt http://quark.itp.tuwien.ac.at/$\sim$grumil/Lloglog.html}}.} The expressions \eqref{eq:Lloglog} contain poles in the weights $\bar h$, $\bar h^\prime$ and $\bar h''=-\bar h-\bar h^\prime$. The correlator between one left and two logarithmic modes does not vanish, even after dropping all contact terms.


The missing correlator between three logarithmic modes is lengthy, even in the limit of large weights. We do not present any formulas for this case, and just mention that we have checked that this correlator is non-vanishing, even after dropping all contact terms, and has the expected poles.\footnote{Details are available at \href{http://quark.itp.tuwien.ac.at/~grumil/logloglog.html}{{\tt http://quark.itp.tuwien.ac.at/$\sim$grumil/logloglog.html}}.}  
If we use real modes instead of complex modes we reproduce all the results above, but with slightly different numerical coefficients. The qualitative features of the correlators do not change. All correlators that vanish above still vanish, including the crucial correlator between a right and two logarithmic modes.

We summarize our results for 3-point correlators with at least two logarithmic insertions:
\begin{subequations}
\label{eq:3cor2log}
\begin{align}
&\lim_{|\textrm{weights}|\to\infty}\langle \psi^R(h,\bar h)\, \psi^{\rm log}(h',\bar h')\,\psi^{\rm log}(h'',\bar h'') \rangle_{\textrm{\tiny CCTMG}} \sim 0  \\
&\lim_{|\textrm{weights}|\to\infty}\langle \psi^L(h,\bar h)\, \psi^{\rm log}(h',\bar h')\,\psi^{\rm log}(h'',\bar h'') \rangle_{\textrm{\tiny CCTMG}} \sim \frac{P^{\rm log}(h,h^\prime,\bar h,\bar h^\prime)}{\bar h \bar h^\prime(\bar h+\bar h^\prime)} \\
&\lim_{|\textrm{weights}|\to\infty}\langle \psi^{\rm log}(h,\bar h)\, \psi^{\rm log}(h',\bar h')\,\psi^{\rm log}(h'',\bar h'') \rangle_{\textrm{\tiny CCTMG}} \sim \frac{{\rm lengthy}}{\bar h \bar h^\prime(\bar h+\bar h^\prime)} \label{eq:logloglog}
\end{align}
\end{subequations}
In particular, we have demonstrated the vanishing of the correlator between a right mode and two logarithmic ones, which is a non-trivial result on the gravity side. We do not provide explicit expressions for the last two correlators in \eqref{eq:3cor2log}, because we do only trust the locations of the poles, but not their residues. The proportionality constants depend on the weights, but do not contain further poles. We see already from the qualitative behavior of the penultimate correlator \eqref{eq:Lloglog} (and similar expressions for the last correlator) that these correlators are non-vanishing and exhibit the required features for LCFT correlators: poles in all three weights and expressions that contain logarithms in appropriate pairs of weights in the large weight expansion.

Let us emphasize the main results of section \ref{se:42}. We have reduced the correlators between three right modes \eqref{eq:3cor0log} and between two left and one logarithmic mode \eqref{eq:3cor1log} to correlators known from Einstein gravity, thence establishing their exact form. Moreover, we have found that all the correlators which should vanish in a LCFT indeed do vanish for CCTMG \eqref{eq:3cor0log}, \eqref{eq:3cor1log}, including the crucial correlator between one right and two logarithmic modes \eqref{eq:3cor2log}. In the latter case we had to invoke a procedure that involved a large weight expansion. While we trust that procedure to generate the correct location of poles (or absence thereof), we do not expect it to yield the correct residues, since we exploited a certain scaling between weights and the coordinate $x$ to evaluate the hypergeometric functions and the integrals. We have not found a simple way to extract the residues of the poles or to avoid the large weight expansion altogether. This will be necessary for obtaining the correlators $\langle \psi^L\, \psi^{\rm log}\,\psi^{\rm log} \rangle$ and $\langle \psi^{\rm log}\, \psi^{\rm log}\,\psi^{\rm log} \rangle$ quantitatively for arbitrary weights. Qualitatively we found that these correlators have poles at the correct places \eqref{eq:3cor2log} and contain the anticipated terms logarithmic in the weights \eqref{eq:withoutlabel}.

\subsection{Comparison with Euclidean logarithmic CFT correlators}\label{se:43}

For completeness we relate now our results  for 2- and 3-point correlators  in sections \ref{se:41} and \ref{se:42} to the more familiar form of various LCFT correlators on the complex plane collected in appendix \ref{app:D}. 
The vanishing correlators on the gravity side also vanish on the LCFT side (and vice versa), so we focus solely on the non-vanishing ones. To this end we transform the results \eqref{eq:correlatorsI}, \eqref{eq:correlatorsII} and \eqref{eq:loglog5} to coordinate space, fixing the integral over time to $2\pi i$ so that we can compare with the Euclidean correlators in appendix \ref{app:D}.\footnote{
We use Fourier transformations with respect to $t,\phi$, not with respect to light cone coordinates $u,v$ or $z,\bar z$. The factor 2 in the relation $\de^{(2)}(t,\phi) = 2\, \de^{(2)}(z,\bar z)$ enters in all formulas \eqref{eq:distcorr}.} 
\begin{subequations}
\label{eq:distcorr}
\begin{align}
& \langle\psi^R(z,\bar z)\,\psi^R(0)\rangle =
\frac{i\pi}{2G_N}\,\frac{\bar \partial^3}{\partial}\,\de^{(2)}(z,\bar z)
\\
& \langle\psi^L(z,\bar z)\,\psi^{\rm log}(0)\rangle =
-\frac{i\pi}{2G_N}\,\frac{\partial^3}{\bar \partial}\,\de^{(2)}(z,\bar z)
\\
& \langle\psi^{\rm log}(z,\bar z)\,\psi^{\rm log}(0)\rangle = -\frac{2 i
\pi}{G_N}\,\ln{\big(m^2\sqrt{-\partial\bar\partial}\big)}\,\frac{\partial^3}{\bar
\partial}\,\de^{(2)}(z,\bar z)
\end{align}
\end{subequations}
Evaluating the distributions in \eqref{eq:distcorr} with
standard methods and keeping only the most singular terms yields
\begin{subequations}
\label{eq:spacecorr}
\begin{align}
& \langle\psi^R(z,\bar z)\,\psi^R(0)\rangle = \frac{c_R}{2\bar z^4} \\
& \langle\psi^L(z,\bar z)\,\psi^{\rm log}(0)\rangle =  \frac{b}{2z^4} \\
& \langle\psi^{\rm log}(z,\bar z)\,\psi^{\rm log}(0)\rangle = -\frac{b\,\ln{(m^2|z|^2)}}{z^4}
\end{align}
\end{subequations}
provided we use the values
\eq{
c_L=0\qquad\qquad c_R=\frac{3}{G_N}\qquad\qquad b = -\frac{3}{G_N}
}{eq:cb}
These are exactly the values for central charges $c_L$, $c_R$ \cite{Kraus:2005zm} and new anomaly $b$  \cite{Skenderis:2009nt} found before. The results \eqref{eq:spacecorr} are equivalent to the LCFT results \eqref{eq:2LCFT}.

We compare now the 3-point correlators.  The vanishing correlators on the gravity side also vanish on the LCFT side (and vice versa), so we focus again on the non-vanishing ones. All non-vanishing 3-point correlators with at most one logarithmic insertion can be reduced to 3-point correlators known from Einstein gravity.
\begin{align}
& \langle \psi^R(z,\bar z)\, \psi^R(z',\bar z')\,\psi^R(0,0) \rangle_{\textrm{\tiny EH}} = \frac{c_{BH}}{\bar z^2\bar z^{\prime\,2}(\bar z-\bar z')^2} 
\label{eq:3p11} \\
& \langle \psi^L(z,\bar z)\, \psi^L(z',\bar z')\,\psi^L(0,0) \rangle_{\textrm{\tiny EH}} = \frac{c_{BH}}{z^2z^{\prime\,2}(z-z')^2} 
\label{eq:3p12}
\end{align}
The results \eqref{eq:3cor0log} and \eqref{eq:3cor1log} then imply
\begin{subequations}
\label{eq:3}
\begin{align}
& \langle \psi^R(z,\bar z)\, \psi^R(z',\bar z')\,\psi^R(0,0) \rangle = \frac{c_R}{\bar z^2\bar z^{\prime\,2}(\bar z-\bar z')^2} \\
& \langle \psi^L(z,\bar z)\, \psi^L(z',\bar z')\,\psi^{\rm log}(0,0) \rangle = \frac{b}{z^2z^{\prime\,2}(z-z')^2} 
\end{align}
\end{subequations}
with the same values of $c_R$ and $b$ as before \eqref{eq:cb}. The results \eqref{eq:3} are equivalent to the LCFT results \eqref{eq:3LCFT0}, \eqref{eq:3LCFT1}. The non-vanishing 3-point correlators with at least two logarithmic insertions were calculated only qualitatively on the gravity side, see for instance the schematic result \eqref{eq:Lloglog}. The appearance of $\ln{(-h\bar h)}$ terms on the gravity side in momentum space is in qualitative agreement with the appearance of $\ln{|z|^2}$ terms in \eqref{eq:3LCFT2}. 

In conclusion all six 2-point correlators and the eight of ten 3-point correlators that we calculated on the gravity side coincide precisely with corresponding LCFT correlators. With the tools provided in this work also the remaining two correlators $\langle\psi^{L/\rm log}\,\psi^{\rm log}\,\psi^{\rm log}\rangle$ can be checked in principle, though we have not found an efficient way to do so. Therefore, we have considered them in the limit of large weights and found qualitative agreement, notably the correct location of poles, so we expect that they coincide as well with corresponding LCFT correlators. It would be of interest to calculate the correlator $\langle\psi^{\rm log}\,\psi^{\rm log}\,\psi^{\rm log}\rangle$ in full detail since this determines another defining parameter of the LCFT, denoted by $a$ in \cite{Kogan:2001ku}.

\section{Discussion}\label{se:5}

\paragraph{Summary} In this paper we confirmed the conjecture \cite{Grumiller:2008qz} that CTMG \eqref{eq:f1} at the chiral point \eqref{eq:f3.5} is dual to a logarithmic CFT: we constructed all regular non-normalizable left, right and logarithmic modes in global coordinates, see section \ref{se:lin} and appendix \ref{app:C}. We plugged these modes into the second \eqref{eq:f69} and third variation of the action \eqref{eq:f73} to evaluate 2- and 3-point correlators on the gravity side in section \ref{se:corr}. We found that they agree with correlators in a logarithmic CFT. This is concurrent with recent calculations by Skenderis, Taylor and van Rees \cite{Skenderis:2009nt}, who constructed 2-point correlators on the gravity side on the Poincar{\'e} patch and also found agreement with logarithmic CFT correlators. 

\paragraph{Generalizations}  It would be interesting to calculate the correlator \eqref{eq:logloglog} between three logarithmic modes not just qualitatively, but in full detail. This will allow to extract another parameter in addition to central charges and new anomaly \eqref{eq:cb}, sometimes denoted by $a$, that determines properties of the LCFT \cite{Kogan:2001ku}. Exploiting our results for the non-normalizable modes also higher order correlators can be calculated. Of course these calculations are rather involved. Some of the tricks we have used simplify these calculations. For example, one can probably show that all boundary terms vanish for all higher order correlators, by analogy to our proof for 3-point correlators. If this is true one just has to vary the bulk action and can partially integrate freely to simplify expressions, e.g.~by applying the on-shell relations \eqref{eq:2p2}, \eqref{eq:2p2i}. It might be of interest to check the 4-point correlators, since they contain the first non-trivial information about the CFT beyond that implied by conformal invariance and the values of central charges $c_{L/R}$, new anomaly $b$ and the parameter $a$.

Our analysis essentially applies also to New Massive Gravity \cite{Bergshoeff:2009hq} at a chiral point: its linearized {\eom} around an AdS$_3$ background for a particular tuning of parameters take the form $({\cal D}^L{\cal D}^R)^2\psi=0$, with the same operators ${\cal D}^{L/R}$ as in \eqref{eq:f22}. The construction of all non-normalizable modes is therefore contained in the present work already: they are given by regular non-normalizable left, right, logarithmic, and flipped logarithmic modes. The latter are constructed from the logarithmic modes in the same way as the right modes are constructed from the left modes: by exchanging the light-like coordinates $u\leftrightarrow v$ and switching the weights $h\leftrightarrow\bar h$. Consistent AdS boundary conditions analog to the ones for CCTMG \cite{Grumiller:2008es} were constructed for New Massive Gravity in \cite{Liu:2009kc}. These boundary conditions include all normalizable left, right, logarithmic and flipped logarithmic modes. 
The main missing ingredient for the calculation of 2- and 3-point correlators in New Massive Gravity is the construction of the second and third variation of the action, which is a straightforward exercise. 

Another interesting generalization is the inclusion of supersymmetry \cite{Becker:2009mk,Andringa:2009yc}. Perhaps an AdS/LSCFT correspondence can be established (see \cite{Caux:1996kq,Khorrami:1998kw,Mavromatos:2001iz} for some LSCFT literature). Of course, once the AdS/L(S)CFT conjecture is taken for granted one can utilize the full power of conformal symmetry to simplify the calculations of correlators. For instance, one can then exploit the conformal Ward identities, which was not possible in the present work whose goal was to substantiate the conjecture by calculating correlators on the gravity side.

\paragraph{Comments on CCTMG as quantum gravity} Since CCTMG apparently is dual to a logarithmic CFT it is neither chiral nor unitary. We address now consequences for its status as a toy model for quantum gravity. One possible option is to truncate the logarithmic modes, either by imposing periodicity conditions \cite{Grumiller:2008qz} or by imposing boundary conditions that are stricter than the requirement of asymptotic AdS, like Brown--Henneaux boundary conditions \cite{Strominger:2008dp}. The dual CFT, if it exists, would be extremal \cite{Witten:2007kt}, chiral \cite{Li:2008dq,Strominger:2008dp}, and avoids the difficulties with holomorphic factorization encountered in \cite{Maloney:2007ud}. That theory contains black holes but no (bulk) gravitons. An indication for the viability of this option is the existence of a consistent modular invariant partition function for $c_R=24k$  with $k\in\mathbb{Z}^+$ \cite{Witten:2007kt,Maloney:2009ck}. 
A counter-indication for the viability of this option is the potential non-existence of extremal CFTs for large central charges \cite{Gaberdiel:2008xb}. 
A related option is that chiral gravity does not have its own dual CFT, but that a chiral truncation of the LCFT dual to CCTMG can be achieved by restricting the latter to a superselection sector with vanishing left charges \cite{Maloney:2009ck}. Since we know many examples of non-unitary theories with zero-norm states (gauge degrees of freedom) and negative norm states (ghosts) that can be truncated consistently to unitary theories by superselecting to ghost number zero, it is conceivable that the same construction is possible for CCTMG.
The assessment of CCTMG and its chiral truncation as a toy model for quantum gravity still remains inconclusive. 

An alternative option not addressed extensively in the literature (see however \cite{Carlip:2008jk}) is the possibility to reverse the sign of the action \eqref{eq:f1}, truncate the black holes by some mechanism and keep the gravitons. We can achieve this by imposing boundary conditions that are stricter than AdS boundary conditions, see the chiral boundary conditions in appendix \ref{app:bc}. This procedure breaks ${\rm Virasoro}_R\times{\rm Virasoro}_L$ to $U(1)_R\times {\rm Virasoro}_L$. TMG (or CTMG with the black holes truncated) may serve as a suitable quantum gravity toy model with an $S$-matrix (or some AdS-analog thereof) for gravity wave scattering. This alternative was not considered in \cite{Witten:2007kt} and subsequent work, mainly because pure Einstein gravity in three dimensions does not provide any degrees of freedom for scattering. By contrast, (C)TMG does provide them in the form of topologically massive gravitons.

It is possible that neither of the truncations mentioned above works consistently at the quantum level, see \cite{Andrade:2009ae} for a recent work that addresses some of these issues. In that case another option could be pursued \cite{Grumiller:2008qz}, namely to unitarily complete the theory. For practical purposes this could mean a lift to a sector of string theory, though it is not obvious if and how this works in practice. While reasonable, this option clearly goes against the original intention \cite{Witten:2007kt} to use pure three dimensional gravity as a suitable quantum gravity toy model.

\paragraph{Prospects for AdS$_3$/LCFT$_2$} CCTMG could serve as a relatively simple gravity dual to certain logarithmic CFTs. To the best of our knowledge this would be the first explicit gravity dual of this type. Besides the AdS/LCFT literature \cite{Ghezelbash:1998rj,Myung:1999nd,Kogan:1999bn,Lewis:1999qv,MoghimiAraghi:2004ds} the only other context where gravity and LCFTs appear together so far seems to be two-dimensional gravity coupled to matter \cite{Bilal:1995rc}. It would be nice to find some condensed matter applications, like some strongly coupled systems described by a logarithmic CFT with $c_L=0$, $c_R>0$. Many logarithmic CFT examples require negative central charge, including the physically interesting examples of turbulence \cite{Flohr:1996ik}, the fractional Quantum Hall effect at filling factor $\nu=5/2$ \cite{Gurarie:1997dw,Cappelli:1998ma,Read:1999fn} or dense polymers \cite{Ivashkevich:1998na}. It is clear that these systems cannot be dual to CCTMG, though they may have other gravity duals, including possibly CTMG with $\mu\ell<1$ or New Massive Gravity. 
Some logarithmic CFT systems have indeed vanishing central charge, like quenched random magnets \cite{Cardy:1999zp} or other critical systems with quenched disorder, dilute self-avoiding polymers, percolation etc.~\cite{Flohr:2001zs,Gaberdiel:2001tr}. It would be of interest to check whether theses systems, at least at strong coupling, allow for a dual description in terms of CCTMG. We have gained now sufficient confidence in the AdS$_3$/LCFT$_2$ correspondence so that we may start looking for applications.

\acknowledgments

We are grateful to Melanie Becker, Steve Carlip, Sean Downes, Matthias Gaberdiel, Gaston Giribet, Roman Jackiw, Niklas Johansson, Per Kraus, Max Kreuzer, David Lowe, Alex Maloney, John McGreevy, Massimo Porrati, Radoslav Rashkov, Stephen Shenker, Kostas Skenderis, Wei Song, Andy Strominger, Marika Taylor and Stefan Theisen for discussions. We thank Ren{\'e} Sedmik for drawing figure \ref{fig:1} on p.~\pageref{fig:1}.

DG was supported by the project MC-OIF 021421 of the European Commission under the Sixth EU Framework Programme for Research and Technological Development (FP6). 
Research at the Massachusetts Institute of Technology is supported in part by funds provided by the U.S. Department of Energy (DoE) under the cooperative research agreement DEFG02-05ER41360.
During the final stage DG was supported by the START project Y435-N16 of the Austrian Science Foundation (FWF). DG thanks the Arnold-Sommerfeld Center for Theoretical Physics for repeated hospitality while part of this work was conceived.

IS was supported by the Transregional Collaborative Research Centre TRR 33, the DFG cluster of excellence ``Origin and Structure of the Universe'' as well as the DFG project Ma 2322/3-1. IS would like to thank the Erwin-Schr\"odinger Institute in Vienna for hospitality and financial support during the workshop ``Gravity in three dimensions'' in April 2009.

\appendix

\section{Boundary conditions}\label{app:bc}


For generic solutions of the linearized {\eom} we summarize here  various asymptotic boundary conditions and regularity conditions at the origin. The full metric $g=g^{\rm AdS}+\psi$ is the sum of the global AdS background \eqref{eq:cg20} and solutions $\psi$ of the linearized EOM \eqref{eq:22.5}. We assume in this appendix that the modes $\psi$ are brought into Gaussian normal coordinates \eqref{eq:GNC}. Modes that are asymptotically AdS$_3$ must have a Fefferman-Graham expansion \eqref{eq:FG}. For the time being we assume that all modes are regular at the origin.

\paragraph{Normalizable modes} By definition these modes are not allowed to modify the boundary metric $\ga^{(0)}_{ij}$ in the expansion \eqref{eq:FG}. We neglect terms that fall off asymptotically ($\rho\to\infty$) and find the following expansions for left, right and logarithmic modes:
\begin{align}
\psi^L_{ij} &= \ga_{ij}^{L\,(2)} + \dots \\
\psi^R_{ij} &= \ga_{ij}^{R\,(2)} + \dots \\
\psi^{\rm log}_{ij} &= \ga_{ij}^{(1)}\,\rho+\ga_{ij}^{{\rm log}\,(2)} + \dots
\end{align}
All normalizable modes are asymptotically AdS, including the bulk graviton encoded in the logarithmic modes.

\paragraph{Non-normalizable modes} Solutions to the linearized {\eom} \eqref{eq:22.5} that are not normalizable according to the criterion above are called ``non-norma\-lizable''. We neglect terms that fall off asymptotically ($\rho\to\infty$) and find the following expansions for left, right and logarithmic modes:
\begin{align}
\psi^L_{ij} &= \ga_{ij}^{L\,(0)}\,e^{2\rho} +  \ga_{ij}^{L\,(2)} +\dots \\
\psi^R_{ij} &= \ga_{ij}^{R\,(0)}\,e^{2\rho} +  \ga_{ij}^{R\,(2)} +\dots \\
\psi^{\rm log}_{ij} &= \ga_{ij}^{(-1)}\, \rho\,e^{2\rho} + \ga_{ij}^{{\rm log}\,(0)}\,e^{2\rho} + \ga_{ij}^{(1)}\,\rho + \ga_{ij}^{{\rm log}\,(2)} + \dots \label{eq:A6}
\end{align}
The left and right non-normalizable modes are asymptotically AdS, while the logarithmic non-normalizable modes 
are not asymptotically AdS.

\paragraph{Brown--Henneaux and beyond} For completeness we mention that in 3-dimensional Einstein gravity (but {\em not} in CCTMG) asymptotically AdS boundary conditions are equivalent to Brown--Henneaux boundary conditions \cite{Brown:1986nw}
\eq{
\psi  \simeq  \left(\begin{array}{l@{\quad}l@{\quad}l} 
\psi_{uu}={\cal O}(1) & \psi_{uv}={\cal O}(1) & \psi_{u\rho}={\cal O}(e^{-2\rho}) \cr 
  & \psi_{vv}={\cal O}(1) & \psi_{v\rho}={\cal O}(e^{-2\rho}) \cr 
  & & \psi_{\rho\rho}={\cal O}(e^{-2\rho}) \end{array}\right)
}{eq:fBH} 
In CCTMG these are replaced by slightly weaker conditions \cite{Grumiller:2008es,Henneaux:2009pw}
\eq{
\psi  \simeq  \left(\begin{array}{l@{\quad}l@{\quad}l} 
\psi_{uu}={\cal O}(\rho) & \psi_{uv}={\cal O}(1) & \psi_{u\rho}={\cal O}(\rho\,e^{-2\rho}) \cr 
  & \psi_{vv}={\cal O}(1) & \psi_{v\rho}={\cal O}(e^{-2\rho}) \cr 
  & & \psi_{\rho\rho}={\cal O}(e^{-2\rho}) \end{array}\right)
}{eq:fGJ} 
We stress again that the boundary conditions \eqref{eq:fGJ} are compatible with asymptotic AdS behavior, cf.~e.g.~\cite{Skenderis:2002wp,Skenderis:2009kd}.

\newcommand{\ivo}{\xi}
\paragraph{Point particle modes} In the body of the paper we consider exclusively modes that are regular at the origin. For sake of completeness, but also because they can be of relevance in other contexts, we address now particular singular modes. We call them `point particle modes' for reasons that will become apparent. The linearized {\eom} (\ref{eq:22.5}) admit solutions $\ivo$ that are locally pure gauge and obey the Brown--Henneaux boundary conditions \eqref{eq:fBH}, 
but which blow up at the origin $\rho=0$. We focus again first on the left modes. These modes have weights $h\leq 1$, $\bar h\geq 0$.
\begin{subequations}
\label{eq:allivo} 
\begin{align}
\ivo^L_{\mu\nu}(h,0)&=e^{-ihu}\,\tanh^h{\!\rho}\,\left(\begin{array}{c@{\quad}c@{\quad}c}
1 & 0 & \frac{2i}{\sinh{(2\rho)}} \\
0 & 0 & 0 \\
\frac{2i}{\sinh{(2\rho)}} & 0 & -\frac{4}{\sinh^2\!\!{(2\rho)}}
\end{array}\right)_{\!\!\!\mu\nu} \\
\ivo^L_{\mu\nu}(h,\bar h)&=\big((\bar L_+)^{\bar h}\ivo^L(h,0)\big)_{\mu\nu}
\end{align}
\end{subequations}
All modes $\ivo^L(h,\bar h)$ are annihilated by ${\cal D}^L$ and $\bar L_-$, transverse $\nabla_\mu \xi^{\mu\nu}_L=0$ and traceless $\xi^\mu_{L\,\mu}=0$. The simplest of these modes $\chi^L=\ivo^L(0,0)$ depends only on the radial coordinate $\rho$: 
\eq{
\chi^L_{\mu\nu} = \left(\begin{array}{c@{\quad}c@{\quad}c}
1 & 0 & \frac{2i}{\sinh{(2\rho)}} \\
0 & 0 & 0 \\
\frac{2i}{\sinh{(2\rho)}} & 0 & -\frac{4}{\sinh^2\!\!{(2\rho)}}
\end{array}\right)_{\!\!\!\mu\nu} 
}{eq:ivo}
We see that some of the components of the mode \eqref{eq:ivo} diverge for $\rho\to 0$. Consequently the perturbative gravitational energy given by the $00$ component of the pseudo 
tensor $ t^\mu_{\;\;\nu}$ obtained from (\ref{eq:f74}) through
\eq{
E=\int \extd^2x \sqrt{-g}\; t^0_{\;\;0}= \int \extd^2x \sqrt{-g} \;\;g^{\mu 0}\,\frac{\delta{\cal{L}}^{(3)}}{\delta g^{\mu 0}} 
}{EG}
diverges.\footnote{We thank Wei Song for pointing this out to us.}   
This divergence is an artifact of perturbation theory.
Indeed these modes correspond to infinitesimal point sources which in
three dimensions cause a conical singularity. This is not a small
deformation of the metric. 

In order to exhibit the relation of the modes $\chi^L$ to point particles in AdS we consider the metric for a point particle with mass ${\cal{M}}$ and angular momentum ${\cal{J}}$ in global AdS-coordinates
\eq{
\extd s^2=\extd\rho^2-\cosh^2\!\rho\,(r_+\extd\tau-r_-\extd\phi)^2+\sinh^2\!\rho\,(r_+\extd\phi-r_-\extd\tau)^2
}{me}
with
\eq{
1-{\cal{M}}=r_+^2+r_-^2\qquad \qquad
-{\cal{J}}= 2r_+r_-
}{}
Global AdS \eqref{eq:cg20} is obtained for ${\cal{M}}={\cal{J}}=0$ while BTZ black holes would correspond to ${\cal{M}}\geq 1$. Let us now consider a perturbation $g\to g+\frac12\,h$ of (\ref{me}) around ${\cal{M}}={\cal{J}}=0$ with small values of mass ${\cal{M}}\ll 1$ and angular momentum ${\cal{J}}\ll 1$. In $(\tau,\phi,\rho)$-coordinates such a perturbation is given by 
\eq{
h_{\mu\nu}=\begin{pmatrix}{2\cal{M}}\cosh^2\!\rho&-{\cal{J}}&0\cr
-{\cal{J}}&-2{\cal{M}}\sinh^2\!\rho&0\cr
0&0&0\end{pmatrix}_{\!\!\mu\nu}
}{dg}
The perturbation $h$ in \eqref{dg} is not in the transverse-traceless gauge \eqref{eq:f16}. The trace is non-zero, $g^{\mu\nu} h_{\mu\nu}=-4{\cal{M}}$. Two components of the divergence vanish, $\nabla_\mu h^{\mu}_\tau= \nabla_\mu h^{\mu}_\phi=0$, but we also have a non-vanishing component:
$\nabla_\mu h^{\mu}_\rho = 4{\cal{M}}\,\coth{(2\rho)}$.
We can bring $h_{\mu\nu}$ into the transverse-traceless form \eqref{eq:f16} by means of an infinitesimal diffeomorphism $\xi^\mu$ with $(\nabla^2-2)\,\xi_\nu=-\nabla_\mu h^{\mu}_\nu$. The solution to these conditions is given by $\xi^\tau=\xi^\phi=0$ and
\eq{
\xi^\rho=\frac{{\cal{M}}}{2}\,\left(\coth\rho+\tanh\rho\right)
}{sold}
Applying $\xi^\mu$ to $h_{\mu\nu}$ we get $h\to\tilde h$ with
\eq{
\tilde h_{\mu\nu}=\begin{pmatrix}\cal{M}&-{\cal{J}}&0\cr
-{\cal{J}}&{\cal{M}}&0\cr
0&0&-\frac{4{\cal{M}}}{\sinh^2(2\rho)}\end{pmatrix}_{\!\!\mu\nu}
}{dg2}
In the $(u,v,\rho)$-coordinates \eqref{eq:cg20} the perturbation $\tilde h$ becomes diagonal, 
\begin{eqnarray}
\tilde h_{\mu\nu}&=& \begin{pmatrix}\frac{1}{2}\left({\cal{M}}-  {\cal{J}}\right)&0&0\cr
& \frac{1}{2}\left({\cal{M}}+{\cal{J}}\right)&0\cr
0&0&-\frac{4{\cal{M}}}{\sinh^2(2\rho)}\end{pmatrix}_{\!\!\mu\nu}\\
&&\nonumber\\
&=& \frac{1}{2}\left({\cal{M}}-  {\cal{J}}\right)\hbox{Re}(\chi^L)+ \frac{1}{2}\left({\cal{M}}+  {\cal{J}}\right)\hbox{Re}(\chi^R)
\label{eq:qwert}
\end{eqnarray}
The quantity $\chi^L$ is given in (\ref{eq:ivo}), and $\chi^R$ is its right handed pendant, which is obtained from $\chi^L$ by exchanging $u\leftrightarrow v$.  The result \eqref{eq:qwert} then establishes the relation between the singular modes  (\ref{eq:ivo}) and localized point sources. Therefore, we call these modes ``point particle modes''. The ADM mass of these perturbations is finite (in fact zero at linear order). 
This result is in agreement with the energy computed in the Hamiltonian formalism. 
 The divergence of the perturbative gravitational energy \eqref{EG} comes about because back-reaction has not been taken into account.

\paragraph{Chiral boundary conditions} With the overall sign of the action as in \eqref{eq:f1} black holes in CCTMG have positive energy \cite{Li:2008dq} and gravitons negative energy \cite{Grumiller:2008qz}. This is problematic for the theory. It is suggestive, therefore, to truncate the theory, either by eliminating negative energy gravitons or by reversing the sign of the action and eliminating the now negative energy black holes. We discuss these two possibilities in some detail and show at the linearized level that one can achieve either of them by imposing suitable boundary conditions, which we call ``chiral boundary conditions''.

We propose now boundary conditions on the fluctuation of the metric $\psi$ that eliminate $\psi^L, \psi^{\rm log}$ and all their descendants. 
\begin{eqnarray}
\psi &\simeq&\left(\begin{array}{l@{\quad}l@{\quad}l}\psi_{uu}=m&\psi_{uv}={\cal O}(1)& \psi_{u\rho}={\cal O}(e^{-2\rho})\cr & \psi_{vv}={\cal O}(1)& \psi_{v\rho}={\cal O}(e^{-2\rho})\cr & & \psi_{\rho\rho}={\cal O}(e^{-2\rho}) \end{array}\right)
\label{eq:f100}
\end{eqnarray}
Here $m$ is a {\em fixed} constant. The boundary conditions \eqref{eq:f100} are more restrictive than the Brown--Henneaux boundary conditions \eqref{eq:fBH} because $\psi_{uu}=m$ cannot vary. Consequently, the asymptotic symmetry group does not consist of two Virasoro copies Virasoro$_L\times$Virasoro$_R$, but is broken to $U(1)_L\times$Virasoro$_R$. This can be shown as usual, by considering the generators of diffeomorphisms $\zeta^\mu$ that preserve the fall-off behavior postulated in \eqref{eq:f100}:
\begin{subequations}
\begin{align}
\zeta^u &= \zeta^u_0 + 2\partial_v^2\zeta^v_0(v)\,e^{-2\rho} + {\cal O}(e^{-4\rho}) \\
\zeta^v &= \zeta^v_0 (v) + {\cal O}(e^{-4\rho}) \\
\zeta^\rho &= -\frac12\,\partial_v \zeta^v_0(v) + {\cal O}(e^{-2\rho})
\end{align} 
\end{subequations}
The $U(1)_L$ is generated by the constant $\zeta^u_0$ and the Virasoro$_R$ is generated by the function $\zeta^v_0 (v)$. All subleading terms are also independent of the light-cone coordinate $u$ because $\partial_u\zeta^\mu=0$ is required to all orders in $e^{-2\rho}$.
It is also easy to see why \eqref{eq:f100} eliminates $\psi^L$ and $\psi^{\rm log}$ but not the right-moving primary $\psi^R$. This is a consequence of requiring $m$ to be fixed. The descendants are obtained by acting (repeatedly) with the remaining Virasoro$_R$ generators, $\bar L_{-n}$ \cite{Brown:1986nw}, on the primaries. The vanishing $uu$-com\-po\-nent of $\psi^R$ changes under Lie-derivative along a vector field $\zeta$ as follows:
\eq{
{\cal L}_\zeta \psi^R_{uu} = 2\psi^R_{u\mu}\,\partial_u\zeta^\mu
}{eq:f109}
The vector fields associated with the generators $\bar L_{-n}$ are all independent of the light-like coordinate $u$. Therefore all descendants of $\psi^R$ have a vanishing $uu$-com\-po\-nent, and we can generalize our conclusions to all descendants: the left and the logarithmic sector are eliminated, while the right sector remains intact.  Thus, at the linearized level we do have a consistent chiral theory where all energies are positive. 

Essentially the same story is true for boundary conditions that eliminate only the right moving sector.
\begin{eqnarray}
\psi&\simeq&\left(\begin{array}{l@{\quad}l@{\quad}l}\psi_{uu}={\cal O}(\rho)&\psi_{uv}={\cal O}(1)&\psi_{u\rho}={\cal O}(\rho\, e^{-2\rho})\cr&\psi_{vv}=m&\psi_{v\rho}={\cal O}(e^{-2\rho})\cr & &\psi_{\rho\rho}={\cal O}(e^{-2\rho}) \end{array}\right)
\label{BC}
\end{eqnarray}
The ensuing theory contains the left-moving boundary graviton $\psi^L$ as well as the logarithmic mode $\psi^{\rm log}$ and their appropriate descendants. Comparison with the Brown--Henneaux case \eqref{eq:fBH} exhibits two differences: The component $\psi_{vv}$ in \eqref{BC} is required to be fixed to some constant of order of unity, while in the Brown--Henneaux case it is allowed to vary at order of unity. This restriction eliminates the right-moving boundary graviton and its descendants, analog to the previous case. In addition, the components $\psi_{uu}, \psi_{u\rho}$ are linearly divergent in $\rho$ as compared to their Brown--Henneaux counterpart. This generalization allows for the logarithmic mode and its descendants.  The asymptotic symmetry group is broken to $U(1)_R\times$Virasoro$_L$. The generators of diffeomorphisms $\zeta^\mu$ that preserve the fall-off behavior postulated in \eqref{BC} are given by
\begin{subequations}
\begin{align}
\zeta^u &= \zeta^u_0(u) + {\cal O}(e^{-4\rho}) \\
\zeta^v &= \zeta^v_0 + 2\partial_u^2\zeta_0^u(u) e^{-2\rho} + {\cal O}(\rho\,e^{-4\rho}) \\
\zeta^\rho &= -\frac12\,\partial_u \zeta^u_0(u) + {\cal O}(e^{-2\rho})
\end{align} 
\end{subequations}
where the subleading terms are such that the conditions $\partial_v\zeta^\mu=0$ hold.
With reversed overall sign of the action as compared to \eqref{eq:f1} all modes have now non-negative energy. 

\section{Some hypergeometric identities}\label{app:hyper}

Frequently we use the Euler transformation
\eq{
_2F_1(a,\,b,\,c;\,z) = (1-z)^{c-a-b}\,_2F_1(c-a,\,c-b,\,c;\,z)
}{eq:hyper0}
Often we use relations between contiguous functions, for instance
\begin{multline}
(a-1-(n-b)z)\,_2F_1(a,b,n+1;z)+(n+1-a)\,_2F_1(a-1,b,n+1;z) \\
= n(1-z)\,_2F_1(a,b,n;z)
\label{eq:contiguous}
\end{multline}
Similar relations can be found in standard literature, like \cite{Erdelyi}.

For integer values $\bar h \leq -1$, $h\geq-1$ we obtain from \eqref{eq:mass6}:
\eq{
F_{uv}=\tilde a (x-1)^{(h-\bar h)/2}(x+1)^{-(h+\bar h)/2}\,_2F_1(-\bar h,\,-\bar h + 1,\,-\bar h+1+h;\,\frac{1-x}{2}\big)
}{eq:hyper1}
The case $\bar h=-1$ is treated in detail in appendix \ref{app:L}. For $\bar h<-1$ identities exist that allow a simple expansion of the hypergeometric function appearing in \eqref{eq:hyper1} in terms of elementary functions. For instance, if $\bar h=-2$ we obtain
\begin{multline}
_2F_1\big(2,\,3,\,3+h;\,\frac{1-x}{2}\big)=-\frac{2}{(x+1)^3}\Big(2h(h+1)(x+h)\,_2F_1\big(1,\,1,\,3+h;\,\frac{1-x}{2}\big)\\
-x(h^2+3h+2)-2h^3-3h^2+h-2\Big)
\label{eq:hyper2}
\end{multline}
and
\eq{
_2F_1\big(1,\,1,\,3+h;z\big)=\frac{(h+2)z}{(z-1)^2}\,\Big(\sum_{k=2}^{h+2}\frac{(1-1/z)^k}{h+3-k}-(1-1/z)^{h+3}\ln{(1-z)}\Big)
}{eq:hyper3}
For other integer values of $\bar h<-2$ similar identities exist.
For integer values $\bar h \leq -1$, $h\geq-1$ we obtain from \eqref{eq:mass5}:
\eq{
F_{vv} = a_2 2^{h+\bar h}\,(x-1)^{(h-\bar h)/2}(x+1)^{-(h+\bar h)/2}\,_2F_1\big(-\bar h-1,\,-\bar h+2,\,h-\bar h+1;\, \frac{1-x}{2} \big)
}{eq:hyper4}
The case $\bar h=-1$ is treated in detail in appendix \ref{app:L}. For $\bar h<-1$ identities exist that allow a simple expansion of the hypergeometric function appearing in \eqref{eq:hyper1} in terms of elementary functions. For instance, if $\bar h=-2$ we obtain
\begin{multline}
_2F_1\big(1,\,4,\,3+h;\,\frac{1-x}{2}\big)=\frac{2}{3(x+1)^3}\,\Big(2h(1-h^2)\,_2F_1\big(1,\,1,\,3+h,\,\frac{1-x}{2}\big)\\
+x^2(2+h)+x(-h^2+h+6)+2h^3+h^2-4h+4\Big)
\label{eq:hyper5}
\end{multline}
and can again use \eqref{eq:hyper3} to evaluate \eqref{eq:hyper5} in terms of elementary functions. For other integer values of $\bar h<-2$ similar identities exist. These identities are consequences of relations between contiguous functions. 

If $h=0,\pm 1$ the hypergeometric function appearing in \eqref{eq:mass5} for $\eps=0$ becomes a (Jacobi) polynomial of degree $-h-1$. If $\bar h \leq -2$ and $h\geq 2$ we can exploit the following representation of the hypergeometric function:
\begin{multline}
_2F_1\big(h-1,\,h+2,\,h+1-\bar h;\,\frac{1-x}{2}\big) = \frac{(-1)^{-\bar h+1}(-2)^{h-1}(h-\bar h)!}{(-2-\bar h)!(1-\bar h)!(h-2)!(h+1)!}\\
\cdot \frac{\extd^{h+1}}{\extd x^{h+1}}\Big((x+1)^{1-\bar h}\,\frac{\extd^{-2-\bar h}}{\extd x^{-2-\bar h}}\Big(\frac{\ln{\frac{x+1}{2}}}{x-1}\Big)\Big)
\label{eq:hyper7}
\end{multline}
This formula allows to express all regular non-normalizable left modes in terms of elementary functions, except for the special cases discussed already. Trivially, we can also express the hypergeometric function in \eqref{eq:mass6} for $\eps=0$ as follows:
\begin{multline}
_2F_1\big(-\bar h,\,1-\bar h,\,-\bar h+h+1;\,\frac{1-x}{2}\big) = \frac{(-1)^{h}(-2)^{-\bar h}(h-\bar h)!}{(-1-\bar h)!(-\bar h)!(h-1)!h!}\\
\cdot \frac{\extd^{-\bar h}}{\extd x^{-\bar h}}\Big((x+1)^h\,\frac{\extd^{h-1}}{\extd x^{h-1}}\Big(\frac{\ln{\frac{x+1}{2}}}{x-1}\Big)\Big)
\label{eq:hyper11}
\end{multline}
if $\bar h\leq -1$ and $h\geq 1$.

Checking regularity of modes requires an expansion around $x=1$. This is straightforward:
\eq{
_2F_1\big(a,\,b,\,c;\,\frac{1-x}{2}\big) = 1 - \frac{ab}{2c}\,(x-1) + {\cal O}(x-1)^2
}{eq:hyper6}
Therefore, the regularity or singularity at the origin is entirely due to the behavior of the polynomial pre-factors in \eqref{eq:mass5} and \eqref{eq:mass6}. 

Checking (non-)normalizability requires an asymptotic expansion. For the massive branch with $h\geq \bar h$ and non-integer values of the Chern--Simons coupling constant $\mu$ we obtain
\eq{
\lim_{x\to\infty}\, _2F_1\big(a,\,b,\,h-\bar h+1;\,\frac{1-x}{2}\big)\propto x^{-c}\qquad c={\rm min}\,\{a,b\} 
}{eq:hyper9} 
For the left branch and integer weights $h\geq \bar h$ the hypergeometric function of interest takes the form
\begin{align}
\!\!\!\!& \lim_{x\to\infty}\, _2F_1\big(-\bar h-1,\,-\bar h+2,\,h-\bar h+1;\,\frac{1-x}{2}\big) \propto x^{\bar h +1} \quad \textrm{if\;} \bar h \leq 1\;\textrm{and}\;h\geq -1\label{eq:hyper7a}\\
\!\!\!\!& \lim_{x\to\infty}\, _2F_1\big(-\bar h-1,\,-\bar h+2,\,h-\bar h+1;\,\frac{1-x}{2}\big) \propto x^{\bar h -2} \quad \textrm{otherwise}
\label{eq:hyper7b}
\end{align}
Thus, for $\bar h\geq 2$ the left modes are normalizable, while for $\bar h\leq 1$ they are non-normalizable as long as $h\geq -1$. However, the three cases $\bar h =0,\pm 1$ have to be treated separately because the solutions of the second order differential equation \eqref{eq:mass3a} are not necessarily compatible with the first order system \eqref{eq:ODEa}, \eqref{eq:ODEb} for these values of $\bar h$ or with the algebraic system \eqref{eq:alg}. The case $\bar h=-1$ is treated in appendix \ref{app:L} and turns out to be consistent. We treat here the other two cases and assume again $h\geq \bar h$.
If $\bar h=0$ then the first three algebraic conditions \eqref{eq:alg} (for $\mu=1$) establish $F_{uv}=0$ and $h=1$. Thus, there is only one non-normalizable left mode if $\bar h=0$. This mode, however, has a singular $\rho\rho$ component, so there are no regular non-normalizable left modes if $\bar h=0$.
If $\bar h=1$ then the first order equation \eqref{eq:ODEb} decouples and yields
$
F_{vv}\propto (x+1)^{(h+1)/2}(x-1)^{(1-h)/2}
$.
The ensuing modes are non-normalizable, but not regular at the origin in the $vv$ component unless $h=1$. Thus, there is no regular non-normalizable left mode if $\bar h=1$, unless $h=1$.
Suppose that $h>1$ and $\bar h<-1$ (otherwise we recover one of the special cases discussed separately above). Then we can exploit the asymptotic expansions
\begin{multline}
_2F_1\big(-\bar h,\,-\bar h+1,\,h-\bar h+1;\,\frac{1-x}{2}\big) = \frac{(h-\bar h)!}{h!(-\bar h)!}\,\big(\frac{x-1}{2}\big)^{\bar h}\,\Big(1 \\
+\frac{2h\bar h}{x}\,\big(\ln{\frac{x}{2}}-\psi(h)-\psi(1-\bar h)+1-2\ga\big)+{\cal O}(\ln{x}/x^2)\Big)
\label{eq:hyper12}
\end{multline} 
and
\begin{multline}
_2F_1\big(h-1,\,h+2,\,h-\bar h+1;\,\frac{1-x}{2}\big) = \frac{2(h-\bar h)!}{(h+1)!(1-\bar h)!}\,\big(\frac{x-1}{2}\big)^{1-h}\,\Big(1 \\
- \frac{(h-1)(1-\bar h)}{x-1} + \frac{h(h-1)(1-\bar h)(-\bar h)}{(x-1)^2} 
-\frac{2(1-\bar h)(-\bar h)(-\bar h-1)(h+1)h(h-1)}{3x^3} \\
\cdot \big(\ln{\frac{x}{2}}-\psi(h+2)-\psi(-\bar h-1)+\frac{11}{6}-2\ga\big)+{\cal O}(\ln{x}/x^4)\Big)
\label{eq:hyper13}
\end{multline} 
Here $\ga$ is the Euler--Mascheroni constant and $\psi(z)=\Ga^\prime(z)/\Ga(z)$ is the digamma-function, with the asymptotic expansion for large weights
\eq{
\psi(h)=\ln{h}-\frac{1}{2h}+{\cal O}(1/h^2)
}{eq:hyper14}

In some considerations the limit for large weights is of interest. To obtain a formula for hypergeometric functions in this limit we proceed as follows. We shift $h\to h+\la$, $\bar h\to\bar h-\la$ and let $\la\to\infty$. The hypergeometric function in \eqref{eq:mass5} leads to the following limit
\eq{
\lim_{\la\to\infty}\,_2F_1\big(\la+h+2,\,\la+h-1,\,2\la+h-\bar h+1;\,\frac{1-x}{2}\big) = \Xi(\la,x)\,\big(1+{\cal O}(1/\la)\big)
}{eq:hyper15}
The function $\Xi(\la,x)$ was derived by Dr.~Watson in 1918 \cite{Watson:1918}:
\eq{
\Xi(\la,x)=f(x)\,\Big(\frac{x-1}{2}\Big)^{-\la}\,\la^{-1/2}\,\frac{\Ga(2\la+h-\bar h+1)}{\Ga(\la+h-1)\,\Ga(\la+2-\bar h)}\,\Big(1+\frac{4-2\sqrt{2(x+1)}}{x-1}\Big)^\la
}{eq:hyper16}
where $f(x)$ is a known $\la$-independent function of $x$. The factor $(x-1)^{-\la}$ cancels with the polynomial pre-factors multiplying the hypergeometric function in \eqref{eq:mass5}. The $\Ga$-functions and powers of $2$ cancel mostly with factors from the normalization constant $a_2$, see footnote \ref{fn:app} on p.~\pageref{fn:app}. In the end a factor $\la^{5/2}$ remains, multiplied by finite terms and by the last bracket in \eqref{eq:hyper16}.  A key observation is that in the range of definition $x\in[1,\infty)$ the last bracket in \eqref{eq:hyper16} is always smaller than 1, and thus goes to zero rapidly as $\la$ tends to $\infty$ so that the whole expression vanishes. The only exception arises in the limit of large $x$, which has to be treated separately. This is an explicit realization of the UV/IR connection \cite{Susskind:1998dq}: the limit of large weights/small distances (UV) on the CFT side implies the limit of large $x$ (IR) on the gravity side.
If $x$ scales like $\lambda^{2+\eps}$ the function $f(x)$ in \eqref{eq:hyper16} behaves as
\eq{
\lim_{\la\to\infty}f(\lambda^{2+\eps} x)\propto \la^{-5(2+\eps)/4} 
}{eq:hyper17}
whereas the last bracket in \eqref{eq:hyper16} behaves as
\eq{
\lim_{\la\to\infty}\Big(1+\frac{4-2\sqrt{2(\lambda^{2+\eps}x+1)}}{\lambda^{2+\eps}x-1}\Big)^\la = \begin{cases} 
1 & {\rm\;if\;} \eps>0,\\ 
0 & {\rm\;if\;} -2<\eps<0,\\
e^{-2\sqrt{2/x}} & {\rm\;if\;} \eps=0 \end{cases}
}{eq:hyper18}

\section{Non-normalizable modes}\label{app:C}

\subsection{Left and right branch}\label{app:L}

The right modes are obtained from the left modes by exchanging $u\leftrightarrow v$ and $h\leftrightarrow \bar h$ in all formulas below. The left modes are obtained as follows. We start from the separation Ansatz
\eq{
\psi_{\mu\nu}^L = e^{-ihu-i\bar h v}\,F_{\mu\nu}(\rho)
}{eq:L}
and solve the {\eom} \eqref{eq:alg}, \eqref{eq:ODEs} with $\mu=1$, assuming regularity at $x=\cosh{(2\rho)}=1$. Below we provide explicit results for the tensor $F_{\mu\nu}$.

\paragraph{Generic case} We summarize here our results for generic regular non-normalizable left modes with weights $h\geq 2$ and $\bar h\leq -2$.\footnote{\label{fn:app} We have chosen the overall normalization such that $F_{vv}=x+{\cal O}(1)$ asymptotically. This choice leads to $a_2=2^{-h}(h+1)!(1-\bar h)!/(h-\bar h)!$ and $\tilde a = 2^{\bar h} (1-h^2) h! (-\bar h)! / (h-\bar h)!$ in \eqref{eq:mass5}, \eqref{eq:mass6}.}
\begin{subequations}
\label{eq:genL}
\begin{align}
F_{vv} &= \frac{(-1)^{h-\bar h}(x-1)^{(h-\bar h)/2}(x+1)^{(h+\bar h)/2}}{2\,(-2-\bar h)!\,(h-2)!}\,\frac{\extd^{h+1}}{\extd x^{h+1}}\Big((x+1)^{1-\bar h}\,\frac{\extd^{-2-\bar h}}{\extd x^{-2-\bar h}}\Big(\frac{\ln{\frac{x+1}{2}}}{x-1}\Big)\Big) \\
F_{uv} &= \frac{(-1)^{h-\bar h}(1-h^2)(x-1)^{(h-\bar h)/2}(x+1)^{-(h+\bar h)/2}}{(-1-\bar h)!\,(h-1)!}\, \frac{\extd^{-\bar h}}{\extd x^{-\bar h}}\Big((x+1)^h\,\frac{\extd^{h-1}}{\extd x^{h-1}}\Big(\frac{\ln{\frac{x+1}{2}}}{x-1}\Big)\Big)\\
F_{uu} &= \frac{h}{\bar h}\,F_{uv} \\
F_{v\rho} &= \frac{2i}{\sqrt{x^2-1}}\,\big(\bar h F_{uv}-h F_{vv}\big)\\ 
F_{u\rho} &= \frac{2i}{h}\,\sqrt{x^2-1}\,\frac{\extd F_{uv}}{\extd x}\\
F_{\rho\rho} &= \frac{4}{x^2-1}\,\big((2x-\frac{h}{\bar h})\,F_{uv} -F_{vv}\big)
\end{align}
\end{subequations}
The results above follow from \eqref{eq:hyper7}, \eqref{eq:hyper11} and \eqref{eq:alg}.

\paragraph{Special case $h=1$} Modes with weights $h=1$ and $\bar h\leq -2$ are given by:
\begin{subequations}
\label{eq:Lspec1}
\begin{align}
F_{vv} &= (x-1)^{(1-\bar h)/2}(x+1)^{(1+\bar h)/2}\\
F_{uv} &= 0\\
F_{uu} &= 0\\
F_{v\rho} &= -2i(x-1)^{-\bar h/2}(x+1)^{\bar h/2}\\
F_{u\rho} &= 0 \\
F_{\rho\rho} &=-4(x-1)^{(-1-\bar h)/2}(x+1)^{-(1-\bar h)/2} 
\end{align}
\end{subequations}

\paragraph{Special case $h=0$} Modes with weights $h=0$ and $\bar h\leq -2$ are given by:
\begin{subequations}
\label{eq:Lspec0}
\begin{align}
F_{vv} &= (x-\bar h)(x-1)^{-\bar h/2}(x+1)^{\bar h/2} \\
F_{uv} &= (x-1)^{-\bar h/2}(x+1)^{\bar h/2} \\
F_{uu} &= 0 \\
F_{v\rho} &= 2i\bar h\,(x-1)^{(-1-\bar h)/2}(x+1)^{-(1-\bar h)/2} \\
F_{u\rho} &= -2i(x-1)^{(-1-\bar h)/2}(x+1)^{-(1-\bar h)/2} \\
F_{\rho\rho} &= 4(x+\bar h)(x-1)^{-1-\bar h/2}(x+1)^{-1+\bar h/2}
\end{align}
\end{subequations}

\paragraph{Special case $h=-1$} Modes with weights $h=-1$ and $\bar h\leq -2$ are given by:
\begin{subequations}
\label{eq:Lspecm1}
\begin{align}
F_{vv} &= (x^2-2\bar h x+2\bar h^2-1)(x-1)^{(-1-\bar h)/2}(x+1)^{-(1-\bar h)/2}\\
F_{uv} &= -4\bar h (x-1)^{(-1-\bar h)/2}(x+1)^{-(1-\bar h)/2} \\
F_{uu} &= 4(x-1)^{(-1-\bar h)/2}(x+1)^{-(1-\bar h)/2} \\
F_{v\rho} &= 2i\,(x^2-2\bar h x-2\bar h^2-1)(x-1)^{-1-\bar h/2}(x+1)^{-1+\bar h/2}\\ 
F_{u\rho} &= 8i\,(x+\bar h)(x-1)^{-1-\bar h/2}(x+1)^{-1+\bar h/2}\\
F_{\rho\rho} &= -4(x^2+6\bar h x+2\bar h^2+3)(x-1)^{(-3-\bar h)/2}(x+1)^{-(3-\bar h)/2}
\end{align}
\end{subequations}

\paragraph{Special case $\bar h=-1$} If $\bar h=-1$ the first order equations \eqref{eq:ODEs} decouple for $\eps=0$. Solving the homogeneous first order equation for $F_{vv}$ yields
$
F_{vv} = (x-1)^{(1+h)/2} (x+1)^{(1-h)/2} 
$.
The first order equation \eqref{eq:ODEa} determines $F_{uv}$ in terms of elementary functions. The integration constant is determined uniquely by the requirement of regularity at $x=1$. We obtain
\begin{subequations}
\label{eq:solLNN}
\begin{align}
F_{vv} &= (x-1)^{(1+h)/2}(x+1)^{(1-h)/2}\, \\
F_{uv} &= (x-1)^{(1+h)/2}(x+1)^{(1-h)/2}\,H(x) \\
F_{uu} &= -h\,(x-1)^{(1+h)/2}(x+1)^{(1-h)/2}\,H(x) \\
F_{v\rho} &= -2i\,(x-1)^{h/2}(x+1)^{-h/2}\,\big(H(x)+h\big)\\
F_{u\rho} &= 2i\,(x-1)^{h/2}(x+1)^{-h/2}\,\big((x+h)\,H(x)+h^2-1\big) \\
F_{\rho\rho} &= 4\, (x-1)^{(-1+h)/2}(x+1)^{-(1+h)/2}\,\big((2x+h)\,H(x)-1\big)
\end{align}
\TABLE[t]{
\begin{tabular}{|cccccccccccc|}\hline
$h$: & -1 & 0 & 1 & 2 & 3 & 4 & 5 & 6 & 7 & 8 & 9  \\
$\bar a$: & -4 & 2 & 0 & 0 & 24 & 100 & 260 & 539 & 4872/5 & 8028/5 & 17316/7 \\ \hline 
\end{tabular}
\caption{Values of the integration constant $\bar a$ in \eqref{eq:HL} for small weights $h$}
\label{tab:1}
}
The function $H(x)$ is given by
\begin{multline}
H(x)=(1-h^2)(x+1)^{h-1}(x-1)^{-h-1} \Big(x+1-2h\, \ln{\frac{x+1}{2}}\\
-\sum_{k=2}^{h}\left(\begin{array}{c} h \\ k \end{array}\right)\frac{(-2)^k}{k-1}\,(x+1)^{1-k}\Big) - \bar a\,(x+1)^{h-1}(x-1)^{-h-1} 
\label{eq:HL}
\end{multline}
with the integration constant $\bar a$ determined uniquely from the requirement of regularity at the origin.\footnote{The form of \eqref{eq:HL} does not make it completely evident that all poles at $x=1$ can be cancelled by a single choice of parameter. However, it is clear that this must be possible since the alternative form of $H(x)$ in terms of a hypergeometric function, \eqref{eq:mass6}, manifestly is regular at $x=1$ for $h\geq\bar h$.} 
\eq{
\bar a = 2(1-h^2)\Big(1-\sum_{k=2}^h\frac{(-1)^k h!}{k!(h-k)!(k-1)}\Big)-4\,\de_{h,-1}
}{eq:barc}
\end{subequations}
See table \ref{tab:1} for small values of the weight $h$. 
Note that in the sums above we have the usual convention that they evaluate to zero if $h<2$.
Near the origin the function $H(x)$ behaves like $(1-h)/2+{\cal O}(x-1)$. Asymptotically we obtain
$\lim_{x\to\infty} H(x) = (1+2h)(1-h^2)/x +{\cal O}(\ln{x}/x^2)$.

\paragraph{Examples with $\bar h=-1$} For convenience some explicit examples are presented below. All modes below have weight $\bar h=-1$.

\paragraph{$h=-1$:}
\begin{subequations}
\label{eq:hm1}
\begin{align}
F_{vv}&=x+1\\
F_{uv}&=\frac{4}{x+1}\\
F_{uu}&=\frac{4}{x+1}\\
F_{v\rho}&=2i(x+3)\sqrt{\frac{x-1}{(x+1)^3}}\\
F_{u\rho}&=8i\sqrt{\frac{x-1}{(x+1)^3}}\\
F_{\rho\rho}&=-\frac{4(x-5)}{(x+1)^2}
\end{align}
\end{subequations}

\paragraph{$h=0$:}
\begin{subequations}
\begin{align}
F_{vv}&=\sqrt{x^2-1} \\
F_{uv}&=\sqrt{\frac{x-1}{x+1}} \\
F_{uu}&=0\\
F_{v\rho}&=-\frac{2i}{x+1}\\
F_{u\rho}&=-\frac{2i}{x+1}\\
F_{\rho\rho}&=4\sqrt{\frac{x-1}{(x+1)^3}}
\end{align}
\end{subequations}

\paragraph{$h=1$:}
\begin{subequations}
\begin{align}
F_{vv}&=x-1 \\
F_{uv}&=0\\
F_{uu}&=0\\
F_{v\rho}&=-2i\sqrt{\frac{x-1}{x+1}}\\
F_{u\rho}&=0\\
F_{\rho\rho}&=-\frac{4}{x+1}
\end{align}
\end{subequations}
The mode $h=1$, $\bar h=-1$ is of particular interest, because it has angular momentum 2 (``non-norma\-lizable boundary graviton''), a property it shares with all primaries. 

\paragraph{$h=2$:}
\begin{subequations}
\label{eq:Lgen}
\begin{align}
F_{vv}&=(x-1)\sqrt{\frac{x-1}{x+1}} \\
F_{uv}&=-3\sqrt{\frac{x+1}{(x-1)^3}}\,(x+1-4\ln{\frac{x+1}{2}}-\frac{4}{x+1})\\
F_{uu}&=6\sqrt{\frac{x+1}{(x-1)^3}}\,(x+1-4\ln{\frac{x+1}{2}}-\frac{4}{x+1})\\
F_{v\rho}&=-\frac{2i (7 + x^2 (-9 + 2 x) + 12 (1 + x)\ln{\frac{x+1}{2}})}{(x-1)^2 (1 + x)}\\
F_{u\rho}&=\frac{6i (5 + (2 - 7 x) x + 4 (1 + x) (2 + x) \ln{\frac{x+1}{2}})}{(x-1)^2 (1 + x)}\\
F_{\rho\rho}&=4\frac{19 + 3 x - 15 x^2 - 7 x^3 + 
     24(1 + x)^2\ln{\frac{x+1}{2}}}{(x-1)^{5/2} (x+1)^{3/2}} 
\end{align}
\end{subequations}
For $h=2$ logarithmic (asymptotically subleading) terms appear in most of the components. This is a generic feature of all modes with $h\geq 2$. It is worthwhile emphasizing that all these modes are regular at $x=1$, despite of the appearance of $1/(x-1)$ factors in various components (e.g.~the factor $1/(x-1)^{5/2}$ in $F_{\rho\rho}$ above). Using the algebraic relations \eqref{eq:algebraic} we can generate all regular non-normalizable left modes from the $(2,-1)$ mode above. 



\subsection{Logarithmic branch}\label{app:log}

The logarithmic modes 
\eq{
\psi^{\rm log}_{\mu\nu}=i(u+v)\psi_{\mu\nu}^L-F_{\mu\nu}^{\rm log}\,e^{-ihu-i\bar h v}
}{eq:log}
are specified uniquely --- up to addition of left modes and overall rescalings --- by providing the tensor $F_{\mu\nu}^{\rm log}$. Once the components $F_{vv}^{\rm log}$ and $F_{uv}^{\rm log}$ are known the remaining ones follow algebraically from the relations \eqref{eq:alglog}. The components $F_{vv}^{\rm log}$ and $F_{uv}^{\rm log}$ are determined from solutions of linear ordinary first order differential equations \eqref{eq:ODElog}.

\paragraph{Generic modes} Using the algebraic relations \eqref{eq:logalgebraic} we can generate all regular non-normalizable logarithmic modes from the $(2,-1)$ mode, see \eqref{eq:loggen} below. We do not list generic modes explicitly.

\paragraph{Special case $\bar h=-1$} We assume here $\bar h=-1$ after taking the limit in \eqref{eq:log1}.
In order to evaluate \eqref{eq:log1} we start with the massive branch solution \eqref{eq:mass2}-\eqref{eq:mass6} and assume that $\eps$ is small and positive. In addition we vary the weights with $\eps$, so that the following two conditions hold
\begin{align}
\bar h &=-1+\eps \\
h - \bar h &= n\qquad n\in\mathbb{N}
\end{align}
The first condition ensures that we have the same periodicity properties as for corresponding normalizable massive modes (see (47) in \cite{Li:2008dq}).\footnote{Unlike normalizable ones there exist non-normalizable solutions with $h,\bar h$ independent from $\eps$.} It implies that the differential equations decouple for any $\eps$, so that we obtain a first order equation for $F_{vv}$. The second condition is necessary since we want to keep periodicity in the angle $\phi$ at each step, which requires the difference $h-\bar h$ to be an integer. 
The assumption $h\geq\bar h$ implies that $n$ is a natural number. The attribute ``non-norma\-lizable'' refers to growth faster than asymptotically AdS, i.e., the modes $\psi$ are not compatible with the Fefferman--Graham expansion \eqref{eq:FG} but violate it logarithmically, see \eqref{eq:A6}. As explained in \cite{Skenderis:2009nt} this is the expected behavior for the source terms of the operators associated with the logarithmic modes.

We obtain the following expression for the massive mode:
\begin{subequations}
\eq{
\psi^\eps_{\mu\nu} = -e^{-i(n-1+\eps)u+i(1-\eps)v}\,(x-1)^{n/2}\,(x+1)^{1-n/2-\eps}\,H^\eps_{\mu\nu}(x)
}{eq:mm1}
where
\begin{align}
H^\eps_{vv} &= 1 \\
H^\eps_{uv} &= H^\eps(x) \\
H^\eps_{uu} &= -\be H^\eps(x) + \eps n(n-2) + {\cal O}(\eps^2)\\
H^\eps_{v\rho} &= -\frac{2i}{\sqrt{x^2-1}}\,\big(H^\eps(x)+n-1+\eps n+{\cal O}(\eps^2)\big)\\
H^\eps_{u\rho} &= \frac{2i}{\sqrt{x^2-1}}\,\big((x+\be)H^\eps(x)+n(n-2)+2\eps n(n-1)+{\cal O}(\eps^2)\big)\\
H^\eps_{\rho\rho} &= \frac{4}{x^2-1}\,\big((2x+\be)H^\eps(x)-1-\eps n(n-2)+{\cal O}(\eps^2)\big)
\end{align}
with $\be=n-1-\eps(x-1)$ and
\eq{
H^\eps(x)=\frac{1-n/2-\eps}{1-\eps}\,_2F_1\big(1,\,2(1-\eps),\,n+1;\,\frac{1-x}{2}\big)
}{eq:H}
\end{subequations}
In the limit $\eps\to 0$ the function $H^\eps(x)$ approaches the function $H(x)$ given in \eqref{eq:HL}, while the parameter $\be$ approaches $h=n-1$. Consequently, the solution \eqref{eq:mm1} up to normalization coincides with the non-normalizable regular left modes \eqref{eq:solLNN} with weights $\bar h=-1$ and $h=n-1$.  

The regular non-normalizable logarithmic modes are determined from \eqref{eq:mm1}-\eqref{eq:H} using the definition \eqref{eq:log1}. Our final result is 
\begin{multline}
\psi^{\rm log}_{\mu\nu}(h=n-1, \bar h=-1) = \big(i(u+v)+\ln{\frac{x+1}{2}}\big)\,\psi^L_{\mu\nu} \\
-e^{-i(n-1)u+iv}\,(x-1)^{n/2}\,(x+1)^{1-n/2}\,\frac{\extd H^\eps_{\mu\nu}}{\extd\eps}\Big|_{\eps=0}
\label{eq:logresult}
\end{multline}
where $\psi^L_{\mu\nu}$ are the regular non-normalizable left modes with weights $h=n-1$, $\bar h=-1$ given in \eqref{eq:solLNN}. 
One can add to $\psi^{\rm log}$ a regular non-normalizable left moving mode with the same weights without changing anything essential. We fix this shift ambiguity in a convenient way in \eqref{eq:logresult}, but in the body of the paper we freely add or subtract such left moving modes to simplify some expressions. This is related to a well-known ambiguity in LCFTs.
The first line of our solution for the logarithmic modes \eqref{eq:logresult} coincides with the relation between normalizable logarithmic and left modes (see Eqs.~(3.1), (3.2) in \cite{Grumiller:2008qz}). The second line contains the function $H^\eps(x)$ \eqref{eq:H} to next to leading order in $\eps$.
In order to obtain $H^\eps(x)$ for arbitrary $n$ we use a relation between contiguous functions \eqref{eq:contiguous} and exploit $_2F_1(0,b,n;z)=1$ to establish a recursion relation
\begin{multline}
_2F_1\big(1,\,2(1-\eps),\,n+1;\,\frac{1-x}{2}\big)\\
=\frac{2n}{(1-x)(n-2(1-\eps))}\,\Big(1-\frac{x+1}{2}\,_2F_1\big(1,\,2(1-\eps),\,n;\,\frac{1-x}{2}\big)\Big)
\label{eq:recursion}
\end{multline}
Thus, we need to evaluate by hand only the starting point $n=1$, which is very simple:
\eq{
_2F_1\big(1,\,2(1-\eps),\,1;\,\frac{1-x}{2}\big) = \big(\frac{x+1}{2}\big)^{-2(1-\eps)} 
}{eq:2F1expand}
Higher values of $n$ are then obtained recursively from \eqref{eq:recursion}.
From our result for regular non-normalizable logarithmic modes \eqref{eq:logresult} we obtain
\eq{
L_0 \psi^{\rm log} = (n-1) \psi^{\rm log} - \psi^L\qquad\bar L_0 \psi^{\rm log} = -\psi^{\rm log} - \psi^L
}{eq:L0log}
This result essentially coincides with the result \eqref{eq:cg31} for normalizable modes for $h=n-1$, $\bar h=-1$. See also the algebraic relations \eqref{eq:logalgebraic}.

\paragraph{Examples with $\bar h=-1$} For convenience some explicit examples are presented below. All modes below have weight $\bar h=-1$. 

\paragraph{$h=-1$:} 
\begin{subequations}
\begin{align}
F^{\rm log}_{vv} &= -(x+1)\,\ln{\frac{x+1}{2}}\\
F^{\rm log}_{uv} &= \frac{4\ln{\frac{x+1}{2}}}{x+1}\\
F^{\rm log}_{uu} &= \frac{4\big(x+\ln{\frac{x+1}{2}}-1\big)}{x+1}\\
F^{\rm log}_{v\rho} &= -\frac{2i\big(x^2+2x+5\big)\ln{\frac{x+1}{2}}}{(x+1)\sqrt{x^2-1}}\\
F^{\rm log}_{u\rho} &=  \frac{8i(x-1)\big(\ln{\frac{x+1}{2}}-1\big)}{(x+1)\sqrt{x^2-1}}\\
F^{\rm log}_{\rho\rho} &= \frac{4\big((x^2+10x-3)\ln{\frac{x+1}{2}}-4(x-1)\big)}{(x+1)^2(x-1)}
\end{align}
\end{subequations}

\paragraph{$h=0$:}
\begin{subequations}
\begin{align}
F^{\rm log}_{vv} &= -\sqrt{x^2-1}\,\ln{\frac{x+1}{2}} \\
F^{\rm log}_{uv} &= \frac{-(x+3)\,\ln{\frac{x+1}{2}}+x-1}{\sqrt{x^2-1}} \\
F^{\rm log}_{uu} &= -2\frac{\sqrt{x^2-1}}{x+1} \\
F^{\rm log}_{v\rho} &= 2i\,\frac{(x+3)\,\ln{\frac{x+1}{2}}-x^2-x+2}{x^2-1}\\
F^{\rm log}_{u\rho} &= -2i\,\frac{(3x+1)\,\ln{\frac{x+1}{2}}-x+1}{x^2-1} \\
F^{\rm log}_{\rho\rho} &= 4\,\frac{-(x^2+6x+1)\,\ln{\frac{x+1}{2}}+2(x^2-1)}{(x^2-1)^{3/2}}
\end{align}
\end{subequations}

\paragraph{$h=1$:}
\begin{subequations}
\begin{align}
F^{\rm log}_{vv} &= -(x-1)\,\ln{\frac{x+1}{2}} \\
F^{\rm log}_{uv} &= -4\,\frac{x-2\ln{\frac{x+1}{2}}-1}{x-1} \\
F^{\rm log}_{uu} &= 4\,\frac{x-2\ln{\frac{x+1}{2}}-1}{x-1} \\
F^{\rm log}_{v\rho} &= 2i\,\frac{(x^2-2x-7)\ln{\frac{x+1}{2}}-2(x-1)(x-3)}{(x-1)\sqrt{x^2-1}} \\
F^{\rm log}_{u\rho} &= 16i\,\frac{(x+1)\ln{\frac{x+1}{2}}-x+1}{(x-1)\sqrt{x^2-1}} \\
F^{\rm log}_{\rho\rho} &= 4\,\frac{(x^2+14x+9)\ln{\frac{x+1}{2}}-8x^2+4x+4}{(x-1)^2(x+1)}
\end{align}
\end{subequations}

\paragraph{$h=2$:}
\begin{subequations}
\label{eq:loggen}
\begin{align}
F^{\rm log}_{vv} &= -\frac{(x-1)^{3/2}\,\ln{\frac{x+1}{2}}}{\sqrt{x+1}} \\
F^{\rm log}_{uv} &= \frac{2(x+1)\ln^2\frac{x+1}{2}-3(x^2+14x+9)\ln\frac{x+1}{2}}{(x-1)\sqrt{x^2-1}} + \frac{2(7x+11)}{\sqrt{x^2-1}} \\
F^{\rm log}_{uu} &= \frac{-4(x+1)\ln^2\frac{x+1}{2}+2(7x^2+42x+23)\ln{\frac{x+1}{2}}}{(x-1)\sqrt{x^2-1}} + \frac{x^2-38x-35}{\sqrt{x^2-1}} \\
F^{\rm log}_{v\rho} &= -2i\,\frac{2(x+1)\,\ln^2{\frac{x+1}{2}}-(2x+1)(x^2-2x+25)\,\ln{\frac{x+1}{2}}}{(x-1)^2(x+1)}-\frac{2i(3x^2+8x+25)}{x^2-1} \\
F^{\rm log}_{u\rho} &=  2i\,\frac{2(x+1)(x+2)\,\ln^2{\frac{x+1}{2}}-(6x^3+47x^2+120x+43)\,\ln{\frac{x+1}{2}}}{(x-1)^2(x+1)} \nonumber \\ & \qquad 
+\frac{4i(14x^2+15x+25)}{x^2-1} \\
F^{\rm log}_{\rho\rho} &= 4\,\frac{4(x+1)^2\,\ln^2{\frac{x+1}{2}}-(5x^3+101x^2+135x+47)\,\ln{\frac{x+1}{2}}}{(x-1)^{5/2}(x+1)^{3/2}} + \frac{4(27x^2+82x+35)}{(x^2-1)^{3/2}}
\end{align}
\end{subequations}
For $h=2$ squared logarithmic (asymptotically subleading) terms appear in most of the components. This is a generic feature of all modes with $h\geq 2$. It is worthwhile emphasizing that all these modes are regular at $x=1$, despite of the appearance of $1/(x-1)$ factors in various components (e.g.~the factor $1/(x-1)^{5/2}$ in $F_{\rho\rho}^{\rm log}$ above). Using the algebraic relations \eqref{eq:logalgebraic} we can generate all regular non-normalizable logarithmic modes from the $(2,-1)$ mode above. 

\section{Correlation functions in Euclidean logarithmic CFT}\label{app:D}

LCFTs arise if there are fields with degenerate scaling dimensions having a Jordan block structure like in \eqref{eq:cg79}. In any LCFT one of these degenerate fields becomes a zero norm state coupled to a logarithmic partner \cite{Caux:1996kq}. For our purposes the following consideration is sufficient. Suppose the operators ${\cal O}^L$ and ${\cal O}^{\rm log}$ form a logarithmic pair with conformal weights $(2,0)$, where ${\cal O}^L$ corresponds to the zero norm state. Suppose that we have an additional operator ${\cal O}^R$ with conformal weights $(0,2)$ that commutes with ${\cal O}^L$ and ${\cal O}^{\rm log}$. With a standard normalization for ${\cal O}^{\rm log}$ and keeping only the most singular pieces in the coincidence limit the correlators between these operators take the form \cite{Gurarie:1993xq,Caux:1996kq,Flohr:2001zs,Gaberdiel:2001tr}
\begin{subequations}
\label{eq:2LCFT}
\begin{align}
& \langle{\cal O}^R(z,\bar z){\cal O}^R(0,0)\rangle = \frac{c_R}{2\bar z^4} \\
& \langle{\cal O}^L(z,\bar z){\cal O}^L(0,0)\rangle = 0 \\
& \langle{\cal O}^L(z,\bar z){\cal O}^R(0,0)\rangle = 0 \\
& \langle{\cal O}^R(z,\bar z){\cal O}^{\rm log}(0,0)\rangle = 0 \\
& \langle{\cal O}^L(z,\bar z){\cal O}^{\rm log}(0,0)\rangle = \frac{b}{2z^4} \\
& \langle{\cal O}^{\rm log}(z,\bar z){\cal O}^{\rm log}(0,0)\rangle = -\frac{b \ln{(m^2|z|^2)}}{z^4} 
\end{align} 
\end{subequations}
The scale $m$ is arbitrary and has no significance. It can be changed by a shift ${\cal O}^{\rm log}\to{\cal O}^{\rm log}+\ga\,{\cal O}^L$. Without loss of generality we set $m=1$. The right central charge $c_R$ and the ``new anomaly'' $b$ have a significant meaning and define key properties of the LCFT. The left central charge $c_L$ is assumed to vanish.

The 3-point correlators without logarithmic insertions are given by \cite{Ginsparg:1988ui}
\begin{subequations}
\label{eq:3LCFT0}
\begin{align}
& \langle{\cal O}^R(z,\bar z){\cal O}^R(z',\bar z'){\cal O}^R(0,0)\rangle = \frac{c_R}{\bar z^2\bar z^{\prime\,2}(\bar z-\bar z')^2} \\
& \langle{\cal O}^L(z,\bar z){\cal O}^R(z',\bar z'){\cal O}^R(0,0)\rangle = 0 \\
& \langle{\cal O}^L(z,\bar z){\cal O}^L(z',\bar z'){\cal O}^R(0,0)\rangle = 0 \\
& \langle{\cal O}^L(z,\bar z){\cal O}^L(z',\bar z'){\cal O}^L(0,0)\rangle = 0 
\end{align} 
\end{subequations}
The 3-point correlators with one logarithmic insertion are given by \cite{Kogan:2002mg}
\begin{subequations}
\label{eq:3LCFT1}
\begin{align}
& \langle{\cal O}^R(z,\bar z){\cal O}^R(z',\bar z'){\cal O}^{\rm log}(0,0)\rangle = 0 \\
& \langle{\cal O}^L(z,\bar z){\cal O}^L(z',\bar z'){\cal O}^{\rm log}(0,0)\rangle = \frac{b}{z^2{z'}^2(z-z')^2} \\
& \langle{\cal O}^L(z,\bar z){\cal O}^R(z',\bar z'){\cal O}^{\rm log}(0,0)\rangle = 0 
\end{align}
\end{subequations}
The 3-point correlators with at least two logarithmic insertion are given by \cite{Kogan:2002mg}
\begin{subequations}
\label{eq:3LCFT2}
\begin{align}
& \langle{\cal O}^R(z,\bar z){\cal O}^{\rm log}(z',\bar z'){\cal O}^{\rm log}(0,0)\rangle = 0 \\
& \langle{\cal O}^L(z,\bar z){\cal O}^{\rm log}(z',\bar z'){\cal O}^{\rm log}(0,0)\rangle = -\frac{2b\ln{|{z'}|^2}+\frac b2}{z^2{z'}^2(z-z')^2} \\
& \langle{\cal O}^{\rm log}(z,\bar z){\cal O}^{\rm log}(z',\bar z'){\cal O}^{\rm log}(0,0)\rangle = \frac{{\rm lengthy}}{z^2{z'}^2(z-z')^2}
\end{align} 
\end{subequations}
Again we have kept only the contributions that are most singular in the coincidence limit.


\begin{thebibliography}{10}

\bibitem{Deser:1982sv}
S.~Deser, {\it {Cosmological Topological Supergravity}},  in {\em Quantum
  Theory Of Gravity} (S.~M. Christensen, ed.), pp.~374--381.
\newblock Adam Hilger, Bristol, 1984.
\newblock \href{http://arXiv.org/abs/Print-82-0692 (Brandeis)}{{\tt
  Print-82-0692 (Brandeis)}}.

\bibitem{Deser:1982vy}
S.~Deser, R.~Jackiw and S.~Templeton, {\it Three-dimensional massive gauge
  theories},  {\em Phys. Rev. Lett.} {\bf 48} (1982) 975--978.

\bibitem{Deser:1982wh}
S.~Deser, R.~Jackiw and S.~Templeton, {\it Topologically massive gauge
  theories},  {\em Ann. Phys.} {\bf 140} (1982) 372--411.
%
{\em Erratum-ibid.} {\bf 185} (1988) 406.

\bibitem{Banados:1992wn}
M.~Banados, C.~Teitelboim and J.~Zanelli, {\it The black hole in
  three-dimensional space-time},  {\em Phys. Rev. Lett.} {\bf 69} (1992)
  1849--1851 [\href{http://arXiv.org/abs/hep-th/9204099}{{\tt
  hep-th/9204099}}].

\bibitem{Kraus:2005zm}
P.~Kraus and F.~Larsen, {\it {Holographic gravitational anomalies}},  {\em
  JHEP} {\bf 01} (2006) 022 [\href{http://arXiv.org/abs/hep-th/0508218}{{\tt
  hep-th/0508218}}].

\bibitem{Li:2008dq}
W.~Li, W.~Song and A.~Strominger, {\it {Chiral Gravity in Three Dimensions}},
  {\em JHEP} {\bf 04} (2008) 082 [\href{http://arXiv.org/abs/0801.4566}{{\tt
  0801.4566}}].

\bibitem{Gurarie:1993xq}
V.~Gurarie, {\it {Logarithmic operators in conformal field theory}},  {\em
  Nucl. Phys.} {\bf B410} (1993) 535--549
  [\href{http://arXiv.org/abs/hep-th/9303160}{{\tt hep-th/9303160}}].

\bibitem{Flohr:2001zs}
M.~Flohr, {\it {Bits and pieces in logarithmic conformal field theory}},  {\em
  Int. J. Mod. Phys.} {\bf A18} (2003) 4497--4592
  [\href{http://arXiv.org/abs/hep-th/0111228}{{\tt hep-th/0111228}}].

\bibitem{Gaberdiel:2001tr}
M.~R. Gaberdiel, {\it {An algebraic approach to logarithmic conformal field
  theory}},  {\em Int. J. Mod. Phys.} {\bf A18} (2003) 4593--4638
  [\href{http://arXiv.org/abs/hep-th/0111260}{{\tt hep-th/0111260}}].

\bibitem{Grumiller:2008qz}
D.~Grumiller and N.~Johansson, {\it {Instability in cosmological topologically
  massive gravity at the chiral point}},  {\em JHEP} {\bf 07} (2008) 134
  [\href{http://arXiv.org/abs/0805.2610}{{\tt 0805.2610}}].

\bibitem{Carlip:2008jk}
S.~Carlip, S.~Deser, A.~Waldron and D.~K. Wise, {\it {Cosmological
  Topologically Massive Gravitons and Photons}},  {\em Class. Quant. Grav.}
  {\bf 26} (2009) 075008 [\href{http://arXiv.org/abs/0803.3998}{{\tt
  0803.3998}}].

\bibitem{Grumiller:2008pr}
D.~Grumiller, R.~Jackiw and N.~Johansson, {\it {Canonical analysis of
  cosmological topologically massive gravity at the chiral point}},  in {\em
  {Fundamental Interactions - A Memorial Volume for Wolfgang Kummer}}.
\newblock World Scientific, 2009.
\newblock \href{http://arXiv.org/abs/0806.4185}{{\tt 0806.4185}}.

\bibitem{Carlip:2008qh}
S.~Carlip, {\it {The Constraint Algebra of Topologically Massive AdS Gravity}},
   {\em JHEP} {\bf 10} (2008) 078 [\href{http://arXiv.org/abs/0807.4152}{{\tt
  0807.4152}}].

\bibitem{Grumiller:2008es}
D.~Grumiller and N.~Johansson, {\it {Consistent boundary conditions for
  cosmological topologically massive gravity at the chiral point}},  {\em Int.
  J. Mod. Phys.} {\bf D17} (2009) 2367--2372
  [\href{http://arXiv.org/abs/0808.2575}{{\tt 0808.2575}}].

\bibitem{Henneaux:2009pw}
M.~Henneaux, C.~Martinez and R.~Troncoso, {\it {Asymptotically anti-de Sitter
  spacetimes in topologically massive gravity}},  {\em Phys. Rev.} {\bf D79}
  (2009) 081502R [\href{http://arXiv.org/abs/0901.2874}{{\tt 0901.2874}}].

\bibitem{Maloney:2009ck}
A.~Maloney, W.~Song and A.~Strominger, {\it {Chiral Gravity, Log Gravity and
  Extremal CFT}},  \href{http://arXiv.org/abs/0903.4573}{{\tt 0903.4573}}.

\bibitem{Skenderis:2009nt}
K.~Skenderis, M.~Taylor and B.~C. van Rees, {\it {Topologically Massive Gravity
  and the AdS/CFT Correspondence}},  \href{http://arXiv.org/abs/0906.4926}{{\tt
  0906.4926}}.

\bibitem{Hotta:2008yq}
K.~Hotta, Y.~Hyakutake, T.~Kubota and H.~Tanida, {\it {Brown-Henneaux's
  Canonical Approach to Topologically Massive Gravity}},  {\em JHEP} {\bf 07}
  (2008) 066 [\href{http://arXiv.org/abs/0805.2005}{{\tt 0805.2005}}].

\bibitem{Li:2008yz}
W.~Li, W.~Song and A.~Strominger, {\it {Comment on 'Cosmological Topological
  Massive Gravitons and Photons'}},  \href{http://arXiv.org/abs/0805.3101}{{\tt
  0805.3101}}.

\bibitem{Park:2008yy}
M.-i. Park, {\it {Constraint Dynamics and Gravitons in Three Dimensions}},
  {\em JHEP} {\bf 09} (2008) 084 [\href{http://arXiv.org/abs/0805.4328}{{\tt
  0805.4328}}].

\bibitem{Sachs:2008gt}
I.~Sachs and S.~N. Solodukhin, {\it {Quasi-Normal Modes in Topologically
  Massive Gravity}},  {\em JHEP} {\bf 08} (2008) 003
  [\href{http://arXiv.org/abs/0806.1788}{{\tt 0806.1788}}].

\bibitem{Lowe:2008ye}
D.~A. Lowe and S.~Roy, {\it {Chiral geometries of (2+1)-d AdS gravity}},  {\em
  Phys. Lett.} {\bf B668} (2008) 159--162
  [\href{http://arXiv.org/abs/0806.3070}{{\tt 0806.3070}}].

\bibitem{Myung:2008ey}
Y.~S. Myung, H.~W. Lee and Y.-W. Kim, {\it {Entropy of black holes in
  topologically massive gravity}},  \href{http://arXiv.org/abs/0806.3794}{{\tt
  0806.3794}}.

\bibitem{Carlip:2008eq}
S.~Carlip, S.~Deser, A.~Waldron and D.~K. Wise, {\it {Topologically Massive AdS
  Gravity}},  {\em Phys. Lett.} {\bf B666} (2008) 272--276
  [\href{http://arXiv.org/abs/0807.0486}{{\tt 0807.0486}}].

\bibitem{Lee:2008gta}
H.~W. Lee, Y.-W. Kim and Y.~S. Myung, {\it {Quasinormal modes for topologically
  massive black hole}},  \href{http://arXiv.org/abs/0807.1371}{{\tt
  0807.1371}}.

\bibitem{Sachs:2008yi}
I.~Sachs, {\it {Quasi-Normal Modes for Logarithmic Conformal Field Theory}},
  {\em JHEP} {\bf 09} (2008) 073 [\href{http://arXiv.org/abs/0807.1844}{{\tt
  0807.1844}}].

\bibitem{Gibbons:2008vi}
G.~W. Gibbons, C.~N. Pope and E.~Sezgin, {\it {The General Supersymmetric
  Solution of Topologically Massive Supergravity}},  {\em Class. Quant. Grav.}
  {\bf 25} (2008) 205005 [\href{http://arXiv.org/abs/0807.2613}{{\tt
  0807.2613}}].

\bibitem{Anninos:2008fx}
D.~Anninos, W.~Li, M.~Padi, W.~Song and A.~Strominger, {\it {Warped AdS$_3$
  Black Holes}},  {\em JHEP} {\bf 03} (2009) 130
  [\href{http://arXiv.org/abs/0807.3040}{{\tt 0807.3040}}].

\bibitem{Giribet:2008bw}
G.~Giribet, M.~Kleban and M.~Porrati, {\it {Topologically Massive Gravity at
  the Chiral Point is Not Unitary}},  {\em JHEP} {\bf 10} (2008) 045
  [\href{http://arXiv.org/abs/0807.4703}{{\tt 0807.4703}}].

\bibitem{Strominger:2008dp}
A.~Strominger, {\it {A Simple Proof of the Chiral Gravity Conjecture}},
  \href{http://arXiv.org/abs/0808.0506}{{\tt 0808.0506}}.

\bibitem{Compere:2008cv}
G.~Compere and S.~Detournay, {\it {Semi-classical central charge in
  topologically massive gravity}},  {\em Class. Quant. Grav.} {\bf 26} (2009)
  012001 [\href{http://arXiv.org/abs/0808.1911}{{\tt 0808.1911}}].

\bibitem{Myung:2008dm}
Y.~S. Myung, {\it {Logarithmic conformal field theory approach to topologically
  massive gravity}},  {\em Phys. Lett.} {\bf B670} (2008) 220--223
  [\href{http://arXiv.org/abs/0808.1942}{{\tt 0808.1942}}].

\bibitem{deHaro:2008gp}
S.~de~Haro, {\it {Dual Gravitons in AdS4/CFT3 and the Holographic Cotton
  Tensor}},  {\em JHEP} {\bf 01} (2009) 042
  [\href{http://arXiv.org/abs/0808.2054}{{\tt 0808.2054}}].

\bibitem{Stevens:2008hv}
K.~A. Stevens, K.~Schleich and D.~M. Witt, {\it {Non-existence of
  Asymptotically Flat Geons in 2+1 Gravity}},  {\em Class. Quant. Grav.} {\bf
  26} (2009) 075012 [\href{http://arXiv.org/abs/0809.3022}{{\tt 0809.3022}}].

\bibitem{Deser:2008rm}
S.~Deser, {\it {Distended Topologically Massive Electrodynamics}},  in {\em
  {Fundamental Interactions - A Memorial Volume for Wolfgang Kummer}}.
\newblock World Scientific, 2009.
\newblock \href{http://arXiv.org/abs/0810.5384}{{\tt 0810.5384}}.

\bibitem{Hotta:2008xt}
K.~Hotta, Y.~Hyakutake, T.~Kubota, T.~Nishinaka and H.~Tanida, {\it {The
  CFT-interpolating Black Hole in Three Dimensions}},  {\em JHEP} {\bf 01}
  (2009) 010 [\href{http://arXiv.org/abs/0811.0910}{{\tt 0811.0910}}].

\bibitem{Quevedo:2008ry}
H.~Quevedo and A.~Sanchez, {\it {Geometric description of BTZ black holes
  thermodynamics}},  {\em Phys. Rev.} {\bf D79} (2009) 024012
  [\href{http://arXiv.org/abs/0811.2524}{{\tt 0811.2524}}].

\bibitem{Oh:2008tc}
J.~J. Oh and W.~Kim, {\it {Absorption Cross Section in Warped AdS$_3$ Black
  Hole}},  {\em JHEP} {\bf 01} (2009) 067
  [\href{http://arXiv.org/abs/0811.2632}{{\tt 0811.2632}}].

\bibitem{Garbarz:2008qn}
A.~Garbarz, G.~Giribet and Y.~Vasquez, {\it {Asymptotically AdS$_3$ Solutions
  to Topologically Massive Gravity at Special Values of the Coupling
  Constants}},  {\em Phys. Rev.} {\bf D79} (2009) 044036
  [\href{http://arXiv.org/abs/0811.4464}{{\tt 0811.4464}}].

\bibitem{Kim:2008bf}
W.~Kim and E.~J. Son, {\it {Thermodynamics of warped AdS$_3$ black hole in the
  brick wall method}},  {\em Phys. Lett.} {\bf B673} (2009) 90--94
  [\href{http://arXiv.org/abs/0812.0876}{{\tt 0812.0876}}].

\bibitem{Mann:2008rx}
R.~B. Mann, J.~J. Oh and M.-I. Park, {\it {The Role of Angular Momentum and
  Cosmic Censorship in the (2+1)-Dimensional Rotating Shell Collapse}},  {\em
  Phys. Rev.} {\bf D79} (2009) 064005
  [\href{http://arXiv.org/abs/0812.2297}{{\tt 0812.2297}}].

\bibitem{Blagojevic:2008bn}
M.~Blagojevic and B.~Cvetkovic, {\it {Canonical structure of topologically
  massive gravity with a cosmological constant}},  {\em JHEP} {\bf 05} (2009)
  073 [\href{http://arXiv.org/abs/0812.4742}{{\tt 0812.4742}}].

\bibitem{Nam:2009dd}
S.~Nam and J.-D. Park, {\it {Hawking radiation from covariant anomalies in 2+1
  dimensional black holes}},  {\em Class. Quant. Grav.} {\bf 26} (2009) 145015
  [\href{http://arXiv.org/abs/0902.0982}{{\tt 0902.0982}}].

\bibitem{Hellerman:2009bu}
S.~Hellerman, {\it {A Universal Inequality for CFT and Quantum Gravity}},
  \href{http://arXiv.org/abs/0902.2790}{{\tt 0902.2790}}.

\bibitem{Sezgin:2009dj}
E.~Sezgin and Y.~Tanii, {\it {Witten-Nester Energy in Topologically Massive
  Gravity}},  \href{http://arXiv.org/abs/0903.3779}{{\tt 0903.3779}}.

\bibitem{Anninos:2009zi}
D.~Anninos, M.~Esole and M.~Guica, {\it {Stability of warped AdS3 vacua of
  topologically massive gravity}},  \href{http://arXiv.org/abs/0905.2612}{{\tt
  0905.2612}}.

\bibitem{Compere:2009zj}
G.~Compere and S.~Detournay, {\it {Boundary conditions for spacelike and
  timelike warped AdS$_3$ spaces in topologically massive gravity}},  {\em
  JHEP} {\bf 08} (2009) 092 [\href{http://arXiv.org/abs/0906.1243}{{\tt
  0906.1243}}].

\bibitem{Hotta:2009zn}
K.~Hotta, Y.~Hyakutake, T.~Kubota, T.~Nishinaka and H.~Tanida, {\it {Left-Right
  Asymmetric Holographic RG Flow with Gravitational Chern-Simons Term}},  {\em
  Phys. Lett.} {\bf B680} (2009) 279--285
  [\href{http://arXiv.org/abs/0906.1255}{{\tt 0906.1255}}].

\bibitem{Anninos:2009jt}
D.~Anninos, {\it {Sailing from Warped AdS$_3$ to Warped dS$_3$ in Topologically
  Massive Gravity}},  \href{http://arXiv.org/abs/0906.1819}{{\tt 0906.1819}}.

\bibitem{Carlip:2009ey}
S.~Carlip, {\it {Chiral Topologically Massive Gravity and Extremal B-F
  Scalars}},  {\em JHEP} {\bf 09} (2009) 083
  [\href{http://arXiv.org/abs/0906.2384}{{\tt 0906.2384}}].

\bibitem{Chow:2009km}
D.~D.~K. Chow, C.~N. Pope and E.~Sezgin, {\it {Exact solutions of topologically
  massive gravity}},  \href{http://arXiv.org/abs/0906.3559}{{\tt 0906.3559}}.

\bibitem{Becker:2009mk}
M.~Becker, P.~Bruillard and S.~Downes, {\it {Chiral Supergravity}},
  \href{http://arXiv.org/abs/0906.4822}{{\tt 0906.4822}}.

\bibitem{Blagojevic:2009ek}
M.~Blagojevic and B.~Cvetkovic, {\it {Asymptotic structure of topologically
  massive gravity in spacelike stretched AdS sector}},  {\em JHEP} {\bf 09}
  (2009) 006 [\href{http://arXiv.org/abs/0907.0950}{{\tt 0907.0950}}].

\bibitem{Vasquez:2009mk}
Y.~Vasquez, {\it {Charged Black Holes in Three Dimensional Einstein Theory with
  Torsion and Chern Simons Terms}},  \href{http://arXiv.org/abs/0907.4165}{{\tt
  0907.4165}}.

\bibitem{Duncan:2009sq}
J.~F. Duncan and I.~B. Frenkel, {\it {Rademacher sums, Moonshine and Gravity}},
   \href{http://arXiv.org/abs/0907.4529}{{\tt 0907.4529}}.

\bibitem{Andrade:2009ae}
T.~Andrade and D.~Marolf, {\it {No chiral truncation of quantum log gravity?}},
   \href{http://arXiv.org/abs/0909.0727}{{\tt 0909.0727}}.

\bibitem{Miskovic:2009kr}
O.~Miskovic and R.~Olea, {\it {Background-independent charges in Topologically
  Massive Gravity}},  \href{http://arXiv.org/abs/0909.2275}{{\tt 0909.2275}}.

\bibitem{Skenderis:2009kd}
K.~Skenderis, M.~Taylor and B.~C. van Rees, {\it {AdS boundary conditions and
  the Topologically Massive Gravity/CFT correspondence}},
  \href{http://arXiv.org/abs/0909.5617}{{\tt 0909.5617}}.

\bibitem{Ertl:2009ch}
S.~Ertl, D.~Grumiller and N.~Johansson, {\it {Erratum to `Instability in
  cosmological topologically massive gravity at the chiral point',
  arXiv:0805.2610}},  \href{http://arXiv.org/abs/0910.1706}{{\tt 0910.1706}}.

\bibitem{Afshar:2009rg}
H.~R. Afshar, M.~Alishahiha and A.~Naseh, {\it {On three dimensional
  bigravity}},  \href{http://arXiv.org/abs/0910.4350}{{\tt 0910.4350}}.

\bibitem{Ghezelbash:1998rj}
A.~M. Ghezelbash, M.~Khorrami and A.~Aghamohammadi, {\it {Logarithmic conformal
  field theories and AdS correspondence}},  {\em Int. J. Mod. Phys.} {\bf A14}
  (1999) 2581--2592 [\href{http://arXiv.org/abs/hep-th/9807034}{{\tt
  hep-th/9807034}}].

\bibitem{Myung:1999nd}
Y.~S. Myung and H.~W. Lee, {\it {Gauge bosons and the AdS(3)/LCFT(2)
  correspondence}},  {\em JHEP} {\bf 10} (1999) 009
  [\href{http://arXiv.org/abs/hep-th/9904056}{{\tt hep-th/9904056}}].

\bibitem{Kogan:1999bn}
I.~I. Kogan, {\it {Singletons and logarithmic CFT in AdS/CFT correspondence}},
  {\em Phys. Lett.} {\bf B458} (1999) 66--72
  [\href{http://arXiv.org/abs/hep-th/9903162}{{\tt hep-th/9903162}}].

\bibitem{Lewis:1999qv}
A.~Lewis, {\it {Logarithmic operators in AdS(3)/CFT(2)}},  {\em Phys. Lett.}
  {\bf B480} (2000) 348--354 [\href{http://arXiv.org/abs/hep-th/9911163}{{\tt
  hep-th/9911163}}].

\bibitem{MoghimiAraghi:2004ds}
S.~Moghimi-Araghi, S.~Rouhani and M.~Saadat, {\it {Correlation functions and
  AdS/LCFT correspondence}},  {\em Nucl. Phys.} {\bf B696} (2004) 492--502
  [\href{http://arXiv.org/abs/hep-th/0403150}{{\tt hep-th/0403150}}].

\bibitem{Ginsparg:1988ui}
P.~H. Ginsparg, {\it Applied conformal field theory},
  \href{http://arXiv.org/abs/hep-th/9108028}{{\tt hep-th/9108028}}.

\bibitem{Liu:1998bu}
H.~Liu and A.~A. Tseytlin, {\it {D = 4 super Yang-Mills, D = 5 gauged
  supergravity, and D = 4 conformal supergravity}},  {\em Nucl. Phys.} {\bf
  B533} (1998) 88--108 [\href{http://arXiv.org/abs/hep-th/9804083}{{\tt
  hep-th/9804083}}].

\bibitem{Arutyunov:1999nw}
G.~Arutyunov and S.~Frolov, {\it {Three-point Green function of the
  stress-energy tensor in the AdS/CFT correspondence}},  {\em Phys. Rev.} {\bf
  D60} (1999) 026004 [\href{http://arXiv.org/abs/hep-th/9901121}{{\tt
  hep-th/9901121}}].

\bibitem{Aharony:1999ti}
O.~Aharony, S.~S. Gubser, J.~M. Maldacena, H.~Ooguri and Y.~Oz, {\it {Large N
  field theories, string theory and gravity}},  {\em Phys. Rept.} {\bf 323}
  (2000) 183--386 [\href{http://arXiv.org/abs/hep-th/9905111}{{\tt
  hep-th/9905111}}].

\bibitem{Bergshoeff:2009hq}
E.~A. Bergshoeff, O.~Hohm and P.~K. Townsend, {\it {Massive Gravity in Three
  Dimensions}},  {\em Phys. Rev. Lett.} {\bf 102} (2009) 201301
  [\href{http://arXiv.org/abs/0901.1766}{{\tt 0901.1766}}].

\bibitem{Boyanovsky:1989jb}
D.~Boyanovsky and C.~M. Naon, {\it An introduction to conformal invariance in
  quantum field theory and statistical mechanics},  {\em Riv. Nuovo Cim.} {\bf
  13N2} (1990) 1--76.

\bibitem{Brown:1986nw}
J.~D. Brown and M.~Henneaux, {\it {Central Charges in the Canonical Realization
  of Asymptotic Symmetries: An Example from Three-Dimensional Gravity}},  {\em
  Commun. Math. Phys.} {\bf 104} (1986) 207--226.

\bibitem{Gurarie:1999yx}
V.~Gurarie and A.~W.~W. Ludwig, {\it {Conformal algebras of 2D disordered
  systems}},  {\em J. Phys.} {\bf A35} (2002) L377--L384
  [\href{http://arXiv.org/abs/cond-mat/9911392}{{\tt cond-mat/9911392}}].

\bibitem{Kogan:2001ku}
I.~I. Kogan and A.~Nichols, {\it {Stress energy tensor in LCFT and the
  logarithmic Sugawara construction}},  {\em JHEP} {\bf 01} (2002) 029
  [\href{http://arXiv.org/abs/hep-th/0112008}{{\tt hep-th/0112008}}].

\bibitem{Kogan:2002mg}
I.~I. Kogan and A.~Nichols, {\it {Stress energy tensor in c = 0 logarithmic
  conformal field theory}},  \href{http://arXiv.org/abs/hep-th/0203207}{{\tt
  hep-th/0203207}}.

\bibitem{Susskind:1998dq}
L.~Susskind and E.~Witten, {\it {The holographic bound in anti-de Sitter
  space}},  \href{http://arXiv.org/abs/hep-th/9805114}{{\tt hep-th/9805114}}.

\bibitem{grtensor}
``{GRTensorII}.''
\newblock This is a package which runs within Maple but distinct from packages
  distributed with Maple. It is distributed freely on the World-Wide-Web from
  the address: {\tt http://grtensor.org}.

\bibitem{Liu:2009kc}
Y.~Liu and Y.-W. Sun, {\it {Consistent Boundary Conditions for New Massive
  Gravity in $AdS_3$}},  {\em JHEP} {\bf 05} (2009) 039
  [\href{http://arXiv.org/abs/0903.2933}{{\tt 0903.2933}}].

\bibitem{Andringa:2009yc}
R.~Andringa {\em et.~al.}, {\it {Massive 3D Supergravity}},
  \href{http://arXiv.org/abs/0907.4658}{{\tt 0907.4658}}.

\bibitem{Caux:1996kq}
J.~S. Caux, I.~Kogan, A.~Lewis and A.~M. Tsvelik, {\it {Logarithmic operators
  and dynamical extension of the symmetry group in the bosonic SU(2)0 and SUSY
  SU(2)2 WZNW models}},  {\em Nucl. Phys.} {\bf B489} (1997) 469--484
  [\href{http://arXiv.org/abs/hep-th/9606138}{{\tt hep-th/9606138}}].

\bibitem{Khorrami:1998kw}
M.~Khorrami, A.~Aghamohammadi and A.~M. Ghezelbash, {\it {Logarithmic N = 1
  superconformal field theories}},  {\em Phys. Lett.} {\bf B439} (1998)
  283--288 [\href{http://arXiv.org/abs/hep-th/9803071}{{\tt hep-th/9803071}}].

\bibitem{Mavromatos:2001iz}
N.~E. Mavromatos and R.~J. Szabo, {\it {D-brane dynamics and logarithmic
  superconformal algebras}},  {\em JHEP} {\bf 10} (2001) 027
  [\href{http://arXiv.org/abs/hep-th/0106259}{{\tt hep-th/0106259}}].

\bibitem{Witten:2007kt}
E.~Witten, {\it {Three-Dimensional Gravity Revisited}},
  \href{http://arXiv.org/abs/0706.3359}{{\tt 0706.3359}}.

\bibitem{Maloney:2007ud}
A.~Maloney and E.~Witten, {\it {Quantum Gravity Partition Functions in Three
  Dimensions}},  \href{http://arXiv.org/abs/0712.0155}{{\tt 0712.0155}}.

\bibitem{Gaberdiel:2008xb}
M.~R. Gaberdiel, S.~Gukov, C.~A. Keller, G.~W. Moore and H.~Ooguri, {\it
  {Extremal N=(2,2) 2D Conformal Field Theories and Constraints of
  Modularity}},  \href{http://arXiv.org/abs/0805.4216}{{\tt 0805.4216}}.

\bibitem{Bilal:1995rc}
A.~Bilal and I.~I. Kogan, {\it {On gravitational dressing of 2-D field theories
  in chiral gauge}},  {\em Nucl. Phys.} {\bf B449} (1995) 569--588
  [\href{http://arXiv.org/abs/hep-th/9503209}{{\tt hep-th/9503209}}].

\bibitem{Flohr:1996ik}
M.~A.~I. Flohr, {\it {Two-dimensional turbulence: A novel approach via
  logarithmic conformal field theory}},  {\em Nucl. Phys.} {\bf B482} (1996)
  567--578 [\href{http://arXiv.org/abs/hep-th/9606130}{{\tt hep-th/9606130}}].

\bibitem{Gurarie:1997dw}
V.~Gurarie, M.~Flohr and C.~Nayak, {\it {The Haldane-Rezayi quantum Hall state
  and conformal field theory}},  {\em Nucl. Phys.} {\bf B498} (1997) 513--538
  [\href{http://arXiv.org/abs/cond-mat/9701212}{{\tt cond-mat/9701212}}].

\bibitem{Cappelli:1998ma}
A.~Cappelli, L.~S. Georgiev and I.~T. Todorov, {\it {A unified conformal field
  theory description of paired quantum Hall states}},  {\em Commun. Math.
  Phys.} {\bf 205} (1999) 657--689
  [\href{http://arXiv.org/abs/hep-th/9810105}{{\tt hep-th/9810105}}].

\bibitem{Read:1999fn}
N.~Read and D.~Green, {\it {Paired states of fermions in two-dimensions with
  breaking of parity and time reversal symmetries, and the fractional quantum
  Hall effect}},  {\em Phys. Rev.} {\bf B61} (2000) 10267
  [\href{http://arXiv.org/abs/cond-mat/9906453}{{\tt cond-mat/9906453}}].

\bibitem{Ivashkevich:1998na}
E.~V. Ivashkevich, {\it {Correlation functions of dense polymers and c = -2
  conformal field theory}},  {\em J. Phys.} {\bf A32} (1999) 1691--1699
  [\href{http://arXiv.org/abs/cond-mat/9801183}{{\tt cond-mat/9801183}}].

\bibitem{Cardy:1999zp}
J.~L. Cardy, {\it {Logarithmic Correlations in Quenched Random Magnets and
  Polymers}},  \href{http://arXiv.org/abs/cond-mat/9911024}{{\tt
  cond-mat/9911024}}.

\bibitem{Skenderis:2002wp}
K.~Skenderis, {\it Lecture notes on holographic renormalization},  {\em Class.
  Quant. Grav.} {\bf 19} (2002) 5849--5876
  [\href{http://arXiv.org/abs/hep-th/0209067}{{\tt hep-th/0209067}}].

\bibitem{Erdelyi}
A.~Erd{\'e}lyi, ed., {\em Higher Transcendental Functions}, vol.~I.
\newblock McGraw-Hill, 1953.

\bibitem{Watson:1918}
G.~N. Watson, {\it Asymptotic expansions of hypergeometric functions},  {\em
  Trans. Cambridge Philos. Soc.} {\bf 22} (1918) 277--308.

\end{thebibliography}

\providecommand{\href}[2]{#2}\begingroup\raggedright\endgroup

\end{document}